\documentstyle[epsfig,11pt]{article}
\def\be{\begin{eqnarray}}
\def\ee{\end{eqnarray}}

\def\half{{\textstyle \frac{1}{2}}}

\def\roughly#1{\mathrel{\raise.3ex\hbox{$#1$\kern-.75em%
\lower1ex\hbox{$\sim$}}}}
\def\lsim{\roughly<}

\def\bfpi{{\mbox{\boldmath $\pi$}}}

\def\bfrho{{\mbox{\boldmath $\rho$}}}
\def\bfalpha{{\mbox{\boldmath $\alpha$}}}
\def\bfeps{{\mbox{\boldmath $\epsilon$}}}
\def\bfSigma{{\mbox{\boldmath $\Sigma$}}}
\def\bfPi{{\mbox{\boldmath $\Pi$}}}
\def\bfGamma{{\mbox{\boldmath $\Gamma$}}}
\def\Tr{{\rm Tr}\,}

\begin{document}

\renewcommand{\thefootnote}{\arabic{footnote}}
\setcounter{footnote}{0}

\vskip 0.4cm
\hfill {\bf FZJ-IKP(TH)-1999-31}

\hfill {\today}
\vskip 1cm

\begin{center}
{\LARGE\bf Generalized Mesons in Dense QCD}

\date{\today}

\vskip 1cm
Mannque Rho$^{a,b}$\footnote{E-mail: rho@spht.saclay.cea.fr},
Edward Shuryak$^b$\footnote{E-mail: shuryak@nuclear.physics.sunysb.edu},
Andreas Wirzba$^{c}$\footnote{E-mail: a.wirzba@fz-juelich.de}
and Ismail  Zahed$^{b}$\footnote{E-mail: zahed@zahed.physics.sunysb.edu}

\end{center}

\vskip 0.5cm

\begin{center}

$^a$
{\it Service de Physique Th\'eorique, CE Saclay,
91191 Gif-sur-Yvette, France}

$^b$
{\it Department of Physics and Astronomy,
SUNY-Stony-Brook, NY 11794, U.\,S.\,A.}

$^c$
{\it FZ J{\"u}lich, Institut f\"ur Kernphysik (Theorie),
D-52425   J{\"u}lich, Germany}

\end{center}

\vskip 0.5cm

\begin{abstract}
QCD superconductors in the color-flavor-locked (CFL) phase
support excitations (generalized mesons)
that can be described as pairs of particles or holes
(rather than particle-hole)  around a
gapped Fermi surface. In weak coupling and to leading logarithm
accuracy the scalar and pseudoscalar excitations are massless and 
the vector and axial-vector excitations are massive and degenerate.
The massless scalar excitations are combined with the longitudinal
gluons leading to the Meissner effect in the CFL phase. 
The  mass of the composite vector excitations is close to twice
the gap in weak coupling, but goes asymptotically to
zero with increasing coupling thereby realizing Georgi's vector limit
in cold and dense matter. We discuss the possible mixing of the composite
scalar and vector excitations with the gluons, their possible coupling to the
modified photons and their decay into light pseudoscalars in the CFL phase. 
The issue of hidden gauge-symmetry in the QCD superconductor is critically
examined. 
The physical implications of our results on soft dilepton and neutrino
emission 
in cold and dense matter are briefly discussed.

\end{abstract}
\newpage

\renewcommand{\thefootnote}{\#\arabic{footnote}}
\setcounter{footnote}{0}

%%%%%%%%%%%%%%%%%%%%%%%%%%%%%%%%%%%%%%%%%%%%%%%%%%%%%%%%%%%%%%%%%%%%%%%%%%%
%%%%%%%%%%%%%%%%%%%%%%%%%% Introduction  %%%%%%%%%%%%%%%%%%%%%%%%%%%%%%%%%%
%%%%%%%%%%%%%%%%%%%%%%%%%%%%%%%%%%%%%%%%%%%%%%%%%%%%%%%%%%%%%%%%%%%%%%%%%%%

\centerline{\bf 1. Introduction}
\vskip 1cm

QCD at large quark density has been discussed in the literature
since the late seventies~\cite{Seventies,LOVE}, but
it has generated an intense activity   especially in the last three
years~\cite{Revival,ALL98,ALL99}, in light of the fact that the
ground state may exhibit a robust superconducting phase, with
novel and nonperturbative phenomena. At high quark chemical potential, 
these phenomena  are accessible by weak coupling analysis.
The QCD superconductor for a number of flavors $N_f\geq 3$ and
a set of degenerate quark masses, breaks color and
flavor symmetry spontaneously, with the excitation of
light Goldstone modes.

Some properties of these light
excitations that we may call ``superpions''~\footnote{We
will use the phrases
``superpion'', ``supervector'' etc.\ for denoting the generalized
mesons in
the zero-size approximation which is behind the ``superqualiton''
point of view of \cite{HRZ}.}
have been addressed recently using effective
Lagrangians~\cite{HRZ,GATTO,SONSTE}. In the latter the
finite size of the pairs is usually ignored, allowing
for a description in terms of point-like excitations
as originally suggested in~\cite{HRZ}. However, in weak coupling,
this approximation need not be invoked since a full
analysis with finite size taken into account is feasible
to a leading-logarithm accuracy.
A direct analysis of the light Goldstone modes in weak
coupling {\em without} using the zero-size approximation,
has been performed  recently~\cite{RWZ}.
It allows a microscopic calculation of the pion
form factor, decay constant and mass in leading
logarithm approximation in the color-flavor-locked (CFL) phase. The
self-generated form factors provide a natural cutoff
to regulate the effective calculations at the Fermi surface.

In this paper, we will pursue the microscopic analysis
for the generalized
scalar, vector and axial-vector mesons viewed as composites of pairs
of quasiparticles or quasiholes in the CFL phase. Throughout, we will
only discuss the octet phase and its associated set 
of generalized mesons. The axial $SU(3)$ singlet is
still expected to be split by the color-flavor triangle-anomaly~\cite{HRZ} 
present in the CFL phase. This issue will be addressed elsewhere.
In section~2, we discuss
the general features of the QCD superconductor in the CFL phase.
In section~3, we show that in the CFL phase both
the pseudoscalar and scalar excitations are massless.
The former are true Goldstone modes, while the latter
are would-be Goldstone modes that combine with the
longitudinal gluons as discussed in section~4. 
In section~5, we show that bound vector
excitations of particles or holes exist in the CFL phase,
and derive an explicit relation for their form factor and
mass. In section~6, we discuss their coupling to currents.
To leading logarithm
accuracy the octet of vectors are degenerate with the
octet of axial-vectors, and decouple from the Noether
currents. In section~7, we show that the composite
vectors do not decay to pions in leading logarithm
accuracy, contrary to their analogues in the QCD vacuum.
In section~8, we show that the composite vectors
decouple from the gluons in the CFL phase as well.
In section~9, we show that issues such as vector
dominance and gauge universality do not {\em immediately} apply to the
generalized mesons of finite size. A hidden local
symmetry can be revealed only for zero size pairs,
which may  not constitute a good approximation
for magnetically bound pairs.
Our conclusions are given in section~10. Some of the
calculations are relegated to the Appendix.

\vskip 1.5cm
\centerline{\bf 2. QCD Superconductor in the CFL Phase}
\vskip 1cm

As shown in~\cite{Revival}, in the CFL phase the quarks have a nonzero
gap. Their propagation is
given in the Nambu-Gorkov formalism by a matrix written in
terms of the two-component Nambu-Gorkov
field ${\bf\Psi}=(\psi, \psi_C)$, where
$\psi$ refers to
quarks  and
$\psi_C(q)=C\bar \psi^T(-q)$ to charge
conjugated quarks,
respectively~\footnote{Here $\bar\psi^T$ is the
transposed and conjugated field and $C\equiv i \gamma^2\gamma^0$.}.
For large $\mu$, the antiparticles  decouple, and the propagator
${\bf S}(q)$ in the chiral limit reads~\cite{PisarskiSuperfluid,PISARSKI}
\begin{eqnarray}
  {\bf S}(q)&\approx& \left(\begin{array}{rr}
     \gamma^0\,(q_0+q_{||}) \Lambda^{-}({\bf q})
    &    -{\bf M}^\dagger\,G^\ast(q)\Lambda^{+}({\bf q})
  \\
     {\bf M}\, G(q)\Lambda^{-}({\bf q})   & \gamma^0\,(q_0-q_{||})\
\Lambda^{+}({\bf q})
\end{array}\right)\,\,\frac{1}{q_0^2-\epsilon_q^2}\; , \nonumber\\
\label{PropPart}
 \end{eqnarray}
where $q_{||}\approx(|{\bf q}|-\mu)$ is the particle or hole momentum
in the direction of the Fermi momentum, such that the particle/hole
energies read
$\epsilon_q \approx \mp\sqrt{q_{||}^2+|G(q)|^2}$ in terms of the gap
function
$G(q)$.
The operators $\Lambda^{\pm}({\bf q})=
\half(1\pm \bfalpha\cdot\hat{{\bf q}})$ are the
positive and negative energy projectors~\footnote{Note that
$\gamma^0\Lambda^{\pm}({\bf q})=\Lambda^{\mp}({\bf q})\gamma^0$,
$\gamma^5\Lambda^{\pm}({\bf q})=\Lambda^{\pm}({\bf q})\gamma^5$
and $\bfalpha\cdot \hat{\bf q} \,\Lambda^{\pm}({\bf q})= \pm \,
\Lambda^{\pm}({\bf q})$.}.  In the CFL phase
${\bf M}= \epsilon_f^a\epsilon_c^a\,\gamma_5= {\bf M}^\dagger$ with
two  antisymmetric tensors $(\epsilon^a)^{bc}=\epsilon^{abc}$,
$a,b,c\in
\{1,2,3\}$, in
flavor (f) and color (c) space, whereas
the charge conjugation operator $C$ is already
incorporated in the definition of the Nambu-Gorkov
field ${\bf\Psi}$. 
The effects of the current quark masses on the quark propagator are
involved in the QCD superconductor. In perturbation theory, we
have to first order in the current quark mass~\cite{RWZ}
\begin{equation}
 \Delta {\bf S}(q) \approx 
%\frac{1}{2\mu\left(q_0^2 - \epsilon_q^2\right)}
    \left( \begin{array}{cc} 
     \frac{ m_{}}{2\mu}\, \frac{q_0+q_{||}}{q_0^2 - \epsilon_q^2} & 
 -\gamma^0 \left( \frac{m_{}{\bf M}^\dagger \Lambda^+({\bf q})}{2\mu} 
   + \frac{{\bf M}^\dagger m_{} \Lambda^-({\bf q})}{2\mu} 
\right)\,
  \frac{G^\ast(q)}{q_0^2-\epsilon_q^2}  \\
  -\gamma^0 \left( \frac{m_{}{\bf M} \Lambda^-({\bf q})}{2\mu} 
   + \frac{{\bf M} m_{}\Lambda^+({\bf q})}{2\mu} \right)
  \frac{G(q)}{q_0^2-\epsilon_q^2} 
      &  \frac{m_{}}{2\mu}\,\frac{-q_0 +q_{||}}{q_0^2 - \epsilon_q^2}
   \end{array}
 \right)
\label{GOR11}
\end{equation}
with $m_{}={\rm diag}(m_u,m_d,m_d)$. Details on the derivation of this
result including ${\cal O}(m_{}^2)$ terms can be found in the 
Appendix-1 and  Appendix-2~\footnote{In what follows, 
``Appendix-$n$" denotes the
item $n$ in the Appendix.}.

Using the color-identity (see  Appendix-4)
\begin{equation}
\sum_a \frac{\lambda^{aT}}2\,\epsilon^A_c \,\frac{\lambda^a}2
=-\frac 23\,\epsilon^A_c\,\,,
\label{identity}
\end{equation}
the gap function $G(q)$ in the CFL phase with massless quarks
satisfies the following gap equation
\begin{eqnarray}
 G(p) &=& \frac {4g^2}3 \,
 \int\frac{d^4q}{(2\pi)^4}\,i{\cal D}(p-q)\,
 \frac{G(q)}{q_0^2-\epsilon_q^2}\nonumber \\
     &=& \frac {4g^2}3 \,
 \int\frac{d^4q_{\mbox{\tiny $E$}}}{(2\pi)^4}\,{\cal D}(p-q)\,
 \frac{G(q)}{q_4^2+q_{||}^2+|G(q)|^2}\; .
 \label{GapEq}
\end{eqnarray}
The second expression follows from Wick rotation to Euclidean space.
For perturbative screening of gluons in the relevant $\omega, \vec{q}$
domain,
the gluon-propagator in Euclidean space reads schematically as~\footnote{
This simplified version was used in~\cite{RWZ,BRWZ} and checked
to be reliable for the leading logarithm results, 
see also Appendix-12.}
\begin{equation}
 {\cal D}(q) = \textstyle\frac12 \displaystyle \frac{1}{q^2 + m_E^2}
  + \textstyle\frac12 \displaystyle \frac{1}{q^2 + m_M^2} \; .
\label{GluonProp}
\end{equation}
Perturbative arguments give $m_E^2/(g\mu)^2=m_D^2/(g\mu)^2\approx
N_f/2\pi^2$ and $m_M^2/m_D^2\approx \pi{|q_4|/|4{\bf q}|}$, where
$m_D$ is the Debye mass, $m_M$ is the magnetic screening due to
Landau damping and $N_f$ the number of flavors~\cite{LeBELLAC}.
To leading logarithm accuracy, the gap equation (\ref{GapEq}) can
be solved using the logarithmic variables
 $x=\ln(\Lambda_*/p_{||})$,
$y=\ln(\Lambda_*/q_{||})$, and
$x_0={\rm ln}(\Lambda_*/G_0)$~\cite{SON1}, where
$\Lambda_*=(4\Lambda_{\perp}^6/\pi m_E^5)$ and
$\Lambda_{\perp}=2\mu$.  
The result 
is 
\begin{equation}
 G(x)=G_0\,\sin \left(\frac {\pi x}{2x_0}\right)
 = G_0 \sin \left(h_* x/\sqrt{3}\right)
 \label{GapSolution}
\end{equation}
 with $h_* x_0/\sqrt{3}=\pi/2$ and $G_0$ 
given by
\begin{eqnarray}
 G_0 \approx \left(\frac{4 \Lambda_{\perp}^6}{\pi m_E^5}\right)\,
  e^{-\frac{\sqrt{3}\pi}{2h_\ast}} \; ,
\label{Gzero}
\end{eqnarray}
where
\begin{equation}
 h^2_*\equiv\frac{g^2}{6\pi^2} \;. \label{h*}
\end{equation}
This result is in agreement with \cite{SON1,SchaferWilczek,BRWZ},
whereas in \cite{PISARSKI,PisarskiNew,Brown} 
there is  an additional prefactor of 2.
Note that $G(q)$ is a real-valued even function of $q_{||}$.

\vskip 1.5cm
\centerline{\bf 3. Generalized Scalar and Pseudoscalar Mesons}
\vskip 1cm

The generalized mesons will refer to excitations
in $qq$ as opposed to the standard mesons which are
excitations in $\overline{q}q$ (see Appendix-9). 
The wavefunctions of the generalized scalar and pseudoscalar excitations
in the QCD superconductor follows from the Bethe-Salpeter equation
displayed in Fig. 1, 
\begin{equation}
\bfGamma^{A} (p, P)=g^2\int\frac{d^4q}{(2\pi)^4}\,
i{\cal D}(p-q)\,
i {\cal V}^a_\mu \,i {\bf S}(q\mbox{+}\frac P2)\,
\bfGamma^{A} (q,P)\,
i{\bf S}(q\mbox{$-$}\frac P2)\,i {\cal V}_a^\mu \; ,
\label{BSeq}
\end{equation}
where the gluon vertex is defined as a Nambu-Gorkov matrix (see Appendix-6)
\begin{equation}
 {\cal V}^a_\mu 
 \equiv \left(\begin{array}{cc} \gamma_\mu\lambda^a/2 & 0 \\
                 0 & C\left(\gamma_\mu\lambda^a/2\right)^TC^{-1}
                 \end{array} \right)
 = \left(\begin{array}{cc} \gamma_\mu\lambda^a/2 & 0 \\
                 0 &- \gamma_\mu{\lambda^{a\,T}}/2
                 \end{array} \right)
 \; .
 \label{gluon-vertex}
\end{equation}
The composite scalar vertex is given by
\begin{eqnarray}
  {\bf \Gamma}^A_{\sigma}\,(p,P)= \frac {1}{F_S}\,\,
\left(\begin{array}{cc} 0 &  i\Gamma_{S}^\ast(p,P)\,
      \left({\bf M}^{A}\right)^\dagger %\,\Lambda^{-}({\bf p})
                  \\
            i\Gamma_{S} (p,P)\,{\bf M}^A % \,\Lambda^{+}({\bf p})
  & 0\end{array}\right)
\label{Svertex}
 \end{eqnarray}
with ${\bf M}^A={\bf M}^{i\alpha}\,( \tau^A)^{i\alpha}$
and ${\bf M}^{i\alpha}=\epsilon_f^i\,\epsilon_c^\alpha\,
\gamma_5$. Note that $\left({\bf M}^A\right)^\dagger = \epsilon_f^i
\epsilon_c^\alpha \gamma_5
({\tau^A}^\ast)^{i\alpha}= \epsilon_f^i
\epsilon_c^\alpha \gamma_5{(\tau^A)^{\alpha i}}$.
The composite pseudoscalar vertex has been discussed in~\cite{RWZ}.
It reads
\begin{eqnarray}
  {\bf \Gamma}^A_{\pi}\,(p,P)= \frac {1}{F_{PS}}\,\,
\left(\begin{array}{cc} 0 &  -i\gamma_5\,\Gamma_{PS}^\ast(p,P)\,
      \left({\bf M}^{A}\right)^\dagger %\,\Lambda^{-}({\bf p})
                  \\
            i\gamma_5\,\Gamma_{PS} (p,P)\,{\bf M}^A % \,\Lambda^{+}({\bf p})
  & 0\end{array}\right) \; .
\label{PSvertex}
 \end{eqnarray}
A thorough discussion of the spin-parity assignment for these vertices
can be found in Appendix-7.

Inserting (\ref{PropPart}) in (\ref{BSeq}) we find, after a few
reductions (see Appendix-8 for details), that both the scalar
and pseudoscalar Bethe-Salpeter vertices obey
\begin{equation}
\Gamma (p, P)=\frac {4g^2}3
\int\frac{d^4q}{(2\pi)^4}\,
i{\cal D}(p-q)\,
\frac{(Q_0+Q_{||})(K_0-K_{||})
-G(Q)G(K)}{(Q_0^2-\epsilon_Q^2)(K_0^2-\epsilon_K^2)}\,
\Gamma (q, P)
\label{CFL444}
\end{equation}
with $Q=q+P/2$ and $K=q-P/2$ and 
$\Gamma(q,P)=\Gamma_S(q,P)=\Gamma_{PS}(q,P)$.
In establishing (\ref{CFL444})
we have made use of the relations
\begin{equation}
\sum_a \frac{\lambda^{aT}}2\,
\left( {\bf M}^A \right)\,
\frac{\lambda^a}2
= -\frac 23\,{\bf M}^A\;,
\qquad \sum_a \frac{\lambda^{aT}}2\,
\left( {\bf M}\,\left({\bf M}^{A}\right)^\dagger\,{\bf M}\right)\,
\frac{\lambda^a}2
\approx -\frac 23\,{\bf M}^A
\label{identity2}
\end{equation}
(see  Appendix-5)
and ignored the symmetric contribution in color-flavor which is
subleading in leading logarithm accuracy. In the meson rest frame
$P=(M,{\bf 0})$, we obtain
\begin{equation}
\Gamma (p, M)=\frac {4g^2}3
\int\frac{d^4q}{(2\pi)^4}\,
i{\cal D}(p-q)\,
\frac{q_0^2-\epsilon_q^2-M^2/4}{(q_0^2-\epsilon_q^2+M^2/4)^2-M^2\,q_0^2}\,
\Gamma (q, M) \;.
\label{PION1}
\end{equation}
This integral equation  can be solved exactly in leading 
logarithm accuracy (see the analogous calculation in section~5). 
The resulting mass $M$  in the present case is
\begin{equation}
M=2G_0\,\left(1-e^{-(\sqrt{3}-\sqrt{3})\pi/h_*}\right)^{\frac 12}=0\;,
\label{GOLD}
\end{equation}
which illustrates the Goldstone nature of the scalar and pseudoscalar
excitations in the QCD superconductor. The pseudoscalar excitations 
are the generalized pions already discussed in~\cite{HRZ,RWZ} with a
form factor $\Gamma(q,0)\propto G(q)$. 
The scalar excitations are would-be
Goldstone modes that get eaten up by the longitudinal components of the
colored gauge fields (see below).

\begin{figure}
 \centerline{\epsfig{file=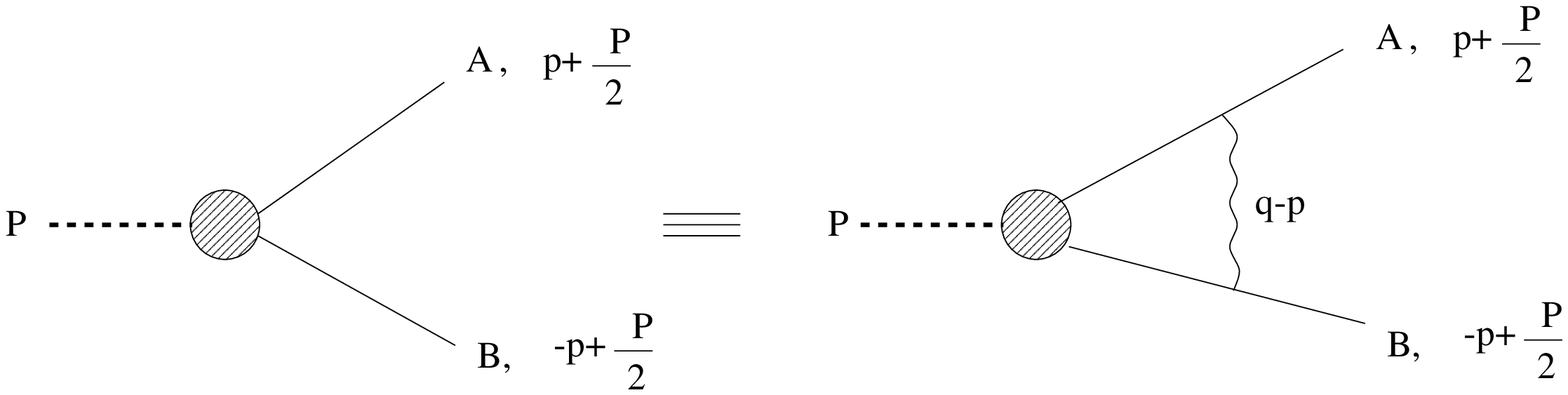,height=1.3in}}
 \caption{Bethe-Salpeter equation for the generalized mesons in the QCD
   superconductor.}
  \label{Fig1}
\end{figure}

Although (\ref{GOLD}) was derived 
using the simplified form (\ref{GluonProp}),
we now show that the outcome is the same, irrespective of that choice.
Indeed, an alternative way of reaching the same result
that is independent of the choice of the gluon propagator
${\cal D} (q)$ in (\ref{PION1}), can be obtained by expanding the
integrand in (\ref{PION1}) in $M^2$. 
For the Goldstone modes, this expansion
is valid and we obtain (see Appendix-10)
\begin{equation}
iF^2\, M^2 \approx \int\,\frac{d^4q}{(2\pi)^4}\,
\frac{\Gamma^2(q)}{q_0^2-\epsilon_q^2} -\frac{3}{4g^2}\,
\int\,d^4x \frac{\Gamma^2(x)}{i{\cal D}(x)} \; ,
\label{PION2}
\end{equation}
where the second integration is over the configuration space
(actually positive semidefinite in Euclidean space). 
Here $M^2$ is the Goldstone  mass squared,
and $F$ $(=F_{PS}=F_{S})$  its decay constant with
\begin{equation}
F^2\approx \frac{i}{4} \int\frac{d^4q}{(2\pi )^4}\,
\frac{q_0^2+3\epsilon_q^2}{(q_0^2-\epsilon_q^2)^3}\,\,
\Gamma^2 (q).
\label{PION3}
\end{equation}
Note that both $M$ and $F$ are functional
of the form-factor $\Gamma (q)$. Minimizing the mass-functional
with respect to $\Gamma (q)$ yields a gap-like equation with
$\Gamma (q)= \kappa G(q)$ as a solution, 
modulo an arbitrary dimensionless constant $\kappa$.
{}From (\ref{GapEq}) we observe in analogy to (\ref{PION2}) that
\begin{equation}
\frac 3{4g^2}\,\int d^4x\, \frac {G^2 (x)}{i{\cal D}(x)}=
\int \frac{d^4q}{(2\pi)^4}\,\frac{G^2(q)}{q_0^2-\epsilon_q^2}\,\,.
\label{CFL8}
\end{equation}
Inserting (\ref{CFL8}) into (\ref{PION2}) yields
$F^2\,M^2\approx 0$, which implies massless Goldstone modes
in the chiral limit, since $F^2$ is nonzero, i.e.
\begin{equation}
F^2\approx \frac{\kappa^2\mu^2}{8\pi^2}\int_0^\infty \, dq_{||}\,\frac{G^2
(q_{||})}{\epsilon_q^3}=\frac{\kappa^2\mu^2}{16\pi^2} \; 
\label{CFL10}
\end{equation}
(see Appendix-10).
For $\kappa=4$  the result for $F^2$ is in 
agreement with the result established in~\cite{RWZ}, where
the axial-vector current normalization has been used.
The dependence of the mass of the generalized pion on 
the current quark mass can be estimated in mass
perturbation theory using an axial-Ward identity~\cite{RWZ}, see 
the Appendix-3. The mass effects are of order $m_{}/\mu$, and 
to leading order, we have~\cite{RWZ}
\begin{equation}
\left( M^2\right)^{\alpha \beta} =
-\frac{\mu G_0}{4 \pi^2 F_T^2}
{\rm Tr}_{cf}
\left( \left[ {m_{}}^2,\tau^\alpha\right]
\left({\bf M}^\dagger {\bf M^\beta }
 \mbox{$-$}{{\bf M}^\beta}^\dagger{\bf M}
 \right) 
+ \left[ {m_{}}^2,{\tau^\alpha}^\ast\right]
\left({\bf M}{\bf M^\beta}^\dagger
 \mbox{$-$}{{\bf M}^\beta}{\bf M}^\dagger
 \right) \right).
\label{MASSX}
\end{equation}
Using the weak coupling values for
$F_T$ and $G_0$, (\ref{MASSX}) becomes
\begin{eqnarray}
\left(M^2\right)^{\alpha\beta}
&=& -\sqrt{\frac{2}{3}}\,\frac{256 \pi^4 }{9  g^5}\, 
\exp\left(-\frac{3\pi^2}{\sqrt{2}\,g} \right)\,
\left\{
{\rm Tr}_{cf}
\left( \left[ m_{}^2 ,\tau^\alpha\right]
\left({\bf M}^\dagger {\bf M^\beta }- {{\bf M}^\beta}^\dagger{\bf M}
 \right) \right) \right.  \nonumber \\
&&\qquad   \left. \mbox{}+
 {\rm Tr}_{cf}\left( \left[ m_{}^2, {\tau^\alpha}^\ast
 \right]
\left({\bf M}{\bf M^\beta}^\dagger- {{\bf M}^\beta}{\bf M}^\dagger
 \right) \right)\right\}\,\,.
\label{GOR13} 
\end{eqnarray}
The color-flavor traces in (\ref{MASSX}-\ref{GOR13}) yield zero. This
is consistent with the fact that
\begin{eqnarray}
{\rm Tr}\left(m\,\rho_0\, {\bf S}(q)\, m\,\rho_0\, {\bf S}(q)\right)
&=&  {\rm Tr}\left(m S_{11}(q) m S_{11}(q)\right) 
   + {\rm Tr}\left(m S_{22}(q) m S_{22}(q)\right) \nonumber\\
&=& \frac{4}{\mu}\,\frac{q_{||}}{q_0^2-\epsilon_q^2}\, 
{\rm Tr}_{cf}\left( {m^2}\right)
  +{\cal O}(1/\mu^2) \; ,
\label{TRASURF}
\end{eqnarray}
to second order in mass perturbation theory, where $\bfrho_0$
is the unit matrix in the Nambu-Gorkov space. Equation (\ref{TRASURF}) 
follows after inserting  the
$S_{11}(q)$  
and $S_{22}(q)$ Nambu-Gorkov components of the full massless propagator
(\ref{PropFull}), 
expanded to next-to-leading
order in  $1/\mu$, and is seen to vanish after the $q_{||}$-integration
is carried out. 
The vanishing of (\ref{TRASURF}) to leading order in $1/\mu$ can be
understood as follows: each mass insertion flips chirality but preserves 
helicity. Hence the quarks must carry opposite energies,  
which is not possible if the antiparticles are absent. 
Indeed, (\ref{TRASURF}) vanishes to leading order 
because of the orthogonality of the 
massless energy projectors occuring in (\ref{PropPart}).

At next-to-leading order in $1/\mu$, however, the vanishing of the
mass is averted by
keeping the antiparticle content of ${\bf S}(q)$ as given in 
(\ref{PropFull}) and using the simplified form (\ref{PSvertex})
of the generalized-pion vertex.
%~\footnote{However, if the sandwiched
%form (\ref{Sandwich}) is used, the result is still zero because
%of the projector structure.\label{footsand}}. 
Note that the antiparticle gap, which
according to \cite{SchaferWilczek} 
is gauge-fixing-term dependent (see also Appendix-12), does not appear at
this order, but first at next-to-next-to-leading order (see
Appendix-1 and Appendix-3).
The mass of the Goldstone modes at  next-to-leading order reads
%(modulo footnote~\ref{footsand})
\begin{eqnarray}
\left(M^2\right)^{\alpha\beta}&\approx& 
-\frac{ G_0^2 \, x_0}{8 \pi^2 F_T^2}\,
\left\{ {\rm Tr}_{cf}\left( [m,\tau^\alpha]_{+}
 \left( {{\bf M}^{\beta}}^\dagger m {\bf M} + {\bf M}^\dagger m {\bf
 M}^\beta\right) \right)
 \right. \nonumber\\ &&\qquad\qquad\qquad  \left.
+ {\rm Tr}_{cf}\left( [m,{\tau^\alpha}^\ast]_{+}
 \left( {{\bf M}^{\beta}} m {\bf M}^\dagger + {\bf M} m {{\bf
 M}^\beta}^\dagger\right)\right) \right\} \nonumber \\
&=&-\frac{2^{18}\, \pi^{10}\,}{3^4\,\sqrt{2}\, g^{11}}\, 
\exp\left({\frac{-3\sqrt{2}\,\pi^2}{g}}\right)
\,
\left\{ {\rm Tr}_{cf}\left( [m,\tau^\alpha]_{+}
 \left( {{\bf M}^{\beta}}^\dagger m {\bf M} + {\bf M}^\dagger m {\bf
 M}^\beta\right) \right)
 \right. \nonumber\\ &&\qquad\qquad\qquad\qquad  \left.
+ {\rm Tr}_{cf}\left( [m,{\tau^\alpha}^\ast]_{+}
 \left( {{\bf M}^{\beta}} m {\bf M}^\dagger + {\bf M} m {{\bf
 M}^\beta}^\dagger\right)\right) \right\}
 \label{M2next}
\end{eqnarray}
in the general case $m_u \lsim m_d \ll m_s$ and with the
weak-coupling values for $F_T$, $G_0$ and $x_0$ (see Appendix-3). 
At next-to-leading order the pion mass relation (\ref{M2next}) is
reminiscent of the quadratic Gell-Mann-Oakes-Renner relation
in the vaccum, as is explicit from the axial-Ward-identity~\cite{RWZ}.
Using the current mass decomposition 
\begin{equation}
m\equiv \frac 13 \,{\rm Tr} (m)\,{\bf 1} + m^\lambda\,\tau^{\lambda}
=\hat{m}\,{\bf 1} + m^\lambda\,\tau^{\lambda}\; ,
\end{equation}
${\bf M}={\bf M}^{aa}\equiv \epsilon_f^a\epsilon_c^a$ 
and ${\bf M}^\beta\equiv {\bf M}^{i\alpha}(\tau^\beta)^{i\alpha}$~\footnote{The
$\gamma_5$ has been removed by the spin trace.} and the identity
\[
[ \tau^\alpha,\tau^\beta]_+ = {\textstyle \frac{4}{3}}
 \delta^{\alpha\beta}\, {\bf 1} 
 +2 d^{\alpha \beta\gamma}\, \tau^\gamma \,\,,
\]
we can unwind the 
color-flavor traces in (\ref{M2next}) to obtain the explicit
mass matrix,
\begin{eqnarray}
\left(M^2\right)^{\alpha\beta}
 &=&\frac{2^{20}\, \pi^{10}\,}{3^4\,\sqrt{2}\, g^{11}}\, 
\exp\left({\frac{-3\sqrt{2}\,\pi^2}{g}}\right)
\, 
\left\{ 
 8\, \hat{m}^2\,\delta^{\alpha\beta}\, 
 +{\textstyle \frac{16}{3}}\, m^\alpha\,m^\beta\, 
 +8\,\hat{m}\,m^\lambda\,d^{\lambda\alpha\beta} 
\right. \nonumber\\ &&  \left.
 +2\left( \hat{m}\,m^\lambda \delta^{\alpha\sigma}
 +  m^\lambda \,m^\gamma\, d^{\gamma\alpha\sigma}\right)\, 
 \epsilon^{Inl}\,
 \epsilon^{Jsm}\,(\tau^\sigma)^{mn} \,(\tau^\beta+\tau^{*\beta})^{IJ}\,
(\tau^\lambda)^{ls}\right\} ,
\label{M2EXPLICIT}
\end{eqnarray}
which is nonzero in the flavor symmetric case. Indeed,
for $m_u=m_d=m_s \equiv m_q$, this result simplifies to
\begin{equation}
M^2 \approx \frac{2^{23}\, \pi^{10}\, m_q^2}{3^4\,\sqrt{2}\, g^{11}}\, 
\exp\left({\frac{-3\sqrt{2}\,\pi^2}{g}}\right) \; ,
 \label{M2SYMMETRIC}
\end{equation}
showing the nonperturbative character of the
Goldstone mass in the gauge coupling $g$. 
We have not checked
whether the corrections to the leading logarithm approximation affects
(\ref{M2EXPLICIT}), since the leading order result (\ref{GOR13}) 
vanishes. 
An expression of similar structure to (\ref{M2SYMMETRIC}) can be found
in \cite{HongLeeMin}.

At this stage, an important remark is in order: 
the non-vanishing of the next-to-leading order 
result depends on our simplification of the
gluon-propagator (\ref{GluonProp}), which leads
to the subsequent simplification in the generalized
meson vertices (\ref{Svertex}-\ref{PSvertex}). Indeed,
if we were to use the  exact gluon-propagator 
(\ref{GluonPropExact}), then the generalized 
meson-vertices (\ref{Svertex}-\ref{PSvertex}) 
(also (\ref{A-vertex-structure})) have to be 
changed to
\begin{equation}
 {\bf \Gamma}_M^A(p,P) \longrightarrow
\widetilde {\bf \Gamma}_M^A(p,P) 
 = \left(\begin{array}{cc} \Lambda^-({\bf p}) & 0\\
                           0 & \Lambda^+({\bf p})         
    \end{array} \right) 
{\bf \Gamma}_M^A(p,P)
\left(\begin{array}{cc} \Lambda^+({\bf p}) & 0\\
                           0 & \Lambda^-({\bf p})         
    \end{array} \right)  \; \,,
  \label{Sandwich}
\end{equation}
to satisfy the pertinent Bethe-Salpeter equations (see Appendix-12).
The additional projectors in (\ref{Sandwich}) cause
the mass of the Goldstone modes to remain massless 
at next-to-leading order as well. We note that the
additional projectors in (\ref{Sandwich}) follow the structure
of the leading quark propagator (\ref{PropPart}),  and
are in general superfluous if each vertex
${\bf \Gamma}^A_M$ is only linked  by the {\em leading} part
of the quark propagator. If the vertex is of a scalar or pseudoscalar
nature, it is already sufficient that each vertex is coupled to at least
{\em one leading} part of the quark propagator (see Appendix-7).
This is the case of
most our results to leading order, hence our simplification. 
The exception is (\ref{M2EXPLICIT}) which is a next-to-leading 
order result, since two subleading propagators (a massive and 
a next-to-leading order massless one) are there attached to one vertex.

\vskip 1.5cm
\centerline{\bf 4. Higgs Mechanism}
\vskip 1cm

The generalized scalar mesons  mix with the longitudinal gluons
through the non-diagonal polarization 
\begin{equation}
{\Pi}_\mu^{aA} (Q) = -ig
\int\,\frac{d^4q}{(2\pi )^4}\,
{\rm Tr}\left(i{\cal V}_\mu^a\, i{\bf S} (q+\frac Q2) \,
i{\bf \Gamma}^A_\sigma (q, Q)\, i{\bf S} (q-\frac Q2) \right)\,\,,
\label{HI3}
\end{equation}
where ${\bf \Gamma}^A_\sigma(q,Q)$ is defined in (\ref{Svertex}).
Since the scalar form factor is ${\bf \Gamma}_S (q, 0) =G(q)/F$, 
equation (\ref{HI3}) can be reduced to
\begin{eqnarray}
  {\Pi}_\mu^{aA} (Q) &=& 
\frac gF\,{\rm Tr}\left(\frac {\lambda^a}2 \,{\bf M^\dagger}\, 
 {\bf M}^A\right)\,
 \int\frac{d^4q}{(2\pi )^4}\, 
\frac{G(q)}{(K_0^2-\epsilon_K^2)(P_0^2-\epsilon_P^2)} \nonumber \\
&& \qquad\times \left\{ \left[ (K_0+K_{||}) G(P)-(P_0+P_{||}) 
 G(K)\right] 
{\rm Tr}\left[\gamma_0 \gamma_\mu\Lambda^+(K)
\Lambda^+ (P)\right] \right.\nonumber \\
&&\qquad\ \mbox{}+\left.\left[ (K_0-K_{||})G(P)-(P_0-P_{||})
G(K)\right]
{\rm Tr}\left[\gamma_0 \gamma_\mu\Lambda^-(K)
\Lambda^- (P)\right]\right\} 
\label{HI4}
\end{eqnarray}
with $K=q +Q/2$ and $P=q-Q/2$.
Clearly ${\Pi}_\mu (0)=0$. In the following, we will use
\[
{\rm Tr}\left(\frac{\lambda^a}{2} {\bf M}^\dagger {\bf M}^A\right)
\equiv {\rm Tr}_f {\rm Tr}_c \left(\frac{\lambda^a}{2} 
{\bf M}^\dagger {\bf M}^A\right)
= - (\lambda^a)^{ji}(\tau^A)^{ji} 
= - {\rm Tr} \left({\tau^a} {\tau^A}\right)= -2\delta^{aA} \; ,
\]
${\rm Tr}[\Lambda^+(K)\Lambda^+(P)]
={\rm Tr}[\Lambda^-(K)\Lambda^-(P)]= 1 +\hat {\bf K}\cdot \hat{\bf
  P}\to 2$
and
${\rm Tr}[\gamma_0\gamma_i\Lambda^+(K)\Lambda^+(P)]
=-{\rm Tr}[\gamma_0\gamma_i\Lambda^-(K)\Lambda^-(P)]
=\hat {\bf K}^i +\hat{\bf P}^i\to 2 \hat{\bf q}^i$
as well as  $K_{||}-P_{||} \approx \hat{\bf q}\cdot {\bf Q}$. In this way,
we obtain to linear order in $Q$,
\begin{eqnarray}
{\Pi}_0^{aA} (Q)&\approx& i\delta^{aA} \,Q_0\,gF_T\; ,\nonumber\\
{\Pi}_i^{aA} (Q) &\approx& i\delta^{aA} \,Q_i\,g\frac{F_S^2}{F_T} 
= i\delta^{aA} \,Q_i\,g v^2 F_T \; .
\label{HI5}
\end{eqnarray}
The temporal and spatial pion decay constants are, 
respectively~\cite{RWZ}
\begin{eqnarray}
F_T^2 &\approx& - 8 i \int \frac{d^4 q}{(2\pi)^4} \, \frac{G^2(q)}
       {(q_0^2-\epsilon_q^2)^2}\; , \\
F_S^2 &\approx& - 8 i \int \frac{d^4 q}{(2\pi)^4} \, 
  \left(\hat {\bf q}\cdot \hat {\bf Q} \right)^2
\frac{G^2(q)}
       {(q_0^2-\epsilon_q^2)^2}\; ,
\label{HI5x}
\end{eqnarray}
where $v^2=F_S^2/F_T^2=1/3$ is the square of the velocity of the
Goldstone modes~\cite{SONSTE,RWZ}. 

The nonvanishing of (\ref{HI5}) implies that the 8 generalized scalars 
in the CFL phase are eaten up by the longitudinal gluons. As a result,
the gluons acquire masses in a manner analogous to the 
familiar Meissner effect. Indeed, if we
denote by $\Sigma^{cf}$ the scalar color-flavor order parameter in the CFL 
phase, then under local color transformations $\Sigma\rightarrow g_c\,\Sigma$.
In the local approximation (leading order in $Q$), the gluon-scalar mixing
is described by the Higgs term
\begin{equation}
{\cal L}_H =  
 \frac 14 \,{\rm Tr}\left| \partial_0 \Sigma-ig\, A_0\Sigma\right|^2
-\frac {v^2}4 \,{\rm Tr}\left|\partial_i\Sigma-ig \, A_i\Sigma\right|^2\,\,.
\label{HX1}
\end{equation}
For the scalar excitations, $\Sigma=F_T\,\lambda^0 +\sigma^A\lambda^A$ and 
(\ref{HX1})
becomes
\begin{eqnarray}
{\cal L}_H &=& \frac 12 \left( \dot{\sigma}^A\dot{\sigma}^A -
v^2\partial_i\sigma^A\,\partial_i\sigma^A\right)
+\frac {g^2F_T^2}2\left(A_0^A\,A_0^A-v^2A_i^A\,A_i^A\right)\nonumber\\
&&-gF_T \left(\dot{\sigma}^AA_0^A-v^2\partial_i\sigma^A\,A_i^A\right)\,\,.
\label{HX2}
\end{eqnarray}
The mixing vertices in (\ref{HX2}) 
are precisely the ones given by (\ref{HI5}).
If we define 
\begin{eqnarray}
\tilde{A}_0^A &=& A_0^A -\frac 1{gF_T}\partial_0\sigma^A \; ,\nonumber\\
\tilde{A}_i^A &=& A_i^A -\frac 1{gF_T}\partial_i\sigma^A \; ,
\label{HX3}
\end{eqnarray}
then (\ref{HX2}) reduces to 
\begin{equation}
{\cal L}_H=\frac {g^2F_T^2}2 \, \tilde{A}_0^A\tilde{A}_0^A
-\frac{g^2F_S^2}2\,\tilde{A}_i^A\tilde{A}_i^A
\label{HX4}
\end{equation}
which is purely a mass term for the new gluon field $\tilde{A}$. 
Originally there are 8 gluons $A$ that are massless with 
two transverse polarizations. After the Higgs mechanism (\ref{HX3}),
the gluons become massive in the CFL phase, with the scalar making
up the longitudinal component. No
scalars are left. The Meissner mass
(\ref{HX4}) refers to the inverse penetration length of {\it static}
colored magnetic fields  in the QCD
superconductor which is unexpectedly small, i.e. $1/g\mu$.
In weak coupling the Meissner 
mass is of the order of the electric screening mass
$m_E\approx g\mu$.  
It is not of the order of $g\,G_0$ as in a conventional 
superconductor with a constant (energy independent) gap. 
It is important to note that the {\em nonstatic
gluonic modes} with $Q_0>G_0$ sense `free quarks' 
for which there is electric
screening but no magnetic screening. A brief analysis of the polarization
function in the CFL phase supporting this is given in the Appendix-11. 
The nonstatic and long-range magnetic effects
are at the origin of the pairing mechanism discussed here, including 
the binding in the  mesonic excitation spectrum.

\vskip 1.5cm
\centerline{\bf 5. Masses of Generalized Vector and Axial-Vector Mesons}
\vskip 1cm

In this section we consider vector mesons
consisting of a pair of (quasi-)quarks or (quasi-)holes
at the Fermi surface with momenta $p_1=-q+P/2$
and $p_2=q+P/2$. In the CFL phase these composites (generalized
vector mesons) have finite size. Since
Lorentz invariance is absent, 
there are electric and magnetic composite mesons.
Their transverse and longitudinal vector
form factors (or wavefunctions) ${\bf\Gamma}^A_{T,L}$ in the
Nambu-Gorkov representation are defined as
\begin{equation}
  {\bf\Gamma}^A_j(q,P) \equiv \gamma_j \,{\bf\Gamma}^A_V(q,P)
= \hat{\bf P}_j \hat{\bf P}^i \gamma_i\,{\bf\Gamma}^A_L (q,P)
    + \left(-g_j^{\ i}-\hat{\bf P}_j\hat{\bf P}^i\right)
     \gamma_i\,{\bf\Gamma}^A_T (q,P)
\label{DEC1}
\end{equation}
with
\begin{eqnarray}
 {\bf\Gamma}^A_L &=& \int d \hat {\bf P}\ \gamma^k\,\hat{\bf P}_k\,
  \hat{\bf P}^j \,{\bf\Gamma}^A_j\; , \label{GV-long} \\
 {\bf\Gamma}^A_T &=& {\textstyle\frac{1}{2}}\int d \hat{\bf P}\
\gamma^k \left(g_k^{\ j}+\hat{\bf P}_k \hat{\bf P}^j\right)\, 
 {\bf\Gamma}^A_j
 \label{GV-transv}
 \; .
\end{eqnarray}
Current conservation implies $P^{\mu}\,{\bf\Gamma}^A_{\mu} (q,P)=0$, 
such that
the temporal form factor ${\bf\Gamma}_0^A(p,P)$ is
not an independent quantity, but
can be expressed  in terms of ${\bf\Gamma}^A_L$ as
\begin{equation}
{\bf\Gamma}^A_0 (p, P)= -\frac {\vec{\gamma}\cdot\vec{\bf P}}{P_0}\,
{\bf\Gamma}^A_L (p, P)\;.
\label{TIME0}
\end{equation}
In the QCD superconductor this implies 1 electric (L) and 2 magnetic
(T) modes for the composite vector mesons. The purpose of this section
is to evaluate their form factors and ``masses'' (or more precisely
excitation energies) in the weak coupling limit.
For that, we note that
the wavefunction (up to a dimensionful normalization)
of the electric and magnetic modes
follow from the Bethe-Salpeter equation displayed
in Fig.~\ref{Fig1}, i.e.
\begin{equation}
\bfGamma_{\nu}^{A} (p, P)=g^2\int\frac{d^4q}{(2\pi)^4}\,
i{\cal D}(p-q)\,
i {\cal V}^a_\mu \,i {\bf S}(q\mbox{+}\frac P2)\,
\bfGamma_{\nu}^{A} (q,P)\,
i{\bf S}(q\mbox{$-$}\frac P2)\,i {\cal V}_a^\mu \; ,
\label{t5}
\end{equation}
where the gluon vertex is defined in (\ref{gluon-vertex}).
As discussed in the Appendix-7,
the composite vector meson
vertices for the transverse and longitudinal modes have the following
structure
%~\footnote{The relation between the off-diagonal
%terms follows, if the general expression
%$\Delta_{12}$=$\gamma^0
%\left(\Delta_{21}\right)^\dagger\gamma^0$,
%see\,\cite{PisarskiSuperfluid}, is applied to $\Gamma^A_j$. Note
%that the vertices of the composite
%axial-vector mesons have the same matrix structure
%as (\ref{CFL2}) with $M^A$ replaced by $i\gamma_5 M^A$.}
\begin{eqnarray}
  {\bf \Gamma}^A_{T,L}\,(p,P)= \frac {1}{F_V}\,\,
\left(\begin{array}{cc} 0 &  \Gamma_{T,L}^\ast(p,P)\,
      \left({\bf M}^{A}\right)^\dagger %\,\Lambda^{-}({\bf p})
                  \\
            \Gamma_{T,L} (p,P)\,{\bf M}^A % \,\Lambda^{+}({\bf p})
  & 0\end{array}\right)\; .
\label{Vvertex}
 \end{eqnarray}
%with ${\bf M}^A={\bf M}^{a\alpha}\,( \tau^A)^{a\alpha}$
%and ${\bf M}^{a\alpha}=\epsilon_f^a\,\epsilon_c^\alpha\,
%\gamma_5$. Note that $\left({\bf M}^A\right)^\dagger = \epsilon_f^a
%\epsilon_c^\alpha \gamma_5
%({\tau^A}^\ast)^{a\alpha}= \epsilon_f^a
%\epsilon_c^\alpha \gamma_5{(\tau^A)^{\alpha a}}$.
Inserting (\ref{PropPart}) in (\ref{t5}) we find, after a few
reductions (see the Appendix-8 for details),
\begin{equation}
\Gamma_{T,L} (p, P)=\frac {4g^2}9
\int\frac{d^4q}{(2\pi)^4}\,
i{\cal D}(p-q)\,
\frac{(Q_0+Q_{||})(K_0-K_{||})
-G(Q)G(K)}{(Q_0^2-\epsilon_Q^2)(K_0^2-\epsilon_K^2)}\,
\Gamma_{T,L}(q, P)
\label{CFL44}
\end{equation}
with $Q=q+P/2$ and $K=q-P/2$. 
%In establishing (\ref{CFL44})
%we have made use of the relations
%\begin{equation}
%\sum_a \frac{\lambda^{aT}}2\,
%\left( {\bf M}^A \right)\,
%\frac{\lambda^a}2
%= -\frac 23\,{\bf M}^A\;,
%\qquad \sum_a \frac{\lambda^{aT}}2\,
%\left( {\bf M}\,\left({\bf M}^{A}\right)^\dagger\,{\bf M}\right)\,
%\frac{\lambda^a}2
%\approx -\frac 23\,{\bf M}^A
%\label{identity2}
%\end{equation}
%(see appendix)
%and ignored the symmetric contribution in color-flavor which is
%subleading in leading logarithm accuracy. The dimensionful
%constant $F_V$ that characterizes the coupling to the vector
%current will be fixed below.
%\begin{figure}
% \centerline{\epsfig{file=fig1.eps,height=1.3in}}
% \caption{Bethe-Salpeter equation for the vectors in the QCD
%   superconductor.}
%  \label{Fig1}
%\end{figure}

In the rest frame of the composite vector meson,
$P=(M_V,{\bf 0})$, equation (\ref{CFL44}) becomes
\begin{equation}
\Gamma_{T,L} (p, M_V)=\frac {4g^2}9
\int\frac{d^4q}{(2\pi)^4}\,
i{\cal D}(p-q)\,
\frac{q_0^2-\epsilon_q^2-M_V^2/4}
{(q_0^2-\epsilon_q^2+M_V^2/4)^2-M_V^2\,q_0^2}\,
\Gamma_{T,L}(q, M_V)\; ,
\label{CFL5}
\end{equation}
where we have used that $\Gamma_{T,L}(q,M_V)$ is an even real
function of $q$. The difference between the vector equation
(\ref{CFL5}) and the scalar equation (\ref{PION1}) is in the
prefactors : $4/9$ versus $4/3$ respectively. As a result,
(\ref{PION1}) admits massless modes, while (\ref{CFL5}) does
not. Indeed, using
(\ref{GluonProp}) in (\ref{CFL5}) and assuming that
$\Gamma_{T,L}(q,M_V)\approx \Gamma_{T,L}(q_{||},M_V)$ with support
only around
the Fermi surface, we get, after a few
integrations~\footnote{The logarithms result from the
$q_\perp$ integration.
The contour integration in $q_0$ is performed
under the assumption that ${\cal D}(p-q)$ is dominated
by nearly static contributions, where
the following identity is  used:
\[\frac{q_0^2-\epsilon_q^2-M_V^2/4}{
(q_0^2-\epsilon_q^2+M_V^2/4)^2-M_V^2q_0^2}
 = \frac{1}{2}\left(\frac{1}{q_0^2-(\epsilon_q -M_V/2)^2}
 + \frac{1}{q_0^2-(\epsilon_q+M_V/2)^2}\right) \; . \]
The remaining steps are analogous to the ones discussed
in \cite{BRWZ}. The coefficient $h_\ast$ is defined in (\ref{h*}).}
\begin{eqnarray}
  \Gamma_{T,L}(p_{||}, M) &\!\!\approx\!\!&
  \frac{h_\ast^2}{36}
\int_0^{\infty}\! dq_{||}\,
   \left(
\frac{1}{\sqrt{q_{||}^2+|G(q_{||})|^2}-M_V/2} +
\frac{1}{\sqrt{q_{||}^2+|G(q_{||})|^2}+M_V/2} \right)
\nonumber\\
  &&\qquad\mbox{}\times
   \,\ln\left\{
   \left(1+\frac{\Lambda_{\perp}^2}
             {(p_{||}\mbox{$-$}q_{||})^2+m_E^2}\right)^3
   \left(1+\frac{\Lambda_{\perp}^3}{|p_{||}\mbox{$-$}q_{||}|^3
 +\frac{\pi}{4}m_D^2|p_{||}\mbox{$-$}q_{||}|}
    \right)^2\right\} \; \nonumber \\
 &&\qquad\mbox{}\times\Gamma_{T,L} (q_{||}, M)\,\,.
 \label{EQFF}
\end{eqnarray}
Since $|p_{||}-q_{||}| \ll m_E =m_D \ll \Lambda_\perp=2\mu$,
equation (\ref{EQFF}) is simplified in leading logarithm accuracy to
\begin{equation}
\Gamma_{T,L} (p_{||}, M_V)\approx
\frac{h_*^2}{18}\,\int_{G_M}^{\Lambda_*}\,\frac{dq_{||}}{q_{||}}\,
{\rm ln}\left(\frac{\Lambda_*^2}{(p_{||}-q_{||} )^2}\right)\,
\Gamma_{T,L} (q_{||}, M_V)
\label{EQFF1}
\end{equation}
with
\begin{equation}
 G_M^2=G_0^2-M_V^2/4\;
 \label{GM2}
\end{equation}
and $\Lambda_\ast= 4 \Lambda_\perp^6/\pi m_E^5$.
 The solution
to (\ref{EQFF1}) is obtained by using the new logarithmic variables
$x={\rm ln}(\Lambda_*/p_{||})$ and $x_M={\rm ln}(\Lambda_*/G_M)$ as
discussed in \cite{BRWZ}. Following this reference, 
the transverse and longitudinal
form factors for the composite vector mesons are found to be
equal to
\begin{equation}
\Gamma_{T,L} (p_{||},M_V)\approx
{G_M}\,{\rm sin}\left(\frac {\pi x}{2x_M}\right)=
\sqrt{G_0^2-\frac {M_V^2}4}\,{\rm
sin}\left(\frac{h_*}{3}\,{\rm ln} \left(\frac{\Lambda_*}{p_{||}}\right)
\right)
\label{FFR}
\end{equation}
with
\begin{equation}
 \frac{h_*}{3}\ln\left(\frac{\Lambda_*}{G_M}\right) = \frac{h_*}{3}x_M=
 \frac{\pi}{2}  \label{x_M}\;.
\end{equation}
Notice the threshold singularity for pair production at $M_V=2G_0$.
%We recall that $h_*=g/(\sqrt{6}\pi)$ and $\Lambda^*=
%256\mu^6/(\pi m_E^5)$.
Using (\ref{GM2}), (\ref{x_M}) and (\ref{Gzero}),
we get the following value for
the mass $M_V$ of the generalized vector meson
\begin{equation}
   M_V=2G_0\,\left(1-e^{-(3-\sqrt{3})\pi/h_*}\right)^{\frac 12} \; .
\label{MVR}
\end{equation}
We recall that $h_*=g/(\sqrt{6}\pi)$. Note that $M_V$ is less than
$2\,G_0$. Thus the composite pairs of particles or holes
are bound exponentially weakly in the CFL phase.
The smaller the coupling,
the smaller
the binding. 
For $g\rightarrow 0$, we reach the breaking of the composite pair,
and their mass asymptotes $2G_0$. For 
$g$ large their mass asymptotes zero which we interpret as
a realization of Georgi's
vector limit~\cite{GEORGI} in dense QCD.
A rerun of these arguments for the
axial-vector composites yield the same mass (see Appendix-8). 
In the CFL superconductor
both the vector and axial-vector 
octets are degenerate in leading logarithm
approximation, in spite of chiral symmetry breaking.

\vskip 1.5cm
\centerline{\bf 6. Vector Meson Coupling to Currents}
\vskip 1cm

\begin{figure}
\centerline{\epsfig{file=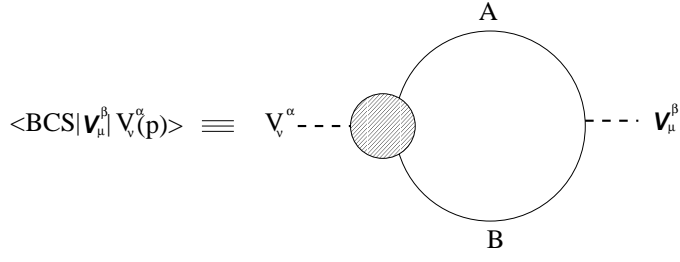,height=1.3in}}
\caption{Vector transition in the QCD superconductor.}
\label{Fig2}
\end{figure}

The dimensionful coupling
$F_V$ of the vector mesons in the QCD superconductor
is defined by
\begin{equation}
\left\langle {\rm BCS} \left| {\bf V}_{\mu}^\alpha (0)
 \right|V_{T,L}^{A} (P)
\right\rangle \equiv
F_V\,M_V\,{\cal E}_{\mu}^{T,L} (P)\,\delta^{\alpha A} \; ,
\label{t7}
\end{equation}
where $\langle {\rm BCS}|$ stands for the CFL ground state
and ${\cal E}_{\mu}^{T,L}$ are the transverse and longitudinal
polarizations.  In terms of the original quark fields ${\bf \Psi}$, the
vector current follows from Noether theorem. Hence (see Appendix-6)
\begin{equation}
{\bf V}_{\mu}^{\alpha} (x) = \overline{\bf \Psi}
\gamma_{\mu}\,\frac 12\,{\bf T}^{\alpha}\bfrho_3{\bf \Psi}
\label{t8}
\end{equation}
with ${\bf T}^A ={\rm diag}\,(\tau^A, \tau^{A*})$
an $SU(3)_{c+F}$ valued generator in the Nambu-Gorkov
representation and $\bfrho_3$ the standard Pauli matrix acting on the
Nambu-Gorkov indices, in accordance with (\ref{gluon-vertex}).
A diagrammatic representation of
(\ref{t7}) is shown in~Fig.~\ref{Fig2}.
Inserting (\ref{t8}) in (\ref{t7}),
we obtain from Fig.~\ref{Fig2}
\begin{equation}
 F_V\,M_V {\cal E}_\mu^{T,L}(P)\,\delta^{\alpha A}
 =-\int \!\!\frac{d^4k}{(2\pi)^4}\,
 \Tr\left(\gamma_{\mu}\,\frac 12\,{\bf T}^\alpha\bfrho_3\,
 i{\bf S}(k+\frac P2)\,i
 {\cal E}_\nu^{T,L}(P) \bfGamma_{V}^{\nu\,A} (k,P)\,
  i {\bf S}(k-\frac P2)\right)
\label{t9}
\end{equation}
with $\bfGamma_V^{\nu\,A}$ as defined in (\ref{DEC1}).
However, because of the spin structure,
\begin{equation}
 {\rm Tr}_s( \gamma_\mu \,\gamma_0 \,\gamma_r\, \alpha^n)=0\; , \qquad
 n=0,1,2,\cdots\;,
  \label{SpinStructure}
\end{equation}
the right hand side 
of (\ref{t9}) vanishes identically, to leading logarithm accuracy
and in the chiral limit.

So, the vector excitations couple to the usual physical
currents only in subleading order if at all. 
An exact and direct assessment of
this coupling is beyond the scope of the present work. Instead, we
will present a variational estimate for $F_V$ (temporal) based on
the variational analysis discussed in section 3 for the scalars and 
pseudoscalars (massless excitations) which turn out to compare well
with the exact results. 
Indeed, a rerun of the variational arguments, i.e.\ the application of 
the steps between (\ref{PION1}) and (\ref{PION2}) to (\ref{CFL5}),
yields 
\begin{eqnarray}
 F_V^2 &\approx& \frac{\mu^2}{8\pi^2} \int_0^\infty d q_{||}\,
 \frac{\Gamma^2_{T,L}(q_{||},M_V)}{\epsilon_q^3}
\approx \frac{\mu^2}{8\pi^2} \int_{G_0}^{\Lambda_\ast} d q_{||}\,
\frac{\kappa^2 G_M^2 \sin^2\left(\frac{h_\ast}{3} 
 \ln\left(\frac{\Lambda_\ast}{q_{||}}\right) \right)}{q_{||}^3} 
\nonumber\\
&=& \frac{\kappa^2\mu^2}{8\pi^2}\,\frac{G_M^2}{G_0^2}\int_0^{x_0} dx\,
 e^{2(x-x_0)} \sin^2\left(\frac{\pi x}{2 x_M}\right) 
= \frac{\kappa^2\mu^2}{8\pi^2}\,\frac{G_M^2}{G_0^2}\,\frac{1-\cos(\pi
 x_0/x_M)}{4}\nonumber\\
&=& \frac{\kappa^2\mu^2(1-\cos(\pi/\sqrt{3})}{32\pi^2}\, 
e^{ -\frac{\pi}{h_\ast}(3-\sqrt{3}\,)} \; .
\end{eqnarray}
The ratio $F_V^2/F^2\ll 1$ is indeed subleading in weak coupling. 
Similar variational arguments for the vector mass yields an upper bound 
of the form (see Appendix-10)
\begin{equation}
 M_V^2 \leq  8\,x_0\, G_0^2  \,\frac{2 \left(1 - ({x_M}/{\pi\,x_0})\,\sin(\pi
     x_0/x_M)\right)}{1-\cos(\pi x_0/x_M)} \; ,
 \label{MV-upper}
\end{equation}
which is generously satisfied by the exact result (\ref{MVR}).

The generalized vector meson coupling to the scalars can be assessed
similarly, by substituting in Fig.~\ref{Fig2} the vector current by the
generalized scalar vertex, i.e.
\begin{equation}
 {\bf \Pi}_i^{aA} (P)
 =-i\int \!\!\frac{d^4k}{(2\pi)^4}\,
 \Tr\left(i{\bf \Gamma}_i^a (k, P)\,
 i{\bf S}(k+\frac P2)\,i
 {\bf \Gamma}_{\sigma}^{A} (k,P)\,
  i {\bf S}(k-\frac P2)\right)\,\,.
\label{MIX1}
\end{equation}
Substituting for the vector and scalar vertex, we obtain
\begin{eqnarray}
 {\bf \Pi}_i^{aA} (P)
 &=&\frac 1{F_V\,F_S}\,{\rm Tr}\left({\bf M}^a{{\bf M}^{A}}^\dagger\right)
\int \!\!\frac{d^4k}{(2\pi)^4}
\frac{(Q_0-Q_{||})(K_0 +K_{||}) -(Q_0+Q_{||})(K_0-K_{||})}
{(Q_0^2-\epsilon_Q^2)(K_0^2-\epsilon_K^2)}\nonumber\\
&& \qquad\qquad\qquad\qquad\qquad\quad\mbox{}\times
\Gamma_V(k,P)\,\Gamma_S(k, P)\,{\rm Tr}\left(\gamma_i\,
\Lambda^+({\bf Q})\Lambda^+({\bf K})\right)
\label{MIX2}
\end{eqnarray}
with $Q=k+P/2$ and $K=k-P/2$.
The spin trace in (\ref{MIX2}) is found to vanish. In leading logarithm
approximation, the generalized vectors and scalars do not mix,
in contrast to the mixing between the generalized scalars and
gluons which is at the origin of the Higgs mechanism discussed
in section 4. Mixing may take place at next-to-leading order with
consequences on leptonic emissivities in dense matter.

\begin{figure}
\centerline{\epsfig{file=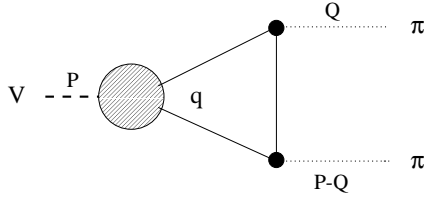,height=1in}}
\caption{$V\rightarrow \pi\pi$ decay in the QCD superconductor.}
\label{Fig3}
\end{figure}

\vskip 1.5cm
\centerline{\bf 7. Vector Meson Coupling to Goldstones}
\vskip 1cm
The composite character of the vector mesons in the CFL phase
allows them to interact with the generalized pions in the QCD
superconductor modulo G-parity. Indeed, the decay process 
$V\rightarrow \pi$ (and in general any odd number of $\pi$)
can be easily seen to vanish in the CFL phase. In this section,
we will estimate the $V\rightarrow \pi\pi$ decay in the CFL phase as 
represented by Fig.~\ref{Fig3} in leading logarithm accuracy and 
in the chiral limit.

The effective vertex associated to Fig~\ref{Fig3}, translates to the
following equation
\begin{eqnarray}
{\cal V}^{ABC}_\mu (P,Q)&=&-\,
\int\,\frac{d^4q}{(2\pi)^4}\,
{\rm Tr}\left(
i{\bf S}(q+\frac P2)\,i\gamma_\mu\,{\bf \Gamma}_V^A (q,P)\,
i{\bf S}(q-\frac P2 )\,i{\bf \Gamma}^B_\pi (q-\frac Q2,P-Q)\,\right.
\nonumber\\
&&\qquad\qquad\qquad\quad\left.\mbox{} \times i
{\bf S}(q-Q+\frac P2 )\,i{\bf \Gamma}^C_\pi(q+\frac{P-Q}{2},Q)
\right)\,\,.
%)\,\,.
\label{DEc11}
\end{eqnarray}
The composite vector and composite pion vertices are given,
respectively, by 
%\begin{equation}
%\begin{array}{lclcl} {\bf \Gamma}_V^A (p) &\approx&
%\Gamma_V^A(p,M_V)&\equiv& \frac {1}{F_V}\,
%\,\Gamma_V (p, M_V)\,\bfrho_1\,{\bf M}_{\bf T}^{\,\,A} %{\bf \Lambda}(p)
%\; , \\
%{\bf \Gamma}^A (p) &\approx& {\bf \Gamma}^A(p,0)&\equiv&
%\frac {1}F\, G(p)\,(-i\bfrho_2)\,{\bf M}_{\bf T}^{\,\,A}\;
%{\bf \Lambda}(p)
%\end{array}
%\label{DEC2}
%\end{equation}
(\ref{Vvertex}) inserted into (\ref{DEC1}) and by (\ref{PSvertex}).
%where
The general structure of the vertex (\ref{DEC1}) and the lack of Lorentz
invariance in
the CFL phase yield eight form factors,
\begin{eqnarray}
{\cal V}^{ABC}_{\mu}(P,Q)&=&+
f^{ABC}\left\{{h}^f_E (P,Q) \,P_0 + {g}^f_E (P,Q)\,
Q_0\right\}\,\delta_{\mu 0}\nonumber\\
&&+f^{ABC}\left\{{h}^f_M (P,Q) \,{\bf P}_i + {g}^f_M (P,Q)\,
{\bf Q}_i\right\}\,\delta_{\mu i}\nonumber\\
&&+d^{ABC}\left\{{h}^d_E (P,Q) \,P_0 + {g}^d_E (P,Q)\,
Q_0\right\}\,\delta_{\mu 0}\nonumber\\
&&+d^{ABC}\left\{{h}^d_M (P,Q) \,{\bf P}_i + {g}^d_M (P,Q)\,
{\bf Q}_i\right\}\,\delta_{\mu i}.
\label{DEC4}
\end{eqnarray}
The electric and magnetic couplings are
${g}_E\approx {g}_E (0,0)$ and
${g}_M\approx {g}_M (0,0)$.
Setting $P=0$ and $Q=(M,0)$, the electric coupling is seen to vanish
because of the mismatch in spin structure (i.e.,
tracing to an uncompensated $\gamma_0$, see (\ref{SpinStructure}))
in the chiral limit, i.e. ${g}_E=0$. Setting $P=0$ and $Q=(0,{\bf Q})$
fixes the magnetic coupling. 
In terms of (\ref{Vvertex}) the composite vertex reads
\begin{eqnarray}
   {\cal V}^{ABC}_{T,L}(P,Q)
   &=&
 \frac{1}{F_V F^2}\int\frac{d^4q}{(2\pi)^4}\,
      \Gamma_{T,L} (q, M_V)\, G^2 (q\mbox{$-$}\frac Q2)\,
  \frac{1}{(q_0^2-\epsilon_q^2)^2}\,
 \frac{1}{q_0^2-\epsilon_{q-Q}^2}\nonumber\\
&&\qquad \times
{\rm Tr}_{fc} \left(\tau^A\left[\tau^B\,,\,\tau^C\right]_+ \right)
\nonumber\\
  &&
 \mbox{}\qquad\times \left\{ G(q)  \left[ q_{-}
   Q_{+}-q_{+}Q_{-}\right]
\right. \nonumber \\
&&\qquad\quad\times
 {\rm  Tr}_s\left[\Lambda^{-}({\bf q}) \gamma_{T,L}
  \Lambda^{-}({\bf q})\Lambda^{-}({\bf {q-Q}})+
 \left(\Lambda^{-}\rightarrow\Lambda^{+}\right)\right]
\nonumber \\
  && \qquad
 \mbox{} - G(q-Q)\left[q_{+}q_{-}-2 G(q)^2\right] \nonumber \\
&&\qquad\quad\times\left.
 {\rm  Tr}_s\left[\Lambda^{-}({\bf q}) \gamma_{T,L}
  \Lambda^{-}({\bf q})\Lambda^{-}({\bf {q-Q}})-
 \left(\Lambda^{-}\rightarrow\Lambda^{+}\right)\right]\right\}\nonumber\\
\label{VERTEXVPP}
\end{eqnarray}
with $q_\pm=q_0\pm q_{||}$,   $Q_\pm =\pm Q_{||}$ and
$\gamma_{L,T}= \gamma_i{\cal E}^{LT}_i$.
Equation (\ref{VERTEXVPP}) is identically
zero, since the spin structure is of the form
\[ {\rm Tr}_s( \gamma^\mu  \alpha^n)=0 \quad \mbox{for}\
\mu=0,1,2,3\quad\mbox{and}\ n=0,1,2,\cdots\;.
\]
In fact, one could have seen this already by inspecting
(\ref{DEc11}) as it contains only one $\gamma^\mu$ and always an even
number of $\gamma^0$'s (either 2 or 0)
from the propagator (\ref{PropPart}) (either one $S_{11}$ and one
$S_{22}$ appear together or only $S_{12}$'s and $S_{21}$'s). This
holds for any value of $P$ and $Q$. Hence $V\rightarrow \pi\pi$ 
vanishes to leading logarithm accuracy and in the chiral
limit in the CFL phase, making the vector excitations real
(zero width).

\vskip 1.5cm
\centerline{\bf 8. Vector Meson Mixing with Gluons}
\vskip 1cm

In the CFL phase the static electric and magnetic gluons are
respectively screened and expelled (Meissner effect). Both the
screening mass and the Meissner mass are of order $g\mu$ which
is large on the scale of the superconductor excitations. So for
all purposes, the static gluons decouple. For the {\em nearly static}
gluons with energy $Q_0\approx G_0$, both the screening and the
Meissner effect in the superconductor weakens substantially. 
Indeed, it is the {\em nearly static magnetic gluons} which are not
screened but only Landau damped that cause the binding of the
composite pairs and their excitations in weak coupling with a
magnetic scale $m_M$. In weak coupling, the vector
mesons are dominant with $M_V\ll m_M$. 

\begin{figure}
\centerline{\epsfig{file=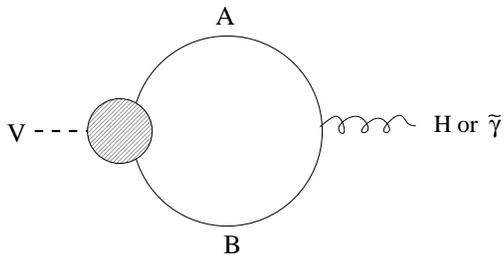,height=1.3in}}
\caption{Composite vector meson transition
into a gluon (H) or a tilde photon ($\tilde\gamma$)
in the QCD superconductor.}
\label{Fig4}
\end{figure}

To analyze the mixing of the transverse composite vector mesons
with the transverse magnetic gluons in the intermediate regime
$M_V\sim m_M$,  we define the 2-component vector fields
$(\rho_i^A, H_i^A)$. Then the mixed propagator for the
transverse modes reads
\begin{eqnarray}
\left(\begin{array}{cc} {\Delta}_V^{-1} (Q)& {\Pi} (Q)\\
{\Pi} (Q)& {\Delta}_H^{-1} (Q)
\end{array}\right)^{\alpha\beta}_{ij} \; ,
\label{RG1}
 \end{eqnarray}
where the diagonal transverse propagators are
\begin{eqnarray}
&&{\Delta}_{V, ij}^{-1, \alpha\beta} (Q)=\delta^{\alpha\beta}
\left(\delta_{ij}-\hat{\bf Q}_i\hat{\bf Q}_j\right)\,
( Q_0^2-v_V^2{\bf Q}^2-M_V^2)^{-1}\,F_{V,T}^2\; ,\nonumber\\
&&{\Delta}_{H, ij}^{-1, \alpha\beta} (Q)=\delta^{\alpha\beta}
\left(\delta_{ij}-\hat{\bf Q}_i\hat{\bf Q}_j\right)\,
( Q_0^2-v_H^2{\bf Q}^2-M_H^2)^{-1}\,F_{H,T}^2
\label{RG2}
\end{eqnarray}
with $v^2=F_S^2/F_T^2$. For the magnetic
gluons $F_{H,T}$ and $F_{H,S}$
are related to the electric color susceptibility
and magnetic permittivity in the superconductor. Their explicit form will
not be needed
for our arguments. The off-diagonal part of the mixed propagator
(\ref{RG1}) follows from Fig.~\ref{Fig4}. Hence
\begin{equation}
{\Pi}_{ij}^{\alpha\beta} (Q) =
-ig \int\frac{d^4q}{(2\pi)^4}\,\,
{\rm Tr}\left(\,{\cal V}_i^{\alpha}\,i{\bf S} (q+\frac Q2) \,
i{\bf \Gamma}^{\beta}_j (q, Q) \, i {\bf S} (q-\frac Q2)\right)
\label{RG3}
\end{equation}
which is found to vanish because one of the $\gamma_0$ from
${\bf S}$ is not compensated at the gluon edge, see also
(\ref{SpinStructure}). In the chiral limit and in leading
logarithm accuracy, therefore, the composite and transverse vector mesons
decouple from the transverse gluons. If any, mixing must occur
at next-to-leading logarithm order or under explicit breaking of 
chiral symmetry.

\vskip 1.5cm
\centerline{\bf 9. Hidden Gauge Symmetry}
\vskip 1cm

We have seen that in weak coupling, 
the composite vector mesons are distinct
from the screened and Higgsed gluons. Could they be the
realization of a hidden local symmetry in the CFL phase,
besides the explicit local color symmetry? Furthermore,
could the hidden local
symmetry of the flavor sector be ``dual''
to the local color symmetry?\footnote{
A similar issue is addressed in \cite{harada} in a different context.}
To answer these particular questions, we recall that in the CFL
phase the color-flavor locking generates multidegenerate
phases~\cite{CFL_ARW,CFL_SW}, 
characterized by the following order parameter
\begin{equation}
\left\langle\overline{\bf \Psi}(x)\,
{\bf M}^{i\alpha}\left(e^{-i\gamma_5\pi^A
{\bf T}^A}\right)^{i\alpha}\,\bfrho_2\,{\bf \Psi}(y)\,\right\rangle
\neq 0 \; .
\label{CFL1}
\end{equation}
Recall ${\bf M}^{i\alpha}=\epsilon^i_f\epsilon^{\alpha}_c\,\gamma_5$,
$\bfrho_2$ a Pauli matrix active on the Nambu-Gorkov entries,
and ${\bf T}^A={\rm diag}\,(\tau^A,\tau^{A,*})$ an $SU(3)_{c+F}$ valued
generator in the Nambu-Gorkov representation.
The CFL phase is invariant under the diagonal
of rigid vector-color plus vector-flavor, i.e. $SU(3)_{c+V}$.

As suggested in \cite{RWZ}, in the CFL phase the generalized pions
can be regarded as bound
states of pairs of particles or holes. Because of the
degeneracy (\ref{CFL1}), they may also be approximately described
by $SU(3)_{c+A}$ valued excitations in the coset
$(SU(3)_c \times SU(3)_L\times SU(3)_R)/SU(3)_{c+V}$,
in the long-wavelength and zero-size limit~\cite{HRZ,GATTO}.
We note that (\ref{CFL1}) can also be rewritten as
\begin{equation}
\left\langle\overline{\bf \Psi}(x)\,
{\bf M}^{i\alpha}\left(\xi_f^{-1}\,\xi_c^{-1}\right)^{i\alpha}
\,\bfrho_2\,{\bf \Psi}(y)\,\right\rangle
=\left\langle\overline{\bf \Psi}(x)\,
\left(\xi_f^T\,\epsilon^\alpha_f\,\xi_f\right)\,
\left(\xi_c^T\,\epsilon^\alpha_c\,\xi_c\right)\,
\gamma_5\,\bfrho_2\,{\bf \Psi}(y)\,\right\rangle
\neq 0 \; ,
\label{CFl2}
\end{equation}
where we have used the unitary gauge
$\xi_c=\xi_f=e^{i\gamma_5\pi^A{\bf T}^A/2}$, and
the identity
\begin{equation}
\left(\epsilon^{c}\right)^{\alpha\beta} \,\left(\xi^{-1}\right)^{cc'}\equiv
\epsilon^{c\alpha\beta} \,\left(\xi^{-1}\right)^{cc'}=
\left(\xi^T\,\epsilon^{c'}\,\xi\right)^{\alpha\beta}\,\,.
\label{CFL3}
\end{equation}
For constant $\pi^A$, the rigid rotations $\xi_f$ and
$\xi_c$ can
be reabsorbed through $\xi_f\,\xi_c{\bf \Psi}\rightarrow
{\bf \Psi}$, leaving invariant the equations in the QCD
superconductor. This rigid degeneracy is at the origin of the
Goldstone modes in the CFL phase.

For composite pairs we observe that (\ref{CFl2})
enjoys a local symmetry through
\begin{equation}
e^{i\gamma_5\pi^A \,{\bf T}^A}
=\xi_f\,\xi_c =\xi_f\, h(x)^{-1}\, h(x)\xi_c \; ,
\label{CFL4}
\end{equation}
where $h(x)$ is an element of local $SU(3)_{c+V}$.
For {\em finite size} pairs, the local invariance (\ref{CFL4})
cannot be transported and reabsorbed in the fermionic
fields ${\bf \Psi}(x)$ and ${\bf \Psi}(y)$ as
they carry different arguments. Hence, strictly speaking,
there is no hidden
symmetry for local $SU(3)_{c+V}$ besides the original local color
symmetry, in general.

However, in the limiting case where $x\rightarrow y$, and the size of the
pair can be ignored~\footnote{We stress that our exact
calculation was rendered possible by the natural cutoff provided
by the finite size of the composite. To what extent the zero-size
approximation can be valid is not clear for the system in
question. This caveat may seem to also apply to effective field
descriptions of hadrons in zero-density environment. The
light-quark hadrons such as $\pi$, $\rho$, $\omega$ etc.\ in the
matter-free vacuum are of course finite-sized but nonetheless can
be given an effective field theory description in terms of local
chiral Lagrangians with hidden gauge symmetry etc. In such a
description, the finite size is naturally accounted for by
higher-order terms in chiral perturbation series. The resolution
of this issue in the present case will have to involve going
beyond the weak coupling and leading-log approximations that are
not addressed here.},
then (\ref{CFL4}) is a hidden symmetry
in the QCD superconductor. Indeed, the effective phases
$\xi_{f,c}$ can be made local, and their corresponding
effective action is invariant under the transformations~\cite{HRZ}
\begin{equation}
\xi_f (x)\rightarrow g_f \,\xi_f (x)\, h^{-1} (x)\; ,\qquad\qquad
\xi_c (x)\rightarrow h(x) \,\xi_c (x)\, g_c^{-1}\; ,
\label{TRANS}
\end{equation}
where $g_f$ and $g_c$ are rigid flavor and color transformations.
The effective action for $\xi_{f}(x)$ and $\xi_c (x)$ or equivalently
their chiral left-right unitary fields,
in the zero size approximation
was originally discussed in~\cite{HRZ}. As a result of the local
invariance (\ref{TRANS}), the
vector mesons composed of pairs of particles or holes
can be regarded as gauge particles of the hidden and local $SU(3)_{c+V}$
in the zero size approximation
\footnote{In~\cite{GATTO}, the hidden gauge symmetry was
identified with the local color gauge group. Here it is clearly
a local symmetry of the QCD superconductor when the pairs are
assumed of zero size. The corresponding gauge particles are composite
pairs of particles and holes as noted in~\cite{HRZ}. 
%and fully characterized here. 
The hidden gauge symmetry arrived at zero-size limit
may be implying a ``dual" relation
between the two as in \cite{harada}.}. They couple minimally to the
generalized
pions, and their properties follow from general principles~\cite{BANDO}.
In particular, their mass is given as $M_V^2=2F^2\,g_{V\pi\pi}^2$ (KSRF-II)
and their coupling to the CFL photon is $g_V=2F^2\,g_{V\pi\pi}$ (KSRF-I),
which are both seen to mix orders in weak coupling. 
In the normal (non-superconducting) phase, the photons only couple through
$\rho$-mesons leading to the concept of vector dominance (VDM), which is
usually manifest through $g_{V\pi\pi}=g_{SU(3)_{c+V}}$ (universality).
In this
limit, the hidden gauge symmetry must be identical to the broken
color symmetry $SU(3)_c$.

Since the pairs in the QCD superconductor have finite size, our
results show that the concepts of hidden gauge symmetry and VDM
are only approximate~\footnote{It is not surprising that such
concepts make precise sense only for modes that can be described
by local fields. This situation is analogous to the role of VDM in
baryon structure. Because of the finite skyrmion size, large $N_c$
effective theories implemented with VDM are not as successful for
baryon electromagnetic properties as they are for mesons.}, and do
not hold in weak coupling and leading-log approximation. We recall
that in weak coupling, the pairs are very close in space
(separation of order $1/\mu$) but far in time (separation of order
$1/m_M\approx 1/(m_E^2G_0)^{1/3}\gg 1/\mu$).

\vskip 1.5cm
\centerline{\bf 10. Conclusions}
\vskip 1cm

We have analyzed the generalized scalar, pseudoscalar, vector and 
axial-vector
excitations in the CFL superconductor in the weak coupling limit. 
We have confirmed that the octet scalar and pseudoscalar excitations
are both massless, and that only the pseudoscalars survive as Goldstone modes,
while the scalars are Higgsed by the gluons leading to the Meissner effect.
We have found that the vectors and axial
vectors are bound and degenerate irrespective 
of their polarization, with a mass
that is less than $2G_0$. Chiral
symmetry is explicitly realized in the vector spectrum in the CFL phase in
leading logarithm approximation, in spite of its breaking in general. 
In the CFL superconductor the vector mesons are characterized by form
factors that are similar but not identical to those of the generalized
pions.

We have explicitly shown that the composite vector mesons can be
viewed as a gauge manifestation of a hidden local $SU(3)_{c+V}$
when their size is ignored (their form factor set to one).
In this limit, the effective Lagrangian description suggested
in~\cite{HRZ,GATTO,SONSTE} is valid with the vector mesons
described as Higgsed gauge bosons. Only in this limit, which
is clearly approximative, do we recover concepts such as vector
dominance and universality. (This is of course what one would expect
in the QCD vacuum as well.) 
In any event, the zero-size
limit is not compatible with the weak-coupling limit, because of
the long-range pairing mechanism at work at large quark chemical
potential. It is an open question whether going beyond the
weak-coupling and leading-log approximations would render the
concepts of effective field theories (e.g., HGS, VDM etc.) more
appropriate.

Although our arguments were exclusive to the CFL phase, it is clear that
they can be minimally changed to accommodate for the case of $N_f=2$,
which shows a qualitatively different form of superconductivity without
color-flavor locking, in particular there are no generalized pions. 
Modulo some color-flavor factors, we have checked
that our results carry over to the vector and axial-vector excitations.
They change minimally for the scalars at the origin of the Higgs mechanism,
also present for two flavors.

The existence of bound light scalar, 
vector and axial-vector mesons in QCD at high
density,  may have interesting 
consequences on dilepton and neutrino emissivities
in dense environments such as the ones encountered in neutron stars. 
For example, in young and hot neutron stars neutrino production via quarks
in the superconducting phase can be substantially modified if the vector
excitations are deeply 
bound with a non-vanishing coupling, a plausible situation 
in QCD in strong coupling.
These excitations may be directly seen by scattering electrons off 
compressed
nuclear matter 
(with densities that allow for a superconducting phase to form) 
and may cause substantial soft dilepton emission in the same 
energy range in ``cold" heavy-ion collisions.

\vskip 1.5cm
\section*{Acknowledgments}
\vskip .5cm
We thank Youngman Kim for help with the figures. MR
thanks Deog Ki Hong, Hyun Kyu Lee and Maciek Nowak for
discussions. This work was supported in part by US-DOE
DE-FG-88ER40388 and DE-FG02-86ER40251.

\begin{appendix}
\vskip 1.5cm
\section*{Appendix}
\vskip 0.5cm
\setcounter{equation}{0}
\renewcommand{\theequation}{A\arabic{equation}}

In this appendix, we give some of the missing steps in
deriving the formulae of the main text.

\vspace{1cm}
\noindent 1. Direct calculation of the propagator (\ref{PropPart})
including mass corrections (\ref{GOR11}):
\vspace{5mm}

According to Ref.~\cite{PisarskiSuperfluid}, the general entries 
for ${\bf S}(q)$ read
\begin{equation}
\begin{array}{lclcl}
 S_{11}(q) &=& - i\left\langle\,\psi(q)\bar\psi(q)\,\right\rangle
           &=& \left\{ \left(G_0^+(q)\right)^{-1}
           -\gamma^0\,\Delta^\dagger(q)\, \gamma^0
           \,G_0^{-}(q)\, \Delta(q) \right\}^{-1}\; ,\\
 S_{12}(q) &=& - i\left\langle\,\psi(q)\bar\psi_C(q)\,\right\rangle
            &=& - G_0^{+}(q)\, \gamma^0 \Delta^\dagger(q) \gamma^0\, 
 S_{22}(q)\; ,
           \\
 S_{21}(q) &=& - i\left\langle\,\psi_C(q)\bar\psi(q)\,\right\rangle
           &=& - G_0^{-}(q)\, \Delta(q)\, S_{11}(q) \; ,\\
 S_{22}(q)  &=& - i\left\langle\,\psi_C(q)\bar\psi_C(q)\,\right\rangle
            &=& \left\{ \left(G_0^-(q)\right)^{-1}
           -\Delta(q)\,
           G_0^{+}(q)\,\gamma^0 \Delta^\dagger(q)\gamma^0 \right\}^{-1}
 \label{PropPartFull}
\end{array}
\end{equation}
with 
\begin{eqnarray}
  \left(G^{\pm}_0(q)\right)^{-1} &=& \gamma_\mu q^\mu \pm \mu \gamma^0 -m
  \nonumber \\
  &=& \gamma^0 \left(q_0 \pm \mu -\bfalpha\cdot {\bf q} \right) - m\; ,
 \label{G0}
\end{eqnarray}
where, in general $m={\rm diag}(m_u,m_d,m_s)$.
Furthermore,
\begin{equation}
  \Delta(q)= {\bf M} G(q) \Lambda^+({\bf q}) + {\bf M} {\overline G}(q) 
\Lambda^{-}({\bf q})
  \; ,
 \label{Delta}
\end{equation}
where ${\bf M}= \epsilon_f^a\epsilon_c^a\,\gamma_5= {\bf M}^\dagger$ with
$(\epsilon^a)^{bc}=\epsilon^{abc}$.
Note that
\begin{equation}
  \gamma^0\Delta^\dagger(q)\gamma^0= 
 -{\bf M}^\dagger G^\ast(q) \Lambda^-({\bf q}) 
 -{\bf M}^\dagger {\overline G}^\ast(q) 
\Lambda^{+}({\bf q})\; 
 \label{Deltadag}
\end{equation}
and
\begin{equation}
   \left({\bf M}^\dagger {\bf M}\right)^{\alpha\beta}_{ij}
 = \delta_{\alpha\beta}\delta_{ij}+
   \delta_{\alpha i}\delta_{\beta j} \; ,
 \label{MM}
\end{equation}
where $\alpha,\beta=1,2,3$ and $i,j=1,2,3$ are color and flavor
indices, respectively.

Inserting (\ref{G0}) and (\ref{Delta}) and (\ref{Deltadag}) into 
$S_{11}(q)$, we get up to second order in $m$: 
% and by neglecting the second  
% term in (\ref{MM}) (which is off-diagonal):
\begin{eqnarray}
S_{11}(q)&\approx & \left\{ \frac{\mbox{}}{\mbox{}}
(\gamma\cdot q + \mu \gamma^0) - m -
  {\bf M}^\dagger 
 \frac{\gamma\cdot q  -\mu \gamma^0}{(q_0\mbox{$-$}\mu)^2\mbox{$-$}|
{\bf q}|^2\mbox{$-$}m^2}
\, {\bf M} \left(|G(q)|^2 \Lambda^{+}({\bf q}) 
+ |{\overline
  G}(q)|^2\Lambda^{-}({\bf q})
  \right) \right. \nonumber \\
   && \left. 
\qquad + {\bf M}^\dagger m \,
 \frac{G^\ast(q) {\overline G}(q)\Lambda^-({\bf q}) 
 +{\overline G}^\ast(q) G(q) 
\Lambda^{+}({\bf q})} 
{(q_0-\mu)^2 -|{\bf q}|^2 - m^2}{\bf M} \right\}^{-1}  \nonumber\\
&=&
\left\{ ( \gamma \cdot q - \mu \gamma^0) (\gamma \cdot q
  + \mu \gamma^0) \right. \nonumber \\
  &&\quad - \left({\bf M}^\dagger {\bf M}
 +\frac{{\bf M}^\dagger m^2 {\bf M}}{(q_0-\mu)^2 -|{\bf q}|^2}\right) 
\,
\left(|G(q)|^2 \Lambda^{+}({\bf q}) + |{\overline G}(q)|^2
  \Lambda^{-}({\bf q})\right)
 \nonumber \\
 &&\quad -\left. (\gamma \cdot q - \mu \gamma^0) \left(m -
{\bf M}^\dagger m {\bf M}\,
 \frac{ G^\ast(q)
  {\overline G}(q) \Lambda^{-}({\bf q}) + {\overline G}^\ast(q)
  {G}(q) \Lambda^{+}({\bf q}) }
{(q_0-\mu)^2 -|{\bf q}|^2} \right)\right\}^{-1} \nonumber \\
&& \qquad\qquad \qquad\qquad\qquad\qquad\qquad\qquad\qquad\qquad \times
  (\gamma\cdot q - \mu \gamma^0)\; . \label{S11h}
\end{eqnarray}
Using 
\[
(\gamma\cdot q - \mu \gamma^0) (\gamma\cdot q +\mu \gamma^0)
= \{q_0^2 - (\mu -|{\bf q}|)^2\} \Lambda^+({\bf q}) 
  + \{q_0^2 - (\mu +|{\bf q}|)^2\}\Lambda^{-}({\bf q}) \; ,
\]
we can transform (\ref{S11h}) into
\begin{eqnarray}
S_{11}(q) &=& 
 \left \{ \Lambda^+({\bf q}) \left[q_0^2- (\mu-|{\bf q}|)^2
  - \left({\bf M}^\dagger {\bf M} + \frac{{\bf M}^\dagger m^2 {\bf M}}
 {(q_0-\mu)^2 - |{\bf q}|^2}\right)\,|G(q)|^2 
 \right] \Lambda^+({\bf q})  \right. \nonumber\\
&&
+  \Lambda^-({\bf q}) \left[q_0^2- (\mu+|{\bf q}|)^2
- \left({\bf M}^\dagger {\bf M} + 
\frac{{\bf M}^\dagger m^2 {\bf M}}{(q_0-\mu)^2 - |{\bf q}|^2}\right)
 \, |{\overline G}(q)|^2
 \right] \Lambda^-({\bf q})  \nonumber\\
&&  -  \Lambda^+({\bf q})\, \gamma^0\left[q^0-\mu +|{\bf q}|\right]
 \left(m-  {\bf M}^\dagger m {\bf M}\,
 \frac{G^\ast(q) {\overline G}(q)}{(q_0-\mu)^2-|{\bf
 q}|^2}\right) \ \Lambda^-({\bf q}) \nonumber \\
&& \left. -  \Lambda^-({\bf q})\,  
 \gamma^0 \left[q^0-\mu -|{\bf q}|\right]
 \left(m- {\bf M}^\dagger m {\bf M}\, 
 \frac{{\overline G}^\ast(q) {G}(q)}{(q_0-\mu)^2-|{\bf
 q}|^2}\right)\  \Lambda^+({\bf q}) \nonumber \right\}^{-1} \nonumber
 \\
&&\qquad\qquad\qquad\qquad\qquad\qquad\qquad\qquad \times
\gamma^0
 (q_0-\mu -\bfalpha\cdot {\bf q}  )\; .
 \label{S11hh}
\end{eqnarray}
Note that (\ref{S11hh}) has the structure
\begin{eqnarray}
 S_{11}(q) &=&\left\{\Lambda^+\, A\, \Lambda^+ 
 + \Lambda^-\, D \, \Lambda^-
 + \Lambda^+\, B\, \Lambda^- + \Lambda^-\, C\, \Lambda^+ \right\}^{-1}
  \,\gamma^0
(q_0-\mu -\bfalpha\cdot {\bf q}  ) \nonumber\\
 &\equiv& \left\{ F \right\}^{-1} \,\gamma^0
 (q_0-\mu -\bfalpha\cdot {\bf q}  ) 
 \;   
\end{eqnarray}
with
\begin{eqnarray}
 A &=& q_0^2- \bfeps_q^2 
            - {\bf M}^\dagger m^2{\bf M}
             \frac{ |G(q)|^2}{(q_0-\mu)^2 -|{\bf q}|^2}\; , 
             \nonumber \\
 D &=& q_0^2- {\overline\bfeps}_q^2 
             - {\bf M}^\dagger m^2 {\bf M}\, 
              \frac{ |{\overline G}(q)|^2}{(q_0-\mu)^2 -|{\bf q}|^2}\;,
             \nonumber \\
 B &=& -\gamma^0 \left[q_0 - \mu +|{\bf q}|\right] 
      \left(m- {\bf M}^\dagger m {\bf M}\,
 \frac{G^\ast(q) {\overline G}(q)}{(q_0-\mu)^2-|{\bf
 q}|^2} \right)\;, \nonumber \\
 C &=& - \gamma^0 \left[q_0 - \mu -|{\bf q}|\right] 
      \left(m- {\bf M}^\dagger m {\bf M}\,
 \frac{{\overline G}^\ast(q) {G}(q)}{(q_0-\mu)^2-|{\bf
 q}|^2} \right) \; , \nonumber \\
 \label{A-D}
\end{eqnarray} 
where 
\begin{eqnarray}
\bfeps_q &\equiv& (\mu -|{\bf q}|)^2 
       + {\bf M}^\dagger {\bf M} | G(q) |^2\; , \label{eq}\\
{\overline \bfeps}_q &\equiv& (\mu +|{\bf q}|)^2 
       + {\bf M}^\dagger {\bf M} |{\overline   G}(q)|^2 \; . 
   \label{eqbar}
\end{eqnarray}
Here the projectors $\Lambda^{\pm}$ are short for 
 $\Lambda^\pm ({\bf q})$, and satisfy 
$\Lambda^\pm \Lambda^{\pm}= \Lambda^{\pm}$ and
$\Lambda^+ + \Lambda^{-} = {\bf 1}$, where the unit matrix refers to
the Dirac space.

We will use now that
the inverse of 
\[
 F\equiv \Lambda^+\, A\, \Lambda^+ 
  + \Lambda^+\, B\, \Lambda^- + \Lambda^-\, C\, \Lambda^+ + \Lambda^-\, D
 \, \Lambda^-
\]
reads
\begin{eqnarray*}
F^{-1}&=&  \Lambda^+\, A^{-1}\,  \Lambda^+  
        +\Lambda^-\, D^{-1}\,\Lambda^- \\
      && \mbox{} -\Lambda^+\, A^{-1}\, B \, D^{-1}\, \Lambda^- 
        -\Lambda^-\, D^{-1}\, C\, A^{-1}\, \Lambda^+\\
      & & \mbox{} + \Lambda^+\, A^{-1}\, B\, D^{-1}\,  C\, A^{-1}\, \Lambda^+
        + \Lambda^-\, D^{-1}\, C\, A^{-1}\, B\, D^{-1}\, \Lambda^-\\
      && \mbox{}+ {\cal O}(m^3) \; 
\end{eqnarray*} 
and furthermore that 
$\gamma^0 \Lambda^{\pm} = \Lambda^{\mp} \gamma^0$
and $\Lambda^{\pm}({\bf q}) \bfalpha\cdot {\bf q} 
= \pm |{\bf q}|\Lambda^{\pm}({\bf q}) $.
Then, we have
\begin{eqnarray}
 S_{11}(q) &\approx& 
 \frac{ \gamma^0 \Lambda^{-}({\bf q}) (q_0 - \mu +|{\bf q}|)}
                 {q_0^2 -\bfeps_q^2}
            +\frac{ \gamma^0 \Lambda^{+}({\bf q}) (q_0 - \mu -|{\bf q}|)}
                 {q_0^2 -{\overline\bfeps}_q^2} 
 \nonumber \\  
           && \mbox{}+ \left((q_0 -\mu)^2 -|{\bf q}|^2\right) 
 \left\{ \frac{1}{q_0^2-\bfeps_q^2} m
          \frac{1}{q_0^2-{\overline\bfeps}_q^2} \, \Lambda^{+}({\bf q})
     + \frac{1}{q_0^2-{\overline\bfeps}_q^2} m
          \frac{1}{q_0^2-{\bfeps}_q^2} \, \Lambda^{-}({\bf q}) \right\}
 \nonumber  \\
 && \mbox{}-G^\ast(q) {\overline G}(q)\,\frac{1}{ q_0^2 -\bfeps_q^2}
   {\bf M}^\dagger m {\bf M} \frac{1}{q_0^2 -{\overline\bfeps}_q^2}
  \Lambda^+({\bf q})
\nonumber \\
&& \mbox{}    -{\overline G}^\ast(q) {G}(q)
 \,\frac{1}{ q_0^2 -{\overline\bfeps}_q^2}
   {\bf M}^\dagger m {\bf M} \frac{1}{q_0^2 -{\bfeps}_q^2}
  \Lambda^-({\bf q})
\nonumber  \\
  &&\mbox{} + \frac{1}{q_0^2-\bfeps_q^2}\,
 {\bf M}^\dagger m^2 {\bf M}\, \frac{|G(q)|^2}
 { (q_0-\mu)^2-|{\bf q}|^2}
\ \frac{ \gamma^0 \Lambda^{-}({\bf q}) (q_0 - \mu +|{\bf q}|)}
                 {q_0^2 -\bfeps_q^2}\nonumber \\
 && \mbox{}+ \frac{1}{q_0^2-{\overline\bfeps}_q^2}\, 
  {\bf M}^\dagger m^2 {\bf M}\, \frac{
    |{\overline G}(q)|^2}{ (q_0-\mu)^2- |{\bf q}|^2}
\ \frac{ \gamma^0 \Lambda^{+}({\bf q}) (q_0 - \mu -|{\bf q}|)}
                 {q_0^2 -{\overline\bfeps}_q^2}
\nonumber \\
 && \mbox{} 
 + \frac{(q_0-\mu)^2-|{\bf q}|^2}{q_0^2-\bfeps_q^2}\,
  \left(m-{\bf M}^\dagger m {\bf M}
  \frac{G^\ast(q) {\overline G}(q)}{(q_0-\mu)^2-|{\bf q}|^2} \right)
  \frac{1}{q_0^2-{\overline\bfeps}_q^2}\, 
\nonumber \\
 && \qquad\qquad \mbox{}\times
\left(m-{\bf M}^\dagger m {\bf M} 
 \frac{{\overline G}^\ast(q) 
 { G}(q)}{(q_0-\mu)^2-|{\bf q}|^2} \right)    
 \,\frac{\gamma^0 \Lambda^{-}({\bf q}) (q_0 - \mu +|{\bf q}|)}
        {q_0^2-\bfeps_q^2} 
\nonumber \\
 && \mbox{} +\frac{(q_0-\mu)^2-|{\bf q}|^2}{q_0^2-{\overline\bfeps}_q^2}
  \left(m-{\bf M}^\dagger m {\bf M}\,
 \frac{{\overline G}^\ast(q) {G}(q)}{(q_0-\mu)^2-|{\bf q}|^2} \right) 
\frac{1}{q_0^2 - \bfeps_q^2}
 \nonumber \\
&& \qquad\qquad \mbox{}\times 
  \left(m -{\bf M}^\dagger m {\bf M}\,
 \frac{{G}^\ast(q) {\overline G}(q)}{(q_0-\mu)^2-|{\bf
                 q}|^2} \right)   
 \,  \frac{\gamma^0 \Lambda^{+}({\bf q}) (q_0 - \mu -|{\bf q}|)}
        { q_0^2 -{\overline\bfeps_q}^2}        
\nonumber \\
&& +{\cal O}(m^3) \; .
 \label{S11h3}
\end{eqnarray}
We apply now the following approximations
\begin{eqnarray}
 \bfeps_q^2 &=& (\mu - |{\bf q}|)^2 + {\bf M}^\dagger {\bf M}
 |G(q)|^2 \approx  (\mu - |{\bf q}|)^2 + |G(q)|^2\equiv 
\epsilon_q^2 \; ,\nonumber \\
{\overline\bfeps}_q^2 &=& (\mu + |{\bf q}|)^2 
 +{\bf M}^\dagger {\bf M}
 |{\overline G}(q)|^2 \approx  (\mu +|{\bf q}|)^2 
 + |{\overline G}(q)|^2\equiv 
{\overline\epsilon}_q^2 \;,
 \label{epsapprox}
\end{eqnarray}
and, since $|{\bf q}| \approx \mu + q_{||}$, 
\begin{eqnarray}
&\epsilon_q^2 \approx q_{||}^2 + |G(q)|^2 \; ,& \label{eqeq} \\
&{\overline
  \epsilon_q^2}\approx 4 \mu^2 \; ,&
 \label{eqeqbar} \\
&(q_0-\mu)^2 -|{\bf q}|^2 \approx -2\mu (q_0 +q_{||})\;. &
 \label{q0minus}
\end{eqnarray}
Then  equation (\ref{S11h3}) simplifies to
\begin{eqnarray}
S_{11}(q) &=& \gamma^0\,\frac{ q_0 +q_{||}}
                 {q_0^2 -\epsilon_q^2}\, \Lambda^{-}({\bf q}) + 
 \frac{\gamma^0 \Lambda^{+}({\bf q})}{2\mu} 
+ \frac{m}{2\mu}\, \frac{q_0+q_{||}}{q_0^2-\epsilon_q^2}
\nonumber \\
&&\mbox{} 
+\gamma^0\, \frac{m^2}{2\mu}\,\left( \frac{ q_0+q_{||}}
 {q_0^2-\epsilon_q^2}\right)^2\,  \Lambda^{-}({\bf q})
 - \gamma^0\,\frac{{\bf M}^\dagger m^2 {\bf M}}{2\mu}
\left( \frac{ |G(q)|}
{q_0^2-\epsilon_q^2}\right)^2\, \Lambda^{-}({\bf q})\, \nonumber \\
&& \mbox{}+
{\cal O}\left(\mu^{-2}, m^3, 
|G(q)|^2\,\mbox{$\frac{{\bf M}^\dagger {\bf M}-{\bf 1}}
{\left(q_0^2-\epsilon_q^2\right)^2}$}\right)\; .
 \label{S11}
\end{eqnarray}
Note that
$S_{22}(q)$ follows from (\ref{S11h3}) under the
substitutions 
\begin{equation}
 \Lambda^\pm({\bf q}) \leftrightarrow \Lambda^\mp({\bf q}) \,,\  \  
 \pm \mu \leftrightarrow \mp \mu\,, \ \ 
 \pm |{\bf q}| \leftrightarrow \mp |{\bf q}|\,,\ \ 
  \pm G^\ast  \leftrightarrow \mp G \,, \ \ 
\pm {\overline G}^\ast \leftrightarrow \mp {\overline G}\,,\ \
 \mbox{and}\; {\bf  M}^\dagger \leftrightarrow {\bf M}\, ,
 \label{substitute}
\end{equation}
which also imply 
$ \pm q_{||} \leftrightarrow \mp q_{||}$.
In fact, these rules can be traced back to 
the replacements  $G_0^+(q) \leftrightarrow G_0^-(q)$ and
$\Delta(q) \leftrightarrow 
\gamma^0\,\Delta^\dagger(q)\, \gamma^0$  which link the various
Nambu-Gorkov components  in (\ref{PropPartFull}) to each other.
Using (\ref{eqeqbar}) and
\begin{equation}
 (q_0+\mu)^2- |{\bf q}|^2 \approx 2\mu (q_0-q_{||})\;,
 \label{q0plus}
\end{equation}
we get
\begin{eqnarray}
S_{22}(q) &=& \frac{ \gamma^0 \Lambda^{+}({\bf q}) (q_0 -q_{||})}
                 {q_0^2 -\epsilon_q^2} 
 -  \frac{\gamma^0 \Lambda^{-}({\bf q})}{2\mu}
 - \frac{m}{2\mu}\, \frac{q_0-q_{||}}{q_0^2-\epsilon_q^2}
 \nonumber \\
&&\mbox{}
- \gamma^0\,\frac{m^2}{2\mu}\, \left(\frac{ q_0-q_{||}}
 {q_0^2-\epsilon_q^2}\right)^2\, \Lambda^{+}({\bf q})
 + \gamma^0\, \frac{{\bf M} m^2 {\bf M}^\dagger}{2\mu}
 \left(\frac{|G(q)|}
{q_0^2-\epsilon_q^2}\right)^2\, \Lambda^{+}({\bf q}) \nonumber \\
&& \mbox{}+
{\cal O}\left(\mu^{-2}, m^3, 
|G(q)|^2\,\mbox{$\frac{{\bf M}^\dagger {\bf M}-{\bf 1}}
{\left(q_0^2-\epsilon_q^2\right)^2}$}\right)\; .
\label{S22}
\end{eqnarray}
Finally,  
\begin{eqnarray}
S_{21}(q) &=& - G_0^{-}(q) \Delta(q) S_{11}(q) \\
          &=& - \frac{\gamma^0(q_0-\mu  -\bfalpha\cdot {\bf q} ) +m}
 {(q_0-\mu)^2-|{\bf q}|^2 -m^2} \times {\bf M} \left[ G(q)
          \Lambda^+({\bf q}) + {\overline G}(q) \Lambda^-({\bf
          q})\right] \times S_{11}(q)
 \nonumber \\
&=& + \left[ \frac{q_0-\mu - |{\bf q}|}
            {(q_0-\mu)^2-|{\bf q}|^2} \,
 {\bf M}\, G(q)  
      \Lambda^{-}({\bf q}) + \frac{q_0-\mu + |{\bf q}|}
            {(q_0-\mu)^2-|{\bf q}|^2} \, {\bf M}\,
{\overline G}(q) \Lambda^{+}({\bf q})\right]\,
 \gamma^0 \,S_{11}(q) \nonumber \\
&& \mbox{} -\frac{m}{(q_0-\mu)^2-|{\bf q}|^2}  
\, {\bf M}\, \left[ G(q)  
      \Lambda^{+}({\bf q}) + {\overline G}(q) \Lambda^{-}({\bf q})\right]\
\, S_{11}(q) \nonumber \\
 &&  \mbox{} +
 m^2
\left[ \frac{q_0-\mu - |{\bf q}|}
            {\left((q_0-\mu)^2-|{\bf q}|^2\right)^2} \,
 {\bf M}\, G(q)  
      \Lambda^{-}({\bf q}) + \frac{q_0-\mu + |{\bf q}|}
            {(q_0-\mu)^2-|{\bf q}|^2} \, {\bf M}\,
{\overline G}(q) \Lambda^{+}({\bf q})\right]\,\nonumber \\
&& \qquad\qquad\mbox{}\times \gamma^0 \,S_{11}(q) 
 + {\cal O}\left(m^3  \right)   \nonumber \\
&=& \frac{ {\bf M} G(q) 
      \Lambda^{-}({\bf q}) }{q_0^2-\epsilon_q^2} - \gamma^0\,
\frac{{\bf M} m}{2\mu}\, \frac{ G(q) \Lambda^{+}({\bf q})}
 {q_0^2-\epsilon_q^2} 
-\gamma^0\,\frac{m {\bf M}}{2\mu}\, \frac{ G(q) \Lambda^{-}({\bf q})}
 {q_0^2-\epsilon_q^2}  \nonumber \\  
&& + \frac{{\bf M} m^2}{2\mu}\,\frac{G(q) \left(q_0+q_{||}\right)
      \Lambda^{-}({\bf q}) }{\left(q_0^2-\epsilon_q^2\right)^2} 
- \frac{{\bf M}{\bf M}^\dagger m^2 {\bf M}}{2\mu}\,
 \frac{G(q)\, |G(q)|^2\, \Lambda^{-}({\bf q}) }
 {\left(q_0+q_{||}\right)\left(q_0^2-\epsilon_q^2\right)^2} 
 \nonumber \\
&& \mbox{}
- \frac{m^2}{2\mu}\,
\frac{ {\bf M} G(q) 
      \Lambda^{-}({\bf q}) }{\left(q_0+q_{||}\right)
\left(q_0^2-\epsilon_q^2\right)}
 \mbox{}  +{\cal O}\left(\mu^{-2}, m^3, 
|G(q)|^2\,\mbox{$\frac{{\bf M}^\dagger {\bf M}-{\bf 1}}
{\left(q_0^2-\epsilon_q^2\right)^2}$}\right)\; . \nonumber 
\end{eqnarray}
The latter equation can be simplified to
\begin{eqnarray}
S_{21}(q)
&=& \frac{ {\bf M} G(q) 
      \Lambda^{-}({\bf q}) }{q_0^2-\epsilon_q^2} - \gamma^0\,
\frac{{\bf M} m}{2\mu}\, \frac{ G(q) \Lambda^{+}({\bf q})}
 {q_0^2-\epsilon_q^2} 
-\gamma^0\,\frac{m {\bf M}}{2\mu}\, \frac{ G(q) \Lambda^{-}({\bf q})}
 {q_0^2-\epsilon_q^2}  \nonumber \\  
&& + \frac{1}{2\mu}\left\{ {\bf M} m^2 \left(q_0+q_{||}\right) 
                         -m^2 {\bf M} \left(q_0-q_{||}\right) \right\}
 \frac{G(q) 
      \Lambda^{-}({\bf q}) }{\left(q_0^2-\epsilon_q^2\right)^2} 
 \nonumber \\
&&\mbox{}  +{\cal O}\left(\mu^{-2}, m^3, 
 |G(q)|^2\,\mbox{$\frac{{\bf M}^\dagger {\bf M}-{\bf 1}}
{\left(q_0^2-\epsilon_q^2\right)^2}$}\right) \; .
 \label{S21}
\end{eqnarray}
Again, $S_{12}$ follows under the substitutions (\ref{substitute})
from $S_{21}$.
Using (\ref{eqeqbar}) and(\ref{q0plus}), we get 
\begin{eqnarray}
S_{12}(q)
&=& -\frac{ {\bf M}^\dagger G^\ast(q) 
      \Lambda^{+}({\bf q}) }{q_0^2-\epsilon_q^2} - \gamma^0\,
\frac{{\bf M}^\dagger m}{2\mu}\, \frac{ G^\ast(q) \Lambda^{-}({\bf q})}
 {q_0^2-\epsilon_q^2} 
-\gamma^0\,\frac{m {\bf M}^\dagger}{2\mu}\, \frac{ G^\ast(q) 
\Lambda^{+}({\bf q})}
 {q_0^2-\epsilon_q^2}  \nonumber \\  
&& + \frac{1}{2\mu}\left\{ {\bf M}^\dagger m^2 \left(q_0-q_{||}\right) 
                         -m^2 {\bf M}^\dagger 
\left(q_0+q_{||}\right) \right\}
 \frac{G^\ast(q) 
      \Lambda^{+}({\bf q}) }{\left(q_0^2-\epsilon_q^2\right)^2} 
 \nonumber \\
&&\mbox{}  +{\cal O}\left(\mu^{-2}, m^3, 
 |G(q)|^2\,\mbox{$\frac{{\bf M}^\dagger {\bf M}-{\bf 1}}
{\left(q_0^2-\epsilon_q^2\right)^2}$}\right) \; .
 \label{S12}
\end{eqnarray}
Note that 
the  above determined  quark propagators (\ref{S11}), (\ref{S22}),
(\ref{S21})
and (\ref{S12}) show neither a 
${\overline G}(q)$ nor a ${\overline G}^\ast (q)$ dependence. 
In fact, such terms first arise 
at order  ${\cal O}(\mu^{-2})$.

\vspace{1cm}
\noindent 2. Perturbative calculation of the mass corrections (\ref{GOR11})
to the propagator (\ref{PropPart}): 
\vspace{5mm}

As a check on the precedent analysis, we now carry a mass
perturbation analysis of the quark propagator in the CFL 
phase. We recall that the massless propagator 
reads~\cite{RWZ}~\footnote{These expressions can easily be derived
with the methods of the former section. Especially, 
they are consistent with
(\ref{S11}), (\ref{S22}), (\ref{S21}) and (\ref{S12}) to order 
${\cal O}(m^0,\mu^{-1})$ and under the approximations (\ref{eqeq}),
(\ref{eqeqbar}), (\ref{q0minus}) 
and (\ref{q0plus}).}
\begin{equation}
\begin{array}{lclcl}
 S_{11}(q)    &=& \left[\Lambda^{+}({\bf q})\,
                \gamma^0\,\frac{q_0-\mu +|{\bf q}| }{q_0^2-\bfeps_q^2}
         \,
 \Lambda^{-}({\bf q})
            +\Lambda^{-}({\bf q})\, 
 \gamma^0\,\frac{q_0-\mu -|{\bf q}|}{q_0^2-{\overline\bfeps}_q^2}
        \, \Lambda^{+}({\bf q})
  \right]
 \; ,
 \\[1mm]
 S_{12}(q)             &=& - 
  \left[\Lambda^{+}({\bf q})\,{\bf M}^\dagger\,
          \frac{G^\ast(q)}{q_0^2-\bfeps_q^2}
     \, \Lambda^{+}({\bf q})
                  +  \Lambda^{-}({\bf q})\,{\bf M}^\dagger\,
\frac{{\overline G}^\ast(q)}
             {q_0^2-{\overline \bfeps}_q^2}\,\Lambda^{-}({\bf q}) 
                \right]\; ,
 \\[1mm]
 S_{21}(q)           &=&   \left[\Lambda^{-}({\bf q})\, {\bf M}\,
                    \frac{ G(q)}
   {q_0^2-\bfeps_q^2}\, \Lambda^{-}({\bf q})
                 + \Lambda^{+}({\bf q})\,{\bf M}
   \frac{{\overline G}(q)}{q_0^2-{\overline\bfeps}_q^2}
   \,\Lambda^{+}({\bf q})
             \right]\; ,
 \\[1mm]
 S_{22}(q)             &=& 
                   \left[ \Lambda^{-}({\bf q}) \,\gamma^0
 \, \frac{q_0+\mu-|{\bf q}|}{q_0^2-\bfeps_q^2}\, \Lambda^{+}({\bf q})
                  +
  \Lambda^{+}({\bf q})   \,\gamma^0\, \frac{q_0+\mu+|{\bf q}|}
 {q_0^2-{\overline\bfeps}_q^2}
  \, \Lambda^{-}({\bf q})             \right] 
\; 
 \end{array}
  \label{PropFull}
\end{equation}
with ${\bfeps}_q^2=(\mu - |{\bf q}|)^2 
 + {\bf M}^\dagger {\bf M}|G(q)|^2$ and
${\bar{\bfeps}}_q^2=(\mu + |{\bf q}|)^2 +{\bf M}^\dagger {\bf M}
 |{\overline  G}(q)|^2$.
Using (\ref{PropFull}), 
we can calculate the mass corrections in perturbation theory, i.e.
\begin{eqnarray}
\Delta_{m} {\bf S}(q) &=& 
 {\bf S}(q) \left(\begin{array}{cc} m & 0 \\
   0 & m \end{array}\right)
 {\bf S}(q)\;, \label{m-lin} \\
\Delta_{m^2} {\bf S}(q) &=& 
 {\bf S}(q) \left(\begin{array}{cc} m & 0 \\
   0 & m \end{array}\right)
 {\bf S}(q) \left(\begin{array}{cc} m & 0 \\
   0 & m \end{array}\right)
 {\bf S}(q) \; , \label{m-quad}
\end{eqnarray}
etc., where, in general, $m={\rm diag} (m_u,m_d,m_s)$.
The remaining task is just to insert the terms (\ref{PropFull}) into
(\ref{m-lin}) and (\ref{m-quad}). One can easily check that
the $m$ and $m^2$ terms of 
(\ref{S11}), (\ref{S22}), (\ref{S21}) and (\ref{S12}) are recovered in
this way.  Hence, the direct and perturbative arguments give the
same result to the order quoted. In retrospect, this is not
surprising. Both approaches use the Dyson expansion of the propagator.
In the perturbative argument, the full massive propagator is 
expanded in terms of the full massless propagator. 
In the direct approach, the same expansion is performed
on the level of the free propagators. The gap functions ${\bf M}G(q)$
and ${\bf M} {\overline G}(q)$ are completely passive with respect
to these expansions. Therefore the results are the same.

The neglect of the color-flavor non-diagonal terms in
(\ref{MM})  through the approximation (\ref{epsapprox}) which 
has also been used in \cite{RWZ} to simplify the denominators, 
can be justified as follows.
The eigenvalues of ${\bf M}$ (=${\bf M}^\dagger$)
read~\cite{BRWZ}
\[
 {\rm eig}\left({\bf M}\right) = +2,+1,+1,+1,-1,-1,-1,-1,-1 \; .
\]
Thus the eigenvalues of ${\bf M}^\dagger {\bf M} = {\bf M}^2$ 
are~\footnote{For the general case $N_c=N_f$, the ``+2'' and ``4'' have
to be replaced by $N_c-1$ and $(N_c-1)^2$, respectively.
Furthermore, there are $\half N_c(N_c-1)$ eigenvalues $+1$ and 
$\half N_c(N_c+1)-1$ eigenvalues $-1$.}
\[
{\rm eig}\left( {\bf M}^\dagger {\bf M} \right) = 4, 1,1,1,1,1,1,1,1 \; .
\]
Note that eight eigenvalues are equal to unit and only the ninth
deviates from this value~\cite{CFL_ARW}. This is related to an 
explicit U(1) degree of freedom in the U(3) color-flavor phase,
whereas the agreement of the other eight eigenvalues corresponds to
the SU(3) sector in the color-flavor phase. Throughout, we have
specialized to the SU(3) phase as indicated in the introduction,
leaving the issue of the additional  U(1) in the presence of the triangle
anomaly for a future discussion.

\vspace{1cm}
\noindent 3. Derivation of the mass formula of the generalized
Goldstone meson:
\vspace{5mm}

Following Ref.~\cite{RWZ}, we consider the chiral Ward identity
implied by the underlying flavor symmetry in the CFL phase.
Indeed, when chiral symmetry is softly broken 
by massive quarks $m_{}={\rm diag}(m_u,m_d,m_s)$, then the pions
are expected to be massive. Hence
\begin{equation}
 0\equiv \int\, d^4x\,\partial^{\mu}_x\,
\left \langle {\rm BCS} \left| 
  T^*\,{\bf A}_{\mu}^\alpha (x) \,\bfpi_B^\beta (0)\,
                   \right|{\rm BCS}\right\rangle   \; ,
\end{equation}
where the axial-vector current ${\bf A}_{\mu}^a$
is given in (\ref{A-vector-vertex})
and the pion field ${\mbox{\boldmath $\pi$}}_B (x)$
in the CFL phase  is defined as (see Appendix-7)
\begin{eqnarray}
  \bfpi_B^\beta (x)= \left(\begin{array}{cc} 0 &
    \overline{\psi}\,\gamma^0\left({\bf M}\,i\tau^\beta\gamma_5
 \right )^\dagger\gamma^0\,\psi_C (x) \\
           \overline{\psi}_C\,{\bf M}\, i\tau^\beta\gamma_5\,\psi (x) & 0
\end{array}\right)\,\,,
\label{a1}
\end{eqnarray}
which is consistent with (\ref{PSvertex}).
The flavor axial-vector current in the CFL phase  (see Appendix-6) obeys
the local divergence equation
\begin{eqnarray}
\partial\cdot {\bf A}^\alpha (x) =
\left(\begin{array}{cc}
    \overline{\psi}\,i  \left[m_{},\half\tau^\alpha\right]_{+}\,
 \gamma_5\,\psi (x) & 0\\
          0 & \overline{\psi}_C\, i \left[m_{}
          ,\half{\tau^{\alpha}}^\ast\right]_{+}\, \gamma_5\,\psi_C (x)
\end{array}\right)\,\,.
\label{a2}
\end{eqnarray}
For massless quarks, the hermitean axial-isovector charge
\begin{eqnarray}
{\bf Q}_5^\alpha \equiv {\bf Q}_5^\alpha (x^0)= \int\, d^3x\,
\left(\begin{array}{cc}
    \overline{\psi}\half\tau^\alpha\gamma^0\gamma_5\,\psi (x) & 0 \\
   0&  \overline{\psi}_C\, \half{\tau^\alpha}^\ast\gamma^0\gamma_5\,\psi_C (x)
\end{array}\right)
\label{a3}
\end{eqnarray}
is conserved and generates axial-vector rotations, e.g.
\begin{equation}
   \left[{\bf Q}_5^\alpha , {\bf \Psi} (x)\right]
     =
   - \gamma_5\,\frac12\,{\bf T}^{\alpha} \,{\bf \Psi} (x)\,\,.
 \label{a4}
\end{equation}
In terms of (\ref{a2}-\ref{a4}), the identity 
(\ref{a1}) yields
the axial Ward-identity
\begin{equation}
 \int d^4x\,
\left 
 \langle {\rm BCS}\left|\,T^*\,
 \left[m_{} ,\half\bfpi^\alpha(x)\right]_{+}\,
  \bfpi_B^\beta (0) \right|{\rm BCS}\right\rangle
 =\,  \left\langle {\rm BCS}\left| \bfSigma^{\alpha\beta}_B(0)
 \right|{\rm BCS}
 \right\rangle
  \; ,
 \label{a5}
\end{equation}
where the diquark field
$\bfSigma^{\alpha\beta}_B(x)$ is defined as
%\begin{eqnarray}
%  \bfSigma^{\alpha\beta}_B (x)= \left(\begin{array}{cc}
%      0 &     \overline{\psi}\,\gamma^0
%  \left[\frac 12 \tau^\alpha,(i\bf M)^\dagger\tau^\beta\right]_+
%   \gamma^0\psi_C(x)\\
%      \overline{\psi}_C
%\left[\frac 12 \tau^\alpha, {i\bf M}\,\tau^\beta\right]_+\psi (x) & 0\\
%\end{array}\right)
%\label{sigmaB}
%\end{eqnarray}
\begin{eqnarray}
 \bfSigma^{\alpha\beta}= {\overline{\bf \Psi}}(x) \left[\, \half 
 {\bf T}^{\alpha}\,,\, \left(\begin{array}{cc}
        0                 & -i\tau^\beta {\bf M}^\dagger \\
       i{\bf M}\tau^\beta & 0 
  \end{array}\right)\,\right]_+ {\bf \Psi} (x)
\label{sigmaB}
\end{eqnarray}
and
$\bfpi (x)$ is the diagonal pion field
\begin{eqnarray}
  \bfpi^\alpha (x)= \left(\begin{array}{cc}
    \overline{\psi}i\tau^\alpha\gamma_5\,\psi (x) & 0\\ 0 &
          \overline{\psi}_C\,i{\tau^\alpha}^\ast\gamma_5\,\psi_C (x)
\end{array}\right)\,\,.
\label{a6}
\end{eqnarray}
%The latter is to be contrasted with the off-diagonal or BCS pion field
%(\ref{a1}). Clearly in the QCD superconductor,
%Clearly in the QCD superconductor,
%$\bfpi (x)$ and $\bfpi_B (x)$ mix
%through (\ref{a5}). This is expected, since particles and/or holes can
%pop up from the superconducting state, thereby changing a normal pion
%to a BCS pion. The true pion is a linear combination of both,
%and the number of pseudoscalar Goldstone modes is only commensurate with
%the dimension of the manifold spanned by (\ref{CFL1}).
 The nonconfining
character of the weak coupling description allows for the occurrence
of the gapped $qq$ and/or $\overline{q}q$ exchange.
Hence,
\begin{eqnarray}
&\left\langle {\rm BCS}\left|\bfSigma^{\alpha\beta}_B(0) 
\right|{\rm BCS}\right\rangle
\approx -\int\!\! \frac{d^4q}{(2\pi)^4} \,{\rm Tr}\left[ i\gamma_5
\half\left[ m_{},{\bf T}^\alpha\right]_{+}\,
i{\bf S}(q)\,\bfPi_B^\beta\,
i{\bf S}(q)\right]\nonumber &\\
&+\left\{\int\!\! \frac{d^4q}{(2\pi)^4} \,{\rm Tr}\left[i\gamma_5
\half\left[m_{},{\bf T}^\alpha\right]_{+}
i{\bf S}(q)\,i\bfGamma^{\xi}_\pi\,i{\bf S}(q)\right]
 \right\}\,\left(\frac{-i}{M^2}\right)^{\xi\xi '}\,
\left\{\int\! \frac{d^4q}{(2\pi)^4} \,{\rm Tr}\left[i\bfGamma^{\xi '}_\pi\,
i{\bf S}(q)\,\bfPi_B^\beta\,
i{\bf S}(q)\right]\right\}\nonumber&\\
\label{a7}
\end{eqnarray}
with
\begin{eqnarray}
  \bfPi_B^\beta \equiv \left(\begin{array}{cc}
      0 &     \gamma^0\left(i\, {\bf M} \tau^\beta\gamma_5 
  \right)^\dagger\gamma^0
  \\
     i\, {\bf M} \tau^\beta\gamma_5  & 0\\
\end{array}\right) \; .
\label{PiB}
\end{eqnarray}
In the chiral limit $m_i\rightarrow 0$, $i\in\{u,d,s\}$, 
the first term in (\ref{a7})
drops out and the identity is fulfilled if
$1/M^2$ is sufficiently singular in $m_i$ 
to match the numerator. The traces
can be evaluated in weak coupling. The result is~\footnote{The use of
$F_T$ instead of $F_S$ in the pion vertex follows from the fact that
the intermediate BCS pion is generated by a chiral rotation of the BCS
ground state. A similar interpretation in matter is made in \cite{TW}.}
\begin{eqnarray}
&&\int \frac{d^4q}{(2\pi)^4}
\,\Tr\left[i\gamma_5\,\half\left[m_{},{\bf T}^\alpha\right]_{+}\,
i{\bf S}(q)\,i\bfGamma^{\xi}_\pi\,i{\bf S}(q)\right]
={\cal O} (m_{}^2)\; ,\label{GOR10a}\\
&&\int \frac{d^4q}{(2\pi)^4} \,\Tr\left[i\bfGamma^{\xi '}_\pi\,
i{\bf S}(q)\,\bfPi_B^\beta\,
i{\bf S}(q)\right]
=\delta^{\xi '\beta}\,\frac {16i}{F_T}\,
\int \frac{d^4q}{(2\pi)^4} \,
\frac{G(q)}{q_0^2-\epsilon_q^2}\; ,
\label{GOR10b}
\end{eqnarray}
which shows that $M^2={\cal O}(m_{}^2)$. To determine the coefficient,
we need to expand the vertices and the propagators in (\ref{a7}) to
leading order in $m_{}$. The ${\cal O} (m_{})$ corrections to both $G(p)$
and $\Gamma (p)$ do not contribute. They trace 
to zero because of a poor
spin structure. Therefore, only the ${\cal O}(m_{})$ 
correction to the
propagator (\ref{PropPart}) is needed, i.e. (\ref{GOR11}).
Inserting  (\ref{PropPart}) {\em together} with 
the mass correction (\ref{GOR11}) into 
(\ref{GOR10a})
yields
\begin{eqnarray}
&& \int\,\frac{d^4q}{(2\pi)^4}
\,\Tr\left[i\gamma_5\,\half\left[m_{},{\bf T}^\alpha\right]_{+}\,
i{\bf S}(q)\,i\bfGamma^{\xi}_\pi\,i{\bf S}(q)\right] \nonumber \\
&&=\ \frac{\mu G_0}{8 \pi^2 F_T}\,
{\rm Tr}_{cf}
\left( \left[ {m_{}}^2,\tau^\alpha\right]
\left({\bf M}^\dagger {\bf M^\beta }- {{\bf M}^\beta}^\dagger{\bf M}
 \right) 
+ \left[ {m_{}}^2,{\tau^\alpha}^\ast\right]
\left({\bf M}{\bf M^\beta}^\dagger- {{\bf M}^\beta}{\bf M}^\dagger
 \right) \right)\; .\nonumber\\
\label{GOR12}
\end{eqnarray}
Here, the resulting integral simplifies under a contour integration
in the
following way:
\begin{eqnarray*}
 i\int \frac{d^4 q}{(2\pi)^4} \, \frac{q_0 \pm q_{||}}
{\left(q_0^2-\epsilon_q^2\right)^2}\, \frac{|G(q_{||})|^2}{2\mu F_T}&=&
 i 2\pi \int_0^{2\mu} \frac{dq_\perp q_\perp}{(2\pi)^2}
\int_{-\infty}^\infty
\frac{dq_{||}}{2\pi}\, \frac{|G(q_{||})|^2}{2\mu F_T}
\int   \frac{i d q_4}{2\pi} \,\frac{iq_4 \pm q_{||}}
{\left(q_4^2+\epsilon_q^2\right)^2 }\, \\
= \mp\frac{\mu^2}{\pi^2} \int_0^\infty d q_{||}\,
\frac{|G(q_{||})|^2}{2\mu F_T} 
\, \frac{q_{||}}
{4 \epsilon_q^3} 
&\approx& \mp \frac{\mu}{8\pi^2 F_T} \int_{G_0}^{\Lambda_\ast} d q_{||}
\,\frac{|G(q_{||})|^2}{q_{||}^2} \\
&\approx& \mp \frac{\mu}{8\pi^2 F_T} \int_{0}^{x_0}
dx\,\frac{e^x}{\Lambda_\ast}
\, G_0^2\,
\sin^2\left(\frac{\pi x}{2 x_0}\right) \\ 
&\approx&  \mp \frac{\mu}{8\pi^2 F_T}  \frac{G_0^2}{\Lambda_\ast} e^{x_0}
= \mp \frac{\mu}{8\pi^2 F_T} G_0\; ,  
\end{eqnarray*}
where the gap solution (\ref{GapSolution}) and 
the logarithmic scales $x=\ln(\Lambda_\ast/q_{||})$ and
$x_0=\ln(\Lambda_\ast/G_0)$ were inserted in the second to last line.
Furthermore, $ \left[m_{}, [m_{}, \tau^\alpha ]_{+} \right] 
=[ m_{}^2,\tau^\alpha]$ was used. 

Inserting (\ref{GOR10b}) 
and (\ref{GOR12}) in (\ref{a7}) and noting that
\begin{equation}
\left\langle \Sigma^{\alpha\beta}_B\right\rangle \equiv
{\rm Tr}\left(
 \left[\, \half 
 {\bf T}^{\alpha}\,,\, \left(\begin{array}{cc}
        0                 & -i\tau^\beta {\bf M}^\dagger \\
       i{\bf M}\tau^\beta & 0 
  \end{array}\right)\,\right]_+
 \,i{\bf S}\right)
=-8\,\delta^{\alpha\beta}\, \int\,
\frac{d^4q}{(2\pi)^4} \, \frac{G(q)}{q_0^2-\epsilon_q^2}\; ,
\label{GOR12x}
\end{equation}
we obtain for the mass of the Goldstone modes
\begin{equation}
\left( M^2\right)^{\alpha\beta}\approx
-\frac{\mu G_0}{4 \pi^2 F_T^2}
{\rm Tr}_{cf}
\left( \left[ {m_{}}^2,\tau^\alpha\right]
\left({\bf M}^\dagger {\bf M^\beta }
 \mbox{$-$}{{\bf M}^\beta}^\dagger{\bf M}
 \right) 
+ \left[ {m_{}}^2,{\tau^\alpha}^\ast\right]
\left({\bf M}{\bf M^\beta}^\dagger
 \mbox{$-$}{{\bf M}^\beta}{\bf M}^\dagger
 \right) \right) \; ,
 \label{MASSXapp}
\end{equation}
which is (\ref{MASSX}), see also Ref.~\cite{RWZ}.
Note that the color-flavor traces, which first appeared in the
transition from (\ref{GOR10a}) to (\ref{GOR12}), 
yield zero.
 
In order to get a non-zero result for the mass matrix of the
 generalized pion, we have to insert in (\ref{GOR10a}) the
 next-to-leading order, ${\cal O}(1/\mu)$, even for the massless terms
 of the propagator, i.e. the second terms on the right hand sides
of (\ref{S11}) and (\ref{S22}) which can be traced back to
the leading terms of  the antiparticle propagator, see
 (\ref{PropFull})~\footnote{The $1/\mu$ expansion of the numerators and
 denominators of the particle propagators only modifies
the leading result (\ref{GOR12}) by overall factors that do not
prevent the vanishing of the color-flavor traces.}.
These terms are in fact antiparticle-gap {\em independent}. The first
antiparticle-gap {\em dependent} piece appears at order ${\cal
 O}(1/\mu^2)$ and is therefore subleading. This is fortunate, since
according to
\cite{SchaferWilczek} the antiparticle-gap is gauge-fixing-term dependent. 
Inserting the above mentioned terms in (\ref{GOR10a}), we get
\begin{eqnarray}
(A37) &=& (A39)+ \frac{i}{4\mu^2 F_T} \int \frac{d^4q}{(2\pi)^4}\,
 \frac{|G(q)|^2}{q_0^2-\epsilon_q^2} \nonumber \\
&& \qquad\qquad \times \left\{ {\rm Tr}_{cf}\left( [m,\tau^\alpha]_{+}
 \left( {{\bf M}^{\xi}}^\dagger m {\bf M} + {\bf M}^\dagger m {\bf
 M}^\xi\right) 
 \right) \right. \nonumber\\
&&\qquad\qquad\quad + \left. {\rm Tr}_{cf}\left( [m,{\tau^\alpha}^\ast]_{+}
 \left( {{\bf M}^{\xi}} m {\bf M}^\dagger + {\bf M} m {{\bf
 M}^\xi}^\dagger\right)\right) \right\} \; .
\end{eqnarray}
Note that
\begin{eqnarray}
i \int \frac{d^4q}{(2\pi)^4}\, \frac{|G(q)|^2}{q_0^2-\epsilon_q^2}
&=& i 2\pi \int_0^{2\mu}\frac{ dq_\perp  q_\perp}{(2\pi)^2} 
\int_{-\infty}^\infty \frac{d q_{||}}{2\pi}\, |G(q_{||})|^2\, 
\int \frac{i d q_4}{2\pi}
\frac{-1}{q_4^2 + \epsilon_q^2} \nonumber \\
&=& \frac{\mu^2}{\pi^2} \int_0^\infty d q_{||}\, 
 \frac{ |G(q_{||})|^2}{2\epsilon_q}
  \nonumber \\
&\approx& \frac{\mu^2}{2\pi^2} \int_{G_0}^{\Lambda_\ast} d q_{||}\,
 \frac{ |G(q_{||})|^2}{q_{||} } \nonumber \\
&=& 
\frac{\mu^2}{2\pi^2} G_0^2 \int_0^{x_0} d x\, \sin^2\left(\frac{\pi x}{2
    x_0} \right) \nonumber \\
&=& \frac{\mu^2  }{2\pi^2}\, G_0^2\, 
 \frac{x_0}{2} = \frac{\mu^2\, G_0^2\, x_0}{4\pi^2}\;,
\end{eqnarray}
where logarithmic scales $x=\ln(\Lambda_\ast/q_{||})$ and
$x_0 = \ln(\Lambda_\ast/G_0)$  were used in the fourth line.
Thus we have for $\Delta(A37)\equiv (A37)-(A39)$:
\begin{eqnarray}
\Delta(A37) &\approx& \frac{ G_0^2 \, x_0}{16 \pi^2 F_T}\,
\left\{ {\rm Tr}_{cf}\left( [m,\tau^\alpha]_{+}
 \left( {{\bf M}^{\xi}}^\dagger m {\bf M} + {\bf M}^\dagger m {\bf
 M}^\xi\right) \right)
 \right. \nonumber\\ &&\qquad\qquad\qquad  \left.
+ {\rm Tr}_{cf}\left( [m,{\tau^\alpha}^\ast]_{+}
 \left( {{\bf M}^{\xi}} m {\bf M}^\dagger + {\bf M} m {{\bf
 M}^\xi}^\dagger\right)\right) \right\} \; .
\end{eqnarray}
Since $G_0$ and $F_T$ are of order ${\cal O}(\mu)$,
we find that  $\Delta(A37)$ is of order ${\cal O}(\mu)$.
This means that the corresponding $(M^2)^{\alpha\beta}$ is
of order ${\cal O}(\mu^0)$, since  there is an additional $1/F_T$ factor
from (\ref{GOR10b}), namely
\begin{eqnarray}
\left(M^2\right)^{\alpha\beta}&\approx& -\frac{ G_0^2 \, x_0}{8 \pi^2 F_T^2}\,
\left\{ {\rm Tr}_{cf}\left( [m,\tau^\alpha]_{+}
 \left( {{\bf M}^{\beta}}^\dagger m {\bf M} + {\bf M}^\dagger m {\bf
 M}^\beta\right) \right)
 \right. \nonumber\\ &&\qquad\qquad\qquad  \left.
+ {\rm Tr}_{cf}\left( [m,{\tau^\alpha}^\ast]_{+}
 \left( {{\bf M}^{\beta}} m {\bf M}^\dagger + {\bf M} m {{\bf
 M}^\beta}^\dagger\right)\right) \right\} \; .
 \label{M2A37}
\end{eqnarray}
In the flavor-symmetric case $m_u=m_d=m_s$ ($\equiv m_q$), 
the color-flavor traces simplify to
\[
2m_q^2 {\rm Tr}_{cf}\left( \tau^\alpha \left\{ {{\bf M}^\xi}^\dagger {\bf
      M} + {\bf M}^\dagger {\bf M}^\xi \right \} \right) 
=2m_q^2 {\rm Tr}_{cf}\left( {\tau^\alpha}^\ast \left\{ {{\bf M}^\xi} {\bf
      M}^\dagger + {\bf M} {{\bf M}^\xi}^\dagger \right \} \right)
= -16 m_q^2\, \delta^{\alpha\xi} \; .
\]
The mass of the generalized pion at next-to-leading order now reads
\begin{eqnarray}
M^2 &\approx& \frac{4 G_0^2\, x_0}{\pi^2 F_T^2}\, m_q^2 
  = \frac{4 m_q^2}{\mu^2}\left\{ \frac{4 \Lambda_\perp^6}{\pi m_E^5} 
\,\exp\left({\frac{-3\pi^2}{\sqrt{2}\,g}}\right) \right \}^2 \,
\frac{3\pi^2}{\sqrt{2}\,
  g}
\nonumber \\
&=& \frac{2^{23} \,\pi^{10}\, m_q^2}{3^4\,\sqrt{2}\, g^{11}}\, 
\exp\left({\frac{-3\sqrt{2}\,\pi^2}{g}}\right) \; ,
\end{eqnarray}
where we have used that
$F_T=\mu/\pi$, eq.~(\ref{Gzero}), $x_0= \frac{\sqrt{3}\,\pi}{2
  h_\ast}$, eq.~(\ref{h*}), $\Lambda_\perp=2\mu$ and $m_E= \sqrt{\frac{N_f}{2}}
\,\frac{g\mu}{\pi}$ with $N_f=3$.

 \vspace{1cm}
\noindent 4. Proof of the color-identity (\ref{identity}):
\vspace{5mm}

Apply the standard identity for $SU(N_c)$ Gell-Mann matrices,
\vspace{5mm}
\begin{equation}
\sum_{a=1}^{N_c^2-1}
      {\textstyle\frac{1}{4}}
 \lambda^a_{\alpha\beta} \lambda^a_{\gamma\delta}
 = \half \delta_{\alpha\delta}\,\delta_{\gamma\beta}
   -{\textstyle\frac{1}{2N}}
      \delta_{\alpha\beta}\,\delta_{\gamma\delta} \; ,
   \label{A-Gell-Mann}
\end{equation}
to the expression
$ \sum_{a=1}^{N_c^2-1}\left(
   \frac{\lambda^{aT}}{2} \epsilon^A
   \frac{\lambda^a}{2}\right)_{\alpha\delta}$,
 i.e.
\begin{eqnarray}
 \sum_{a=1}^{N_c^2-1}
   {\textstyle\frac{1}{4}} \lambda^{aT}_{\alpha\beta}
 \epsilon^{A\beta\gamma}
 \lambda^a_{\gamma\delta} &=&
\sum_{a=1}^{N_c^2-1} {\textstyle\frac{1}{4}}
\lambda^{a}_{\beta\alpha}
  \lambda^a_{\gamma\delta}\epsilon^{A\beta\gamma}
= \left(\half \delta_{\beta\delta}\,\delta_{\gamma\alpha}
-{\textstyle\frac{1}{2N_c}}
\delta_{\beta\alpha}\,\delta_{\gamma\delta}\right)
\epsilon^{A\beta\gamma}\nonumber\\
& =& \half\left( \epsilon^{A\delta\alpha}
-{\textstyle\frac{1}{N_c}} \epsilon^{A\alpha\delta}\right)
= -\frac{N_c+1}{2N_c}\epsilon^{A\alpha\delta}
 \label{A-identity}
\end{eqnarray}
which is identical to  (\ref{identity}) for the case $N_c=3$.
It is easy to check that also
\begin{equation}
\sum_a \frac{\lambda^{a}}2\,\epsilon^A_c \,\frac{\lambda^{a T}}2
=-\frac {N_c+1}{2N_c}\,\epsilon^A_c
\label{A-identity-T}
\end{equation}
holds.
Finally, by replacing $\beta \leftrightarrow \alpha$ in the bracket of
the third term of
(\ref{A-identity}) one can easily derive 
\begin{equation}
\sum_a \frac{\lambda^{a}}2\,\epsilon^A_c \,\frac{\lambda^{a}}2
=-\frac {1}{2N_c}\,\epsilon^A_c \; .
\label{A-identity-noT}
\end{equation}

\vspace{1cm}
\noindent 5. Proof of the relations (\ref{identity2}):
\vspace{5mm}

The first identity follows immediately from (\ref{identity})
or (\ref{A-identity}), if one
inserts
\begin{eqnarray}
 \sum_a \frac{\lambda^{aT}}{2}\,
 \left( {\bf M}^A \right)\,
 \frac{\lambda^a}{2}
&=& \sum_a \frac{\lambda^{aT}}{2}\,
 \left( \epsilon^I_f \epsilon^J_c \gamma_5
\left(\tau^A\right)^{I J} \right)\,
 \frac{\lambda^a}{2}\nonumber\\
&=& -\frac{N_c+1}{2N_c}\, \left(
 \epsilon^I_f {\epsilon^J_c} \gamma_5 \left(\tau^A\right)^{I
 J} \right)= -\frac{N_c+1}{2N_c}  {\bf M}^A \;.
 \label{A-identity21}
\end{eqnarray}
It is easy to see that the same relation holds,
if ${\bf M}^A$ is replaced by
${{\bf M}^A}^\dagger$.

In order to show the second relation of (\ref{identity2}), we use that
\[
 {\bf M} {\bf M}^A {\bf M}= \gamma_5\, \epsilon^{I}_f\,\epsilon^{I}_c\,
\epsilon^{J}_f \left(\tau^A\right)^{JK} \epsilon^{K}_c\,
\epsilon^{L}_f\,\epsilon^L_c
\]
and that
\begin{eqnarray*}
\left(\epsilon_c^{I}\epsilon^{K}_c\epsilon^L_c\right)^{\alpha\beta}
&=& - \epsilon^{I\alpha\beta}_c\, \delta^{KL}+\epsilon^{I\alpha L}_c
\,\delta^{K\beta}\; ,\\
\left(\epsilon_f^{I}\epsilon^{J}_f\epsilon^L_f\right)^{ij}
&=& - \epsilon^{Lij}_f\, \delta^{IJ}+\epsilon^{LIj}_f\,
\delta^{iJ} \; .
\end{eqnarray*}
Therefore
\begin{eqnarray*}
\left({\bf M} {\bf M}^A {\bf M}\right)^{\alpha i,\beta j}
&=& \gamma_5\, \epsilon_f^{Kij} \epsilon_c^{J\alpha\beta}
\left(\tau^A \right)^{JK} +\cdots \\
&=&  \gamma_5\, \epsilon_f^{Kij} \epsilon_c^{J\alpha\beta}
\left({\tau^A}^T \right)^{KJ}+\cdots = \left({{\bf M}^A}^\dagger
 \right)^{\alpha i, \beta j}+\cdots\;,
\end{eqnarray*}
where the dots refer to terms which are symmetric in color-flavor
and finally subleading to leading logarithm order.
Naturally, also $\left({\bf M} {{\bf M}^A}^\dagger {\bf M}\right)
= {\bf M}^A+\cdots$ holds.
Therefore the second relation of (\ref{identity2})
approximately follows from the first one, if the above mentioned
subleading terms are neglected.

\vspace{1cm}
\noindent 6. The structure of the vector and axial-vector currents:
\vspace{5mm}

The two-component Nambu-Gorkov field
\[{\bf \Psi} =
 \left(\begin{array}{c} \psi \\ C \bar\psi^T \end{array}\right)
\]
transforms under vector and
axial-vector transformations as follows
\begin{equation}
{\bf U}_V {\bf \Psi}=\left(\begin{array}{cc}
 e^{i\half\tau^A \alpha^A} & 0 \\
 0  & e^{-i\half{\tau^A}^\ast \alpha^A} \end{array}\right) {\bf \Psi}
\quad \mbox{and} \quad
  {\bf U}_A {\bf \Psi}=\left(\begin{array}{cc}
 e^{i\half\gamma_5\tau^A \beta^A} & 0 \\
 0  & e^{i\half\gamma_5{\tau^A}^\ast \beta^A}
\end{array}\right) {\bf \Psi}
\;.
\end{equation}
The vector and axial-vector currents are diagonal in the Nambu-Gorkov
formalism, since they result as Noether currents from the diagonal
kinetic term
\[ \overline{\bf \Psi} i \gamma^\mu \partial_\mu\, \bfrho_0 {\bf \Psi}
=
\left(\begin{array}{cc}
\bar \psi \gamma^\mu\partial_\mu \psi & 0 \\
0 & \psi^T  {\gamma^\mu}^T \partial_\mu
\bar\psi^T\end{array}\right)
= \overline{\bf \Psi}
\left(\begin{array}{cc}
 \gamma^\mu\partial_\mu  & 0 \\
0 &   -C {\gamma^\mu}^TC^{-1}  \partial_\mu\
\end{array}\right){\bf \Psi}
\; ,
\]
with $\bfrho_0$ the unit matrix in the Nambu-Gorkov space.
Alternatively, they can be derived by a prescription
from \cite{PisarskiSuperfluid} which generalizes
the standard current
structure $\bar\psi\,\Gamma \,\psi$ to the charge conjugated sector as
$\bar \psi_C \,C \Gamma^T C^{-1}\, \psi_C$.
Because of $\gamma_\mu^T= - C^{-1}\gamma_\mu C$
and $C^{-1}\gamma_5 C = \gamma_5^T=\gamma_5$, we have
\begin{eqnarray}
 {\bf V}^A_\mu &\equiv& \overline{\bf \Psi}\left(\begin{array}{cc}
 \gamma_\mu\half {\tau^A} & 0 \\
                 0 & C\left(\gamma_\mu\half{\tau^A}\right)^TC^{-1}
                 \end{array} \right) {\bf \Psi}
= \overline{\bf \Psi}
  \left(\begin{array}{cc} \gamma_\mu\half{\tau^A} & 0 \\
                 0 &-\gamma_\mu \half{{\tau^A}^\ast}
                 \end{array} \right){\bf \Psi}
\nonumber \\
&=& \overline{\bf \Psi} \gamma_\mu \half {\bf T}^A\bfrho_3{\bf \Psi}
 \; ,
 \label{V-vector-vertex}
\end{eqnarray}
and
\begin{eqnarray}
 {\bf A}^a_\mu &\equiv& \overline{\bf \Psi}
\left(\begin{array}{cc}
 \gamma_\mu\gamma_5\half\tau^A & 0 \\
                 0 & C\left(\gamma_\mu\gamma_5\half\tau^A\right)^TC^{-1}
                 \end{array} \right){\bf \Psi}
= \overline{\bf \Psi}\left(\begin{array}{cc} \gamma_\mu\gamma_5
 \half\tau^A & 0 \\
                 0 &\gamma_\mu\gamma_5
                 \half{\tau^A}^\ast
                 \end{array} \right){\bf \Psi}\nonumber\\
&=&
 \overline{\bf \Psi} \gamma_\mu\gamma_5 \half {\bf T}^A{\bf \Psi}
\; ,
 \label{A-vector-vertex}
\end{eqnarray}
with ${\bf T}^A ={\rm diag} (\tau^A, {\tau^A}^\ast)$ and $\bfrho_3$ the
standard Pauli matrix.
In (\ref{A-vector-vertex}) it was used that
$C \gamma_5^ T \gamma_\mu^T C^{-1}=
 C C^{-1} \gamma_5 C (- C^{-1} \gamma_\mu
C) C^{-1} = - \gamma_5\gamma_\mu = \gamma_\mu \gamma_5$.
Furthermore, note that the derivation of the
gluon-vertex (\ref{gluon-vertex}) is totally analogous to
 (\ref{V-vector-vertex}).

\vspace{1cm}
\noindent 7. The structure of the vertices for the generalized mesons:
\vspace{5mm}

Because of the particle-particle (or hole-hole) substructure, all
generalized mesons have the vertex structure
\begin{equation}
\overline{\bf \Psi}\, {\bf \Gamma}_M \,{\bf \Psi}=
\overline{\bf \Psi} \left( \begin{array}{cc}\
   0 & \left({\Gamma_M}\right)_{12} \\
 \left({\Gamma_M}\right)_{21} & 0 \end{array}\right) {\bf \Psi}
 \label{A-meson-vertex}\; .
\end{equation}
By conjugating the 21-component $\bar\psi_C
 \left({\Gamma_M}\right)_{21}
 \psi = \psi^T C \left({\Gamma_M}\right)_{21} \psi$,
namely~\footnote{The minus sign results from the Grassman property
of the fermion spinors. Note that $C=-C^\dagger=-C^{-1}$ and
$\gamma_0 C^{-1}=- C{\gamma^0}^T$.}
\begin{equation}
 -\psi^\dagger \left\{\left( \Gamma_M\right)_{21}\right\}^\dagger
C^\dagger {\psi^T}^\dagger= -\bar\psi \gamma^0 \left\{\left(
 \Gamma_M\right)_{21}\right\}^\dagger \gamma^0 \gamma^0 C^{-1}
{\psi^\dagger}^T = \bar\psi \gamma^0 \left\{\left(
 \Gamma_M\right)_{21}\right\}^\dagger \gamma^0 C (\bar\psi)^T\; ,
 \label{A-qqc-hermiticity}
\end{equation}
one can derive a general
rule (see~\cite{PisarskiSuperfluid}) 
which links the 12 and 21 components of
${\bf \Gamma}_M$
\begin{equation}
  \left({\Gamma_M}\right)_{12} =
\gamma^0 \left\{\left( {\Gamma_M}\right)_{21} \right\}^\dagger
  \gamma^0 \; .
\label{12-21-rule}
\end{equation}
Thus, in order to determine the structure of the generalized vertices,
we only have to determine the structure of the 21-component.

As in the standard case,  the structure of the (generalized)
meson vertices follows from the transformation properties
of the (hermitean)
bilinears $\overline{\bf \Psi}\, {\bf \Gamma}_M \, {\bf \Psi}$
under proper (ordinary continuous) Lorentz transformations
$\psi(x) \to \psi'(x')=S(\Lambda) \psi(\Lambda x)$
with $S(\Lambda)= \exp(-\frac{i}{4}
\sigma_{\alpha\beta} \omega^{\alpha \beta})$ and under the (discrete)
parity
transformation $\psi(x) \to \psi'(x')=\gamma_0 \psi(t,-{\bf x})$.
Using that
$S(\Lambda)^T = C S(\Lambda)^{-1} C^{-1}$ and ${\gamma^0}^T C=
\gamma^0 C = - C \gamma^0$ , we can easily derive the following
transformation properties of the bilinears $\bar\psi_C \,
\left(\Gamma_M\right)_{21}\, \psi$ under proper Lorentz and parity
transformations (which generalize the transformation properties
of the standard bilinears $\bar\psi\, \Gamma\, \psi$, see
e.g.~\cite{ItzyksonZuber}):
\begin{equation}
 \begin{array}{rclcll}
   \bar\psi_C'(x')  \gamma_5 \psi'(x')
   &=& \bar\psi_C(x)  \gamma_5 \psi(x)
     &\quad& \mbox{scalar} & 0^+\; , \\
   \bar\psi_C'(x')  \psi'(x')
   &=& \det(\Lambda) \bar\psi_C(x)\psi(x)
     &\quad& \mbox{pseudoscalar} & 0^- \; , \\
   \bar\psi_C'(x')  \gamma^\mu\gamma_5 \psi'(x')
   &=& \Lambda^\mu_{\ \nu}\bar\psi_C(x)  \gamma^\nu\gamma_5 \psi(x)
     &\quad& \mbox{vector}& 1^- \; , \\
   \bar\psi_C'(x')  \gamma_\mu \psi'(x')
   &=& \det(\Lambda)\Lambda^\mu_{\ \nu}\bar\psi_C(x)  \gamma^\nu \psi(x)
     &\quad& \mbox{axial-vector}& 1^+ \; .
\end{array}
\end{equation}
Note the appearance of the extra $\gamma_5$ relative to the
standard rules of e.g.~\cite{ItzyksonZuber}. Taking the flavor matrix
in the color-flavor-locked way
into account, we have the following 21 components of the generalized
meson vertices:
\begin{equation}
\begin{array}{rclcl}
 \left(\Gamma_{\sigma}(p,P)\right)_{21}
      &=& i {\bf M}^A\, \Gamma_{S}(p,P)/F_{S} & \quad &
\mbox{generalized sigma}\;, \\
 \left(\Gamma_{\pi}(p,P)\right)_{21}
      &=& i \gamma_5\, {\bf M}^A\, \Gamma_{PS}(p,P)/F_{PS} & \quad &
\mbox{generalized pion}\;, \\
 \left(\Gamma_\mu (p,P)\right)_{21}
      &=&  \gamma_\mu \,{\bf M}^A\, \Gamma_V(p,P)/F_V & \quad &
\mbox{generalized vector meson}\; , \\
 \left(\Gamma_{5\,\mu} (p,P)\right)_{21}
      &=& i\gamma_\mu\gamma_5 \,{\bf M}^A\, \Gamma_{AV}(p,P)/F_{AV}
      & \quad &
\mbox{generalized axial-vector meson}\; , \\
\end{array}
\label{A-Gamma-21}
\end{equation}
where ${\bf M}^A \equiv \gamma_5 \epsilon_f^a \epsilon_c^\alpha
(\tau^A)^{a\alpha}$ and $(\epsilon^a)^{bc}=\epsilon^{abc}$. The
form factors and decay constants have been introduced in \cite{RWZ} for
the pionic case and in (\ref{Vvertex}) for the vector case.
Note thay after inserting  unity ${\bf 1}= 
\Lambda^+({\bf p})+\Lambda^{-}({\bf  p})$ to the left and right of the
$ \left(\Gamma_M (p,P)\right)_{21}$'s, we can project these vertices onto
the particle-particle, particle-antiparticle, antiparticle-particle
and antiparticle-antiparticle sectors in analogy to (\ref{Delta}): 
\begin{eqnarray}
\left(\Gamma_M (p,P)\right)_{21}
 &=& \Lambda^+({\bf p}) \left(\Gamma_M (p,P)\right)_{pp} \Lambda^+({\bf p})
  +\Lambda^+({\bf p}) \left(\Gamma_M (p,P)\right)_{pa} \Lambda^-({\bf p})
 \nonumber \\
 && 
 +\Lambda^-({\bf p}) \left(\Gamma_M (p,P)\right)_{ap} \Lambda^+({\bf p})
  +\Lambda^-({\bf p}) \left(\Gamma_M (p,P)\right)_{aa} \Lambda^-({\bf p}) 
 \label{pp-21-projections} \; . 
\end{eqnarray}
In the scalar and pseudoscalar case, the mixed 
(particle-anti\-particle and anti\-particle-particle) vertices 
vanish 
identically since
$\Lambda^\pm({\bf p})\gamma_5 \Lambda^\mp({\bf  p})=0
= \Lambda^\pm({\bf p}) \Lambda^\mp({\bf  p})$. The sum of the remaining
particle-particle and antiparticle-antiparticle has exactly 
the structure of (\ref{Delta}).
In the vector and axialvector case, the mixed
terms $\left(\Gamma_M (p,P)\right)_{pa}$ and 
$\left(\Gamma_M (p,P)\right)_{ap}$
survive. However,
in the leading-logarithm approximation only the particle-particle
parts $\left(\Gamma_M (p,P)\right)_{pp}$
of the vertices are needed.

Contrary to the standard case, the phases cannot be 
determined from the hermiticity property of the quark bilinears, since
the hermiticity is automatically satisfied under the condition 
(\ref{12-21-rule}). In general, the
form factors are complex-valued, such that the phases can be chosen at
will. Our phase choice corresponds to real-valued form factors with
attractive Bethe-Salpeter kernels 
(see (\ref{t5}) and (\ref{A-GV3})). 
Taking the (\ref{12-21-rule}) rule into account, we  have
\begin{equation}
\begin{array}{rclcl}
 {\bf \Gamma}_{\sigma}^{A}(p,P)
      &=& i \bfrho_1\,{\bf M}_{\bf T}^{\,\,A}\, 
 \Gamma_{S}(p,P)/F_{S}
& \quad &
\mbox{generalized sigma}\;, \\
 {\bf \Gamma}_{\pi}^{A}(p,P)
      &=& \gamma_5 \,\bfrho_2\,{\bf M}_{\bf T}^{\,\,A}\,
      \Gamma_{PS}(p,P)/F_{PS}
& \quad &
\mbox{generalized pion}\;, \\
 {\bf \Gamma}_{\mu}^{A} (p,P)
      &=&  \gamma_\mu \,\bfrho_1\, {\bf M}_{\bf T}^{\,\,A} \,
 \Gamma_V(p,P)/F_V & \quad &
\mbox{generalized vector meson}\; , \\
 {\bf \Gamma}_{5\,\mu}^{A} (p,P)
      &=& \gamma_\mu\gamma_5 \,\bfrho_2\, {\bf M}_{\bf T}^{\,\,A}\,
  \Gamma_{AV}(p,P)/F_{AV}
      & \quad &
\mbox{generalized axial-vector meson}\; , \end{array}
 \label{A-vertex-structure}
\end{equation}
where
\begin{eqnarray}
  {\bf M}_{\bf T}^{\,\,A}
\equiv \left(\begin{array}{cc} {\bf M}^A & 0\\ 0 & {{\bf M}^A}^\dagger
\end{array}\right)= \left(\begin{array}{cc} {\bf M}^{i\alpha}
\left(\tau^A\right)^{i\alpha}& 0\\ 0 & {\bf M}^{i\alpha}
\left({\tau^A}^\ast\right)^{i\alpha}
\end{array}\right) = {\bf M}^{i\alpha}
 \left({\bf T}^A\right)^{i\alpha}
\label{TM}
\end{eqnarray}
and
$\bfrho_1$ and $\bfrho_2$ are the standard Pauli matrices acting on
the Nambu-Gorkov indices. 
As mentioned above,
the form factors
$\Gamma_{S}(p,P)$, $\Gamma_{PS}(p,P)$, 
$\Gamma_V(p,P)$ and $\Gamma_{AV}(p, P)$
are now assumed to be
real,
where $\Gamma_{PS}(p,0)= G(p)$.
The vertex structure of (\ref{A-vertex-structure})
is in agreement with (\ref{Svertex}) and (\ref{PSvertex}) 
as well as (\ref{Vvertex})  as used in (\ref{DEC1}).

If $\left(\Gamma_M (p,P)\right)_{21}$ is projected onto the 
particle-particle, particle-antiparticle, anti\-particle-particle and
anti\-particle-anti\-particle sectors as in (\ref{pp-21-projections}), it
follows from the  (\ref{12-21-rule})-rule and 
$\gamma^0  \Lambda^{\pm}({\bf p})\gamma^0=  \Lambda^{\mp}({\bf p}) $ 
that  the corresponding contributions of the 1-2 component read
\begin{eqnarray}
\left(\Gamma_M (p,P)\right)_{12}
 &=& \Lambda^-({\bf p})\gamma^0 \left(\Gamma_M
 (p,P)\right)_{pp}^\dagger \gamma^0
  \Lambda^-({\bf p})
  +\Lambda^-({\bf p})\gamma^0  \left(\Gamma_M (p,P)\right)_{pa}^\dagger 
 \gamma^0\Lambda^+({\bf p}) \nonumber \\
&&
 \!\!\!\! +\Lambda^+({\bf p})\gamma^0 \left(\Gamma_M (p,P)\right)_{ap}^\dagger
 \gamma^0   \Lambda^-({\bf p})
  +\Lambda^+({\bf p})\gamma^0   \left(\Gamma_M (p,P)\right)_{aa}^\dagger 
 \gamma^0  \Lambda^+({\bf p})\nonumber \\  
 \label{pp-12-projections} 
\end{eqnarray}
in analogy to (\ref{Deltadag}).
In the leading logarithm approximation, only the first term, i.e.\
the particle-particle term, is needed. In this case
the vertex structure (\ref{A-vertex-structure}) has to be modified
by the ``sandwich-rule'' (\ref{Sandwich}) which is compatible with
(\ref{pp-21-projections}) and (\ref{pp-12-projections}). 
As mentioned in section~3 (see also Appendix-12), 
the simplified vertex structure (\ref{A-vertex-structure}) can be
used, if
only {\em leading} propagators (\ref{PropPart})~\footnote{In the
  scalar and pseudoscalar case, it already is sufficient that only {\em one}
  leading propagator is coupled per vertex.} 
are coupled to the vertex and
if the simplified form (\ref{GluonProp}) 
of the gluon propagator is used. 
%Otherwise, the vertices should
%be modified by the ``sandwich-rule'' (\ref{Sandwich}). 
%which is 
%compatible with (\ref{12-21-rule}), since 
%$\gamma^0 \Lambda^{\pm}({\bf p}) \gamma^0 = \Lambda^{\mp}({\bf p})$.  

\vspace{1cm}
\noindent 8. Derivation of equation (\ref{CFL444}) from (\ref{BSeq}) and 
equation (\ref{CFL44}) from (\ref{t5}):
\vspace{5mm}

Defining $Q\equiv q+P/2$ and $K\equiv q-P/2$,
the 12 component of (\ref{t5}) reads:
\begin{eqnarray}
\left(\Gamma_j^A(p,P)\right)_{12}&=& g^2\int
\frac{d^4q}{(2\pi)^4}\, i {\cal D}(p\mbox{$-$}q) \gamma_\mu
\frac{\lambda^a}{2}
\left[ S_{11}(Q) \left(\Gamma_j^A(q,P)\right)_{12}
 S_{22}(K) \right.\nonumber\\
&& \qquad\qquad\qquad\left.\mbox{}
       +S_{12}(Q)\left(\Gamma_j^A(q,P)\right)_{21}  S_{12}(K)\right]
 (-\gamma^\mu )\frac{{\lambda^a}^T}{2}\; .
 \label{A-GV1}
\end{eqnarray}
The expression for $\left(\Gamma^A_j(p,P)\right)_{21}$ follows from
(\ref{A-GV1}) with the replacements $1\leftrightarrow 2$ and
$\lambda^a\leftrightarrow {\lambda^a}^T$.
We now write
$\left(\Gamma_j^A\right)_{12}
= \frac{1}{F_V} \gamma_j \Gamma_V {{\bf M}^A}^\dagger$ where
we assumed $\Gamma_{V}$ to be real.
Inserting then (\ref{PropPart}) for the propagators,
we can transform  (\ref{A-GV1}) to~\footnote{In the following, we
will use the simplified form (\ref{A-vertex-structure}) of the
generalized-meson vertex. The results for the sandwiched form 
(\ref{Sandwich}) will be discussed at the end of this
section.}
\begin{eqnarray}
\gamma_j \Gamma_V(p,P) {{\bf M}^A}^\dagger&\!\!=\!\!& -g^2\int
\frac{d^4q}{(2\pi)^4}\, \frac{ i {\cal D}(p\mbox{$-$}q)\, \Gamma_V(q,P)}
                {(Q_0^2 -\epsilon_Q^2)(K_0^2-\epsilon_K^2)}
                \nonumber\\
&&\quad \mbox{}\times \left\{
Q_+ K_{-}
\left[ \gamma_\mu \gamma^0
 \Lambda^-({\bf Q}) \gamma_j \gamma^0\Lambda^+({\bf K}) \gamma^\mu\right]
\left[\frac{\lambda^a}{2} {{\bf M}^A}^\dagger
 \frac{{\lambda^a}^T}{2}\right]
\right. \nonumber\\
&&\quad \left.\mbox{}
 +G(Q)G(K)
 \gamma_\mu
 \Lambda^+({\bf Q}) \gamma_j \Lambda^+({\bf K})
\gamma^\mu \left[\frac{\lambda^a}{2} {\bf M} {\bf M}^A {\bf M}
\frac{{\lambda^a}^T}{2}
 \right]\right\}\!,
 \label{A-GV2}
\end{eqnarray}
where it was used that both ${\bf M}$ (=${\bf M}^\dagger$) and ${\bf M}^A$
contain a $\gamma_5$ matrix.
Note that the corresponding expression for the composite axial-vector
meson differs from (\ref{A-GV2}) by the replacement
${\bf M}^A\to i\gamma_5 {\bf M}^A$ and by an additional minus sign in
front of
the second term on
the
right hand side.
Using (\ref{identity2}) for the flavor contractions and moving the
$\gamma^0$ matrices through,
we finally get
\begin{eqnarray}
\gamma_j \Gamma_V(p,P) &=& -\frac{ 2g^2}{3}
 \int \frac{d^4q}{(2\pi)^4}\,
  \frac{ i {\cal D}(p\mbox{$-$}q)\, \Gamma_V(q,P)}
                {(Q_0^2 -\epsilon_Q^2)(K_0^2-\epsilon_K^2)}
                \nonumber\\
&&\qquad \mbox{}\times  \left\{Q_+ K_{-}-G(Q)G(K)\right\}
 \left[\gamma_\mu
 \Lambda^+({\bf Q}) \gamma_j \Lambda^+({\bf K}) \gamma^\mu\right] \; ,
 \label{A-GV3}
\end{eqnarray}
where we ignored a symmetric contribution in color-flavor which
is subleading to logarithmic accuracy.
This equation will be simplified by applying the longitudinal and
transverse projection (\ref{GV-long}) and (\ref{GV-transv}),
respectively, and then taking the Dirac trace.
The left hand side becomes $4\Gamma_{L,T}(p,P)$, whereas
on the right hand side the Dirac structure becomes
\begin{eqnarray}
{\textstyle \frac{1}{3}}
 \Tr\left[ \gamma^j\gamma_\mu \Lambda^+({\bf Q}) \gamma_j  \Lambda^+({\bf
 K}) \gamma^\mu \right] &=& -{\textstyle \frac{2}{3}}
\Tr\left[ \gamma^j\Lambda^+({\bf Q}) \gamma_j  \Lambda^+({\bf K})\right]
 \nonumber\\
&=& -\frac{2}{3}\Tr\left[\left(3\Lambda^{+}({\bf Q})
  -\hat{\bf Q}^k\alpha^k\right)
\Lambda^{+}({\bf  K})
\right] \; ,
 \label{A-GV4}
\end{eqnarray}
where we summed over the index $j$ and
used the Dirac matrix identity
$\gamma_\mu \gamma_j \gamma^\mu = -2\gamma_j$.
In the rest frame, the trace (\ref{A-GV4}) is just $-8/3$.
Inserting this back into
(\ref{A-GV3}) and dividing by four, we get the quoted result
(\ref{CFL44}) which identically holds in the axial-vector case.

For deriving the corresponding expression of the generalized pion
from (\ref{BSeq}), the $\gamma_j$ matrices in
(\ref{A-GV2}) and (\ref{A-GV3})
have to be replaced by $-i\gamma_5$.
After multiplying both sides with $i\gamma_5$ and then taking
the Dirac trace,
we still get $4 \Gamma_{PS}(p,P)$ on the left hand side, whereas
on the right hand the Dirac trace
\[
 \Tr[\gamma_5\gamma_\mu \Lambda^+({\bf Q})\gamma_5\Lambda^{+}({\bf
K})\gamma^\mu]=
-\Tr[\gamma^\mu\gamma_\mu \Lambda^+({\bf Q})\Lambda^{+}({\bf K})]
=-4\Tr[ \Lambda^+({\bf Q})\Lambda^{+}({\bf K})]
\]
reduces to  a factor $-8$ instead of $-8/3$ in the rest frame.
This is the reason why the prefactor in the Bethe-Salpeter kernel
of the generalized pion (see (\ref{CFL444}) or
(\ref{GapEq}) for the gap itself)  is three times bigger than the one of
the generalized vector and axial-vector (see (\ref{CFL44})).

For the generalized sigma, the $\gamma_j$ in (\ref{A-GV2}) and 
(\ref{A-GV3}) has to be replace by the i times the unit matrix. 
Furthermore, there is an additional minus sign in front of the second
term on the right hand side 
of (\ref{A-GV2}) and an additional overall minus sign on the right hand
side of (\ref{A-GV3}). After multiplying both sides with $-i$ and
taking the Dirac trace, there is still $4 \Gamma_{S}(p,P)$ on the
left hand side, whereas on the right hand side the Dirac trace
\[
 \Tr[\gamma_\mu \Lambda^+({\bf Q}) \Lambda^+({\bf K})\gamma^\mu]
  = 4\Tr[\Lambda^+({\bf Q}) \Lambda^+({\bf K})]
\]
reduces to $+8$ in the rest frame. This opposite sign, relative to the
pion case, cancels  against the above mentioned opposite sign 
on the right hand side of (\ref{A-GV3}). Thus the Bethe-Salpeter
equation
of the generalized sigma and pion are the same, see (\ref{CFL444}). 

If the sandwiched form of the generalized mesons is
used, equation (\ref{A-GV3}) must be first projected from the left and
right with
$\Lambda^{-}(p)$, before we can multiply with $\gamma_j$
(or correspondingly with $-i\gamma_5$ in the pion 
or $i\bf{1}$ in the scalar case) and  take
the Dirac trace. This modification induces a factor $\half$
on the left {\em and} right hand side of the projected (\ref{A-GV3})
and
therefore does not change the final answer (\ref{CFL44}) and
(\ref{CFL444}) 
for
the gap equations for the (axial-)vector and (pseudo-)scalar case,
respectively.

The result of (\ref{A-GV3}) is valid for the phase choice of 
(\ref{A-Gamma-21}),
i.e.\ for real-valued form factors. If the opposite phase-choice
had been
made, i.e.\ if the form factors were assumed to be 
purely imaginary-valued, the  
$Q_{+} K_{-}-G(Q)G(K)$ term in (\ref{A-GV3}) would have to read
$Q_{+}K_{-} +G(Q)G(K)$ instead. The error which this choice will
induce can be estimated by perturbation theory by inserting 
(\ref{GapSolution}) and (\ref{FFR}) or an analogous form factor  into
\begin{equation}
 \frac{h_\ast^2}{18}\int_{G_M}^{\Lambda_\ast} \frac{d q_{||}}{q_{||}}
\,\frac{-2 G^2(q_{||})}{q_{||}^2}
\,\ln\left(\frac{\Lambda_\ast^2}{(p_{||}-q_{||})^2}\right) 
 \Gamma(q_{||},M_V) \; 
\end{equation}
as very small, i.e. ${\cal O}(h_\ast^2)$ relatively to
$\Gamma(q_{||},M_V)$.

\vspace{1cm}
\noindent 9. The structure of the vertices for the standard mesons:
\vspace{5mm}

The Bethe-Salpeter kernels of the diagonal {\em standard} 
``$\bar q\ q$'' -type mesons in the CFL phase are
\begin{equation}
\begin{array}{rclcl}
 {\widetilde {\bf\Gamma}}_{\sigma}^{A}(p,P)
      &=& \bfrho_0 {\bf N}_{\widetilde T}^{\,\,A}\, 
 \widetilde \Gamma_{S}(p,P)/\widetilde F_{\Sigma}
& \quad &
\mbox{standard sigma}\;, \\
 {\widetilde {\bf\Gamma}}_{\pi}^{A}(p,P)
      &=& i\gamma_5 \,\bfrho_0\,{\bf N}_{\widetilde T}^{\,\,A}\, 
 {\widetilde{\Gamma}}_{PS}(p,P)/\widetilde F
& \quad &
\mbox{standard pion}\;, \\
 {\widetilde {\bf\Gamma}}_{\mu}^{A} (p,P)
      &=&  \gamma_\mu \,\bfrho_3\, {\bf N}_{\widetilde T}^{\,\,A} \,
 {\widetilde\Gamma}_V(p,P)/\widetilde F_V & \quad &
\mbox{standard vector meson}\; , \\
 {\widetilde{\bf \Gamma}}_{5\,\mu}^{A} (p,P)
      &=& \gamma_\mu\gamma_5 \,\bfrho_0\, {\bf N}_{\widetilde T}^{\,\,A}\,
  \widetilde\Gamma_{AV}(p,P)/\widetilde F_{AV}
      & \quad &
\mbox{standard axial-vector meson}\; , \end{array}
 \label{A-standard-structure}
\end{equation}
where ${\bf N}_{\widetilde T}^{\,\,A}= \epsilon^a_f \epsilon^\alpha_c
(\widetilde T^A)^{a\alpha}$. Thus ${\bf N}_{\widetilde T}^{\,\,A}$
is of the same form as 
${\bf M}_{\bf T}^{\,\,A}$, without the $\gamma_5$, however, and
with ${\bf T}^A={\rm diag}(\tau^A,{\tau^A}^\ast)$ 
replaced by ${\widetilde T}^A={\rm diag}({\widetilde\tau}^A,
{{\widetilde\tau}^A})$, where
$\widetilde\tau_{1,3}= \tau_{1,3}$ and
$\widetilde\tau_2 =i\tau_2$. Furthermore $\bfrho_3$ is the usual Pauli
matrix, whereas $\bfrho_0$ is the corresponding unit matrix. We have
checked that the Bethe-Salpeter equations 
resulting from (\ref{A-standard-structure})
vanish identically. Standard scalar, pseudoscalar, vector and axial-vector
excitations are not supported by the QCD superconductor in leading logarithm
approximation.

\vspace{1cm}
\noindent 10. {}From equation (\ref{PION1}) to equation (\ref{CFL10})
and other variational approximations:
\vspace{5mm}

After multiplying (\ref{PION1}) with $3/4g^2$ and using  the 
Fourier-transformations
\begin{eqnarray*}
  \Gamma(p,M)&=& \int d^4 x\, e^{ipx}\, \Gamma(x)\; ,\\
{\cal  D}(p-q) &=& \int d^4x\, e^{i(p-q) x}\, {\cal D}(x)\; ,
\end{eqnarray*}
we can transform (\ref{PION1}) into
\[
\frac{3}{4g^2} \Gamma(x) = \int \frac{d^4 q}{(2\pi)^4} \, e^{-i qx}\, i
    {\cal D}(x)\, \frac{q_0^2-\epsilon_q^2 -M^2/4}
       {(q_0^2-\epsilon_q^2+M^2/4)^2-M^2 q_0^2}\, \Gamma(q) \; ,
\] 
where $\Gamma(q)\equiv \Gamma(q,M)$.
Multiplying  both sides with $\Gamma(x)/ i{\cal D}(x)$ and integrating
over $x$, we get
\begin{eqnarray*}
 \frac{3}{4 g^2} \int d^4 x\, \frac{\Gamma^2(x)}{i{\cal D}(x)} 
 &=&
 \int \frac{d^4 q}{(2\pi)^4} 
 \left\{ \int d^4 x\, e^{-i qx}\, \Gamma(x) \right\}
 \frac{q_0^2-\epsilon_q^2 -M^2/4}
       {(q_0^2-\epsilon_q^2+M^2/4)^2-M^2 q_0^2}\, \Gamma(q) \\
 &=& 
  \int \frac{d^4 q}{(2\pi)^4}\, \Gamma(-q)\,
 \frac{q_0^2-\epsilon_q^2 -M^2/4}
       {(q_0^2-\epsilon_q^2+M^2/4)^2-M^2 q_0^2}\, \Gamma(q) \; .
\end{eqnarray*}
Assuming that $\Gamma(q)$ is an even function in $q$ in analogy to
$G(q)$ and
Taylor-expanding the right-hand side in $M^2$, we  get (\ref{PION2}),
i.e.
\begin{eqnarray*}
 \frac{3}{4 g^2} \int d^4 x\, \frac{\Gamma^2(x)}{i{\cal D}(x)} 
 &\approx& \int \frac{d^4 q}{(2\pi)^4} \frac{\Gamma^2(q)}
  {q_0^2-\epsilon_q^2}
    + \frac{M^2}{4}\int \frac{d^4 q}{(2\pi)^4} 
 \frac{ q_0^2+3 \epsilon_q^2}{(q_0^2-\epsilon_q^2)^3}\, \Gamma^2(q) \\
 &=& \int \frac{d^4 q}{(2\pi)^4} \frac{\Gamma^2(q)}
  {q_0^2-\epsilon_q^2}\  - i M^2 F^2 \;,
\end{eqnarray*} 
where definition  (\ref{PION3}) was used.
By assuming that $\Gamma(q)=\kappa G(q)$ is an even real-valued function of 
$q_{||}$, by
Wick-rotating to Euclidean space and evaluating the resulting
$q_4$ integration
as a contour integration,
we can calculate $F^2$ defined in the last equation as follows:
\begin{eqnarray*}
F^2 &\equiv& \frac{i}{4} \int \frac{d^4 q}{(2\pi)^4}\,
\frac{q_0^2+3 \epsilon_q^2}{(q_0^2-\epsilon_q^2)^3}\, \Gamma^2(q)\\
&=& \frac{1}{4}\, 2\pi 
 \, \int_0^{2\mu} \frac{q_\perp d q_\perp}{(2\pi)^2}\  2
\int_0^{\infty} \frac{dq_{||}}{2\pi} \
 \Gamma^2(q_{||})\int_{-\infty}^{+\infty} \frac{dq_4}{2\pi}\,
\frac{-q_4^2+\epsilon_q^2}{(q_4-i\epsilon_q)^3(q_4+i\epsilon_q)^3} \\
&=& \frac{1}{4}\,\frac{\mu^2} {\pi^2}   
\int_0^{\infty} dq_{||} \  \Gamma^2(q_{||})\  \frac{2\pi i}{2\pi}\,
\frac{1}{i 2 \epsilon_q^3} \\
&=& \frac{\mu^2}{8\pi^2} \int_0^{\infty} dq_{||}  \, \
\frac{\Gamma^2(q_{||})}{\epsilon_q^3}= 
\frac{\kappa^2\mu^2}{8\pi^2} \int_0^{\infty} dq_{||}  \, \
\frac{G^2(q_{||})}{\epsilon_q^3}\approx
\frac{\kappa^2\mu^2}{8\pi^2} \int_{G_0}^{\Lambda_\ast} dq_{||}  \, \
\frac{G^2(q_{||})}{q_{||}^3}
\; .
\end{eqnarray*}
After inserting (\ref{GapSolution}) into the last equation and
shifting to the logarithmic scales $x=\ln(\Lambda_\ast/q_{||})$
and $x_0=\ln(\Lambda_\ast/G_0)$, one finally arrives at
(\ref{CFL10}), i.e.
\begin{equation}
F^2\approx\frac{\kappa^2\mu^2}{8\pi^2} \int_0^{x_0} d x \, e^{2(x-x_0)}
 \,\sin^2(\pi x/2x_0) 
=\frac{\kappa^2\mu^2}{8\pi^2}\ \frac{8 x_0^2 +\pi^2 - e^{-2x_0}\pi^2}
 {16 x_0^2 +4 \pi^2}\approx
\frac{\kappa^2\mu^2}{16\pi^2}\;,
\end{equation}
since $x_0\gg 1$, see~\cite{RWZ}.

Finally note the variational approximation
\begin{eqnarray*}
  \int \frac{d^4 q}{(2\pi)^4}\, 
 \frac{\Gamma_{T,L}^2(q)}{q_0^2-\epsilon_q^2}
 &=& 2\pi \int_0^{2\mu}\frac{q_\perp dq_\perp}{(2\pi)^2} \
 2\int_0^\infty \frac{d q_{||}}{2\pi}\, \Gamma_{T,L}^2(q_{||})
 \int \frac{i d q_4}{2\pi} \frac{1}{-q_4^2-\epsilon_q^2} \\
 &=& -i\frac{\mu^2}{\pi^2} \int_0^\infty dq_{||}\, 
  \frac{\Gamma_{T,L}^2(q_{||})}{2\epsilon_q}\\ 
 &\approx& -i\frac{\kappa^2 \mu^2}{2\pi^2}
 \int_{G_0}^{\Lambda_\ast} dq_{||}
 \,\frac{G_M^2\sin^2\left(\frac{h_\ast}{3}
        \ln\left(\frac{\Lambda_\ast}{q_{||}}\right)\right)}{q_{||}}
 \\ &=& -i\frac{\kappa^2 \mu^2 G_M^2}{2\pi^2}
 \int_0^{x_0} dx\,\sin^2\left(\frac{\pi x}{2x_M}\right) \\ 
 & =& - i \frac{\kappa^2 \mu^2 G_M^2}{4 \pi^2}\, x_0
 \left(1-\frac{x_M}{\pi x_0} \sin\left(\frac{\pi
       x_0}{x_M}\right)\right)
  \; ,
\end{eqnarray*}
which is used to derive (\ref{MV-upper}) from the vector-meson
analog of (\ref{PION2}).

%
%We may note that
%\begin{eqnarray*}
%\left(\tau^\alpha \,{{\bf M}^A}^\dagger \,{\bf M}\right)^{aa}_{ii}
%&=& (\tau^\alpha)^{ab}\,\epsilon^{Ibc}_f\, (\tau^A)^{JI}\,
%\epsilon^{J i j}_c \,\epsilon_f^{Kca}\,\epsilon^{Kji}_c\\
%&=& (\tau^\alpha)^{ab}\left(
%  \delta_{Ia}\delta_{bK}-\delta_{IK}\delta_{ab}\right)\, (\tau^A)^{JI}
%   \,\epsilon_c^{J\alpha\beta}\,\epsilon_c^{K\beta\alpha}\\
%&=&
%-2 (\tau^\alpha)^{ab}(\tau^A)^{ba}
%=-2{\rm Tr}_f \,\,(\tau^\alpha\tau^A)= -4 \delta^{\alpha A}\\
%&=&-2 (\tau^\alpha)^{ba}(\tau^A)^{ab}
%= ({\tau^\alpha}^T)^{ab}\,\epsilon^{Ibc}_f\, (\tau^A)^{IJ}\,
%\epsilon^{J i j}_c \,\epsilon_f^{Kca}\,\epsilon^{Kji}_c\\
%&=& \left({\tau^\alpha}^\ast \,{\bf M}^A\, {{\bf
%      M}}\right)^{aa}_{ii}\; ,
%\end{eqnarray*}
%so that ${\rm Tr}_{f}{\rm Tr}_{c}\,
%({\tau^\alpha} {{\bf M}^A}^\dagger {\bf M})=
%{\rm Tr}_{f}{\rm Tr}_{c}\,( {\tau^\alpha}^\ast {\bf M}^A{\bf M}^\dagger)=
%-2\, {\rm Tr}_f(\tau^\alpha\tau^A)$.
%Also, we note that
%\begin{eqnarray*}
%{\rm Tr}_s\left[ \Lambda^+({\bf Q})\Lambda^+({\bf K})\right] &=&
%{\rm Tr}_s\left[ \Lambda^-({\bf Q})\Lambda^-({\bf K})\right]
% =1+\hat{\bf Q}\cdot
% \hat{\bf K}\\
%{\rm Tr}_s\left[ \gamma_0\gamma^i \Lambda^+({\bf Q})\Lambda^+({\bf K})
% \right] &=&
%- {\rm Tr}_s\left[\gamma_0\gamma^i
% \Lambda^-({\bf Q})\Lambda^-({\bf K})\right]=
%\hat{\bf Q}^i +\hat{\bf  K}^i\; .
%\end{eqnarray*}

\vspace{1cm}
\noindent 11. Gluon polarization function in the CFL phase:
\vspace{5mm}

In the CFL phase, the gluon polarization function for the
bare gluon ($AA$) fields is
\begin{equation}
{\Pi}_{\mu\nu}^{ab} (Q) =-g^2\int\,\frac{d^4q}{(2\pi )^4}\,
{\rm Tr}\left(i{\cal V}_\mu^a\, i{\bf S} (q+\frac Q2) \,
i{\cal V}_\nu^b \, i{\bf S} (q-\frac Q2) \right)\; .
\label{HI1}
\end{equation}
Using ${\rm Tr}_{cf}(\lambda^a\lambda^b)=6\,\delta^{ab}$ and
\[
  {\rm Tr}_{cf} (\lambda^a {\bf M} {\lambda^b}^T {\bf M} )
 = - 2{\rm Tr}_c(\lambda^a \lambda^b) = - 4 \,\delta^{ab} \; , 
\]
we may write (\ref{HI1}) in the following form
\begin{eqnarray}
{\Pi}_{\mu \nu}^{ab} (Q)&=& -g^2\, \delta^{ab}\int \frac{d^4 q}{(2\pi)^4}\,
\frac{1}{(K_0^2-\epsilon_K^2)(P_0^2-\epsilon_P^2)}\nonumber \\
&&\qquad \times
\left\{\frac{\mbox{}}{\mbox{}}
 3\left(K_0 P_0+K_{||}P_{||}\right)\,{\rm Tr}\left[
\gamma_\mu\gamma^0\Lambda^-(K)\gamma_\nu\gamma^0\Lambda^-(P)\right]
 \right. \nonumber\\
&& \qquad\quad \left.
+2G(K)G(P)\, 
 {\rm Tr}\left[\gamma_\mu\Lambda^+(K)\gamma_\nu\Lambda^-(P)\right]
\frac{\mbox{}}{\mbox{}}\right\}
\; 
\label{HI11}
\end{eqnarray}
with $K=q+Q/2$ and $P=q-Q/2$. 

For the temporal polarization, we have
\begin{equation}
{\Pi}_{00}^{ab} (Q)= -g^2 \,\delta^{ab}\,\int \frac{d^4 q}{(2\pi)^4}\,
\frac{3(K_0P_0+K_{||}P_{||})+2G(K)G(P)}
{(K_0^2-\epsilon_K^2)(P_0^2-\epsilon_P^2)} 
\left(1+\hat{K}\cdot\hat{P}\right)\,\,.
\label{HI12x}
\end{equation}
For $Q=0$, this simplifies to
\begin{eqnarray}
{\Pi}_{00}^{ab} (0)&=& - g^2 \delta^{ab} \int \frac{d^4q}{(2\pi)^4}\,
 \frac{6 (q_0^2 +q_{||}^2) + 4 G^2(q)}{(q_0^2-\epsilon_q^2)^2} \nonumber\\
 &=& -\delta^{ab} \frac{g^2 \mu^2}{\pi^2} \int_0^\infty d q_{||}\,
 \int_{-\infty}^{+\infty} \frac{i\,d q_4}{2\pi}
\, \frac{ -6 q_4^2+ 6 \epsilon_q^2 -2 G^2(q)}
 { (q_4^2+\epsilon_q^2)^2} \nonumber\\
 &=& \delta^{ab} \frac{g^2 \mu^2}{\pi^2} \int_0^\infty d q_{||}\,
  \frac{i G^2(q)}{2\epsilon_q^3} = i \delta^{ab}
 \,\left(\frac{g\mu}{2\pi}\right)^2\; .
\label{HI12}
\end{eqnarray}
Here we have used  that the angle and $q_\perp$ integrations contribute
a factor $2\pi 2\mu^2$ and, after a Wick rotation to Euclidean space
and contour integration, that $\int_0^\infty d q_{||}\,
G^2(q)/\epsilon_q^3 = 1/2$ (see above). 

For the spatial polarization, we obtain
\begin{equation}
{\Pi}_{ij}^{ab} (Q)= g^2 \,\delta^{ab}\,\int \frac{d^4 q}{(2\pi)^4}\,
\frac{3(K_0P_0+K_{||}P_{||})-2G(K)G(P)}
{(K_0^2-\epsilon_K^2)(P_0^2-\epsilon_P^2)} 
\left(g_{ij} (1-\hat{K}\cdot\hat{P}) -\hat{K}_i\hat{P}_j-
\hat{K}_j\hat{P}_i\right)
\label{HI12xx}
\end{equation}
after using the spin trace
\begin{equation}
\Tr[ \gamma^i \Lambda^{\pm}({\bf K}) \gamma^j \Lambda^{\mp}({\bf P})]
= g^{ij} -(g^{im} g^{jn}-g^{ij}g^{mn} +g^{i n}g^{jm}) 
\hat {\bf K}^m \hat{\bf  P}^n\,.
\end{equation}
For $Q=0$, this simplifies
\begin{eqnarray}
{\Pi}_{ij}^{ab} (0)&=&-g^2\delta^{ab} \int \frac{d^4 q}{(2\pi)^4}\,
 \frac{6 (q_0^2+ q_{||}^2) - 4 G^2(q)}
{(q_0^2-\epsilon_q^2)^2} \, \hat{\bf q}_i \hat{\bf q}_j\nonumber \\
&=& -{\textstyle\frac{1}{3}}
g^2\delta^{ab}\delta_{ij} \int \frac{d^4 q}{(2\pi)^4}\,
 \frac{6 q_0^2+ 6\epsilon_q^2 - 10 G^2(q)}
{(q_0^2-\epsilon_q^2)^2} \nonumber \\
&=& {\textstyle\frac{5}{6}}
\delta^{ab}\delta_{ij} \frac{g^2\mu^2}{\pi^2} \int_0^\infty d q_{||}\,
\frac{i G^2(q)}{\epsilon_q^3} = i
\delta^{ab}\delta_{ij}
\frac{5 g^2\mu^2}{12\pi^2} \; .
\label{HI1x4}
\end{eqnarray}
The lack of transversality in the $AA$ polarization, which is manifest
in (\ref{HI12}) and (\ref{HI1x4}), is fixed by the mixing with the 
scalars (Higgs mechanism) and the additional contribution from the modes
within the Fermi surface (nonsurface modes).

\vspace{1cm}
\noindent 12. The comparison of the exact and
simplified in-medium gluon propagator:

\vspace{5mm}

According to eq.~(6.51) of \cite{LeBELLAC} and eq.~(10) of
\cite{SchaferWilczek}, 
the in-medium retarded (Minkowski-space) 
gluon-propagator in a general covariant gauge
reads (modulo an overall phase factor)
\begin{equation}
 {\cal D}_{\mu\nu}(q)= i\,\frac{P_{\mu\nu}^T}{q^2-G}
                       +i\,\frac{P_{\mu\nu}^L}{q^2-F}
                       -i\,\xi\, \frac{ P_{\mu\nu}^{GF}}{q^2}\; ,
 \label{GluonPropExact}
\end{equation}
where $F\equiv m_D^2 = \frac{N_f}{2\pi^2} g^2\mu^2$ ($\equiv m_E^2$) 
and $G=
\frac{\pi}{4}\, \frac{-iq_0}{|{\bf q}|}\, m_D^2$ ($\equiv m_M^2$). 
The propagator 
contains the gauge parameter $\xi$, which must not appear in physical
results.
The projectors appearing in (\ref{GluonPropExact}) read
\begin{eqnarray}
P_{\mu\nu}^T &=& (1-g_{\mu 0})(1-g_{\nu 0})\left(-g_{\mu \nu} 
 -\frac{q_\mu q_\nu}{{\bf q}^2} \right) \label{Ptransv}\; , \\
P_{\mu \nu}^L &=& -g_{\mu \nu} +\frac{q_\mu\, q_\nu}{q^2} - P_{\mu
 \nu}^T \; ,
 \label{Plong} \\
P_{\mu \nu}^{GF} &=& \frac{q_\mu \,q_\nu}{q^2}\; , \label{Pgfix}
\end{eqnarray}
where $q^2=(q^0)^2 - {\bf q}^2$ and $g_{\mu\nu} = g^{\mu\nu}= {\rm
  diag}(1,-1,-1,-1)$ for $\mu,\nu=0,1,2,3$.
The transverse projector can be written as
\begin{eqnarray}
 P_{\mu 0}^T &=& P_{0\nu}^T = 0 \; , \\
 P_{i j}^T   &=& P_{j i}^T  = \delta_{i j} - \hat {\bf q}_i \hat {\bf
 q}_j \;,
\end{eqnarray}
where $\hat {\bf q}_i \equiv q_i /|{\bf q}|$.
Eq.~(\ref{GluonPropExact}) should be compared with the simplified form
\begin{equation}
 {\cal D}_{\mu\nu}(q)= i\,\frac{1}{2}\, \frac{-g_{\mu\nu}}{q^2-G}
                     +  i\,\frac{1}{2}\, \frac{-g_{\mu\nu}}{q^2-F}
 \label{GluonPropUs}\;,
\end{equation}
which is the analog in Minkowski space of the screened perturbative
gluon
propagator in Euclidean space used here  
in (\ref{GluonProp}), see also \cite{BRWZ,RWZ}.
We now proceed to show that (\ref{GluonPropExact}-\ref{GluonPropUs})
yield equivalent results in the context of our analysis.

The three-momentum can be split into the Fermi-momentum ${\bf q}_F$
 and a momentum $\vec l$ measured relative to the Fermi surface, 
\begin{eqnarray}
 |{\bf q}| = | {\bf q}_F +{\vec l} | = \mu + l_{||}
 +\frac{l^2_\perp}{2\mu} + {\cal O}(1/\mu^2) \; ,
 \label{qlarge}
\end{eqnarray}
where $l_{||}$ and $l_\perp$ are the projections of the 
relative momentum in the direction of and orthogonal to 
the Fermi-momentum ${\bf P}$, respectively.
Because of the decomposition (\ref{qlarge}), we have, modulo $1/\mu^2$ 
corrections, 
$q^2 = q_0^2 -{\bf q}^2 \approx -{\bf q}^2$ and  
we can simplify the longitudinal 
projector as follows (see \cite{SchaferWilczek}))
\begin{eqnarray}
 P_{\mu \nu}^L &\approx& -g_{\mu 0}\, g_{\nu 0} \; .
\end{eqnarray}
Finally we will use that 
\begin{eqnarray}
\delta^{ij}  P_{i j}^T &=& 2\;, 
 \label{h1}
\\
  \hat {\bf q}\cdot \hat {\bf Q} \, \hat {\bf k}\cdot \hat {\bf Q}
 &=& -\half\,\left(1- \hat {\bf q} 
 \cdot \hat {\bf k}\right) +{\cal O}(1/\mu) ,
 \label{h2}
\end{eqnarray}
where $Q\equiv k-q$.
The second formula can be derived with the help of eq.~(\ref{qlarge})
applied to $|{\bf k}|$, $|{\bf q}|$ and  $|{\bf Q}|=|{\bf k}-{\bf q}|$, 
i.e., $|{\bf Q}|^2=|{\bf k}- {\bf q}|^2 \approx
2\mu^2 \left(1-\hat{\bf q}\cdot \hat{\bf k} \right)$
and $\hat{\bf q}\cdot {\bf Q}\, \hat{\bf k}\cdot {\bf Q} 
= (|{\bf k}|\,\hat{\bf q}\cdot \hat{\bf k}- |{\bf q}|) 
(|{\bf k}|- |{\bf q}|\,\hat{\bf q}\cdot \hat{\bf k}) 
$ $\approx -\mu^2(1-\hat{\bf q}\cdot \hat{\bf k} )^2$.

The Dirac structure of the gap equation of \cite{SchaferWilczek}
(and of the Bethe-Salpeter equation (\ref{BSeq})
{\em in case} the generalized-pion-vertex follows the sandwich rule
(\ref{Sandwich}), i.e., the
projectors $\half(1 \pm \bfalpha\cdot \hat{\bf q})$
are kept in the generalized-pion-vertex  (\ref{PSvertex})) reads:
\begin{eqnarray}
A^{\mu\nu} &\equiv& \half \Tr \left[ \gamma^\mu 
                    \half(1 -s_R \gamma^0 \gamma^m \hat{\bf q}^m)
                    \gamma^\nu 
                    \half(1 +s_L \gamma^0 \gamma^n \hat{\bf k}^n) \right ]
                    \nonumber \\
 &&= \half g^{\mu \nu} 
    + \frac{s_L}{2} \left( g^{\mu n} g^{\nu 0}-g^{\mu 0} g^{\nu n}\right)
                    \hat{\bf k}^n
  + \frac{s_R}{2} \left( g^{\mu m} g^{\nu 0}-g^{\mu 0} g^{\nu m}\right)
                    \hat{\bf q}^m \nonumber  \\
 && \  +\frac{s_R s_L}{2} \left( 2 g^{\mu 0} g^{\nu0} \delta^{mn}
                      -g^{\mu \nu} \delta^{mn}
                           - g^{\mu m} g^{\nu n} - g^{\mu n} g^{\nu m}
                     \right) \hat{\bf q}^m \hat{\bf k}^n \; ,
 \label{tracewith}
\end{eqnarray}
where $s_L = s_R = \pm 1$ for the gap and $s_L= -s_R = \pm 1$ 
for the anti-gap, see eq.~(9) of \cite{SchaferWilczek}.

The Dirac structure of the Bethe-Salpeter equation (\ref{BSeq})
{\em
  without}  the projectors $\half(1\pm\bfalpha\cdot \hat{\bf q})$ 
in the generalized-pion-vertex (\ref{PSvertex}) reads
\begin{eqnarray}
   B^{\mu\nu} &=& 
 {\textstyle\frac{1}{4}} \Tr[ \gamma^\mu \half(1 \mp \gamma^0 \gamma^m
 \hat{\bf q}^m ) \gamma^\nu] \nonumber \\
  &=& \half g^{\mu \nu} \mp \half\left(g^{\mu 0} g^{m \nu} -g^{\mu m} g^{\nu
      0}\right) \hat{\bf q}^m \; .
 \label{tracewithout}
\end{eqnarray}
Here, the projector $ \half(1 \mp \gamma^0 \gamma^m
\hat{\bf q}^m )$ results solely from the leading parts of the quark
propagators (\ref{PropPart}) and not from the
generalized-pion vertex. 
The prefactor $\frac{1}{4}$ in (\ref{tracewithout}) 
versus the prefactor $\half$ in 
(\ref{tracewith}) can be traced back to the division by 
the Dirac trace on the l.h.s.\ of the Bethe-Salpeter equation, namely
to the division by 
$\Tr[1] =4 $ in (\ref{tracewithout}) 
versus $\Tr[\half(1+s_L \gamma^0 \gamma^n \hat{\bf
  k}^n)]=2$ in (\ref{tracewith}).

Using the projectors of ${\cal D}_{\mu\nu}(k-q)$ as given in 
(\ref{GluonPropExact}),
we get
\begin{eqnarray}
 A^{\mu\nu } P^L_{\mu \nu} =- A^{00} = -\half \left(1 +s_R s_L\, \hat
 {\bf q} \cdot \hat{\bf k} \right) \; .
 \label{AL}
\end{eqnarray}
Assuming as in \cite{SchaferWilczek} 
that $\hat{\bf q} \cdot \hat{\bf k}  \equiv \cos(\theta)
\approx 1$ in the {\em numerators} of the gap-equation, we get the
weight $-1$ for the longitudinal contribution to the gap and 0 for the
longitudinal contribution to the anti-gap.
This should be compared with
\begin{equation}
 B^{\mu \nu} P^L_{\mu \nu} = - \half 
 \label{BL}
\end{equation}
for both the gap and the anti-gap. Note that (\ref{BL}) is just the
average of (\ref{AL}).

Furthermore, we have using (\ref{h2})
\begin{eqnarray}
 A^{\mu\nu } P^T_{\mu \nu}&=& \half g^{ij} P^T_{ij}
+ \frac{s_Rs_L}{2} \left( - 2 P_{mn}^T +\delta^{ij} P_{ij}^T
  \delta^{mn}\right ) \hat{\bf k}^n \hat{\bf q}^m \nonumber \\
&=& - 1 +s_Rs_L\left( \hat{\bf k}\cdot \hat{\bf q}- \hat{\bf k}\cdot
  \hat{\bf q}
 +  \hat{\bf k}\cdot \hat{\bf Q}\,  \hat{\bf q}\cdot \hat{\bf Q}
\right) \nonumber \\
&\approx& -1 -\half s_R s_L +\half s_R s_L \, \hat{\bf k}\cdot \hat{\bf
  q} \;.
\end{eqnarray}
Under the approximation $  \hat{\bf k}\cdot \hat{\bf q} \approx 1$, we
get for the gap-case $ -\frac{3}{2} +\half \hat{\bf k}\cdot \hat{\bf q}
 \approx -1$ and for the anti-gap
case $ -\half -\half \hat{\bf k}\cdot \hat{\bf q} \approx -1$.
This agrees with the unprojected case 
\begin{equation}
B^{\mu \nu} P_{\mu \nu}^T = - \half \delta^{ij} P_{ij}^T = -1 
 \label{BT}
\end{equation}
for both the gap and the anti-gap.

Finally,  
\begin{eqnarray}
  A^{\mu\nu } P^{GF}_{\mu \nu}
 &=& \half \frac{g^{\mu\nu}Q_\mu Q_\nu}{Q^2} +\frac{s_R s_L}{2}
\left( 2 \frac{Q^0 Q^0}{Q^2} \,\hat{\bf k}\cdot \hat{\bf q} 
- 2 \frac{{\bf Q} \cdot \hat{\bf k}\, {\bf Q}\cdot \hat{\bf q}}{Q^2}-
\frac{Q^2}{Q^2}\,  \hat{\bf k}\cdot \hat{\bf q} \right) \nonumber \\
&=& \half - \frac{s_Rs_L}{2} \, \hat{\bf k}\cdot \hat{\bf q} 
+s_R s_L \left( \hat{\bf k}\cdot \hat{\bf q} \, \frac{{Q^0}^2}{Q^2}-
  \frac{{\bf Q} \cdot \hat{\bf k}\, {\bf Q}\cdot \hat{\bf q}}{Q^2}
 \right)
 \nonumber \\
&\approx&  \half - \frac{s_Rs_L}{2} \, \hat{\bf k}\cdot \hat{\bf q} 
+s_R s_L \left( - \frac{{\bf Q} \cdot \hat{\bf k}\, {\bf Q}\cdot
 \hat{\bf q}}
 {-{\bf Q}^2}
 \right) \nonumber \\
&\approx& \half -\frac{s_Rs_L}{2} \, \hat{\bf k}\cdot \hat{\bf q} 
 +s_R s_L \left( \half \hat{\bf k}\cdot \hat{\bf q}  -\half \right)
 \nonumber \\
&=& \half\left(1 -s_R s_L\right) \; ,
 \label{AGF}
\end{eqnarray}
where $Q^2 \approx - {\bf Q}^2$ and (\ref{h2}) was used. 
Note that the gauge-fixing dependence vanishes for the gap and gives a
weight factor $+1$ for the anti-gap.   

This should be compared with 
\begin{equation}
B^{\mu \nu} P_{\mu \nu}^{GF} = +\half 
 \label{BGF}
\end{equation}
for both the gap and the anti-gap. Again, the  result 
(\ref{BGF}) of the unprojected case is the average 
of the sandwiched one, (\ref{AGF}).

If $A^{\mu\nu}$ is contracted  with $-\half g_{\mu\nu}$ as in the
gluon-propagator (\ref{GluonPropUs}) and in the Euclidean analog 
(\ref{GluonProp}), 
we get
\begin{eqnarray}
 A^{\mu\nu}\,\frac{-1}{2}\,g_{\mu\nu} 
&=& - 1 + \frac{s_R s_L}{2} \left( -g^{00}\, \hat{\bf q}\cdot\hat{\bf k}
 +2 \, \hat{\bf q}\cdot\hat{\bf k} + \half g^{mn}  \hat{\bf q}^m
 \hat{\bf k}^n + \half g^{mn}  \hat{\bf q}^m
 \hat{\bf k}^n\right) \nonumber \\
&=& -1 +\frac{s_R s_L}{2} \left(-\hat{\bf q}\cdot\hat{\bf k}+2\,
  \hat{\bf q}\cdot\hat{\bf k}- \hat{\bf q}\cdot\hat{\bf k}\right)
\nonumber\\
&=& -1
\end{eqnarray}
for both the gap and the anti-gap.
  
If $B^{\mu\nu}$ is contracted with $-\half g_{\mu\nu}$,
then 
\begin{equation}
 B^{\mu\nu}\,\frac{-1}{2}\,g_{\mu\nu} =-{\textstyle\frac{1}{4}}
\,g^{\mu\nu}g_{\mu\nu}=-1 
 \label{Bwithout}
\end{equation}
for both the gap and the anti-gap.

In summary: the use of the simplified propagator (\ref{GluonProp})
together with the {\em unprojected} meson vertices, i.e. (\ref{Sandwich})
without the additional projectors, yields the same results as the use
of the exact propagator (\ref{GluonPropExact}) (or even the simplified
propagator)
together with the  {\em sandwiched}  meson vertices (\ref{Sandwich}),
to leading order. At higher order, the simplifications 
require amendments. 
To leading order, the differences resulting from (\ref{tracewith})
and (\ref{tracewithout}) can be traced back to the presence 
or absence of the projector $\Lambda^{\pm}({\bf k})$ 
on the l.h.s.\ of the Bethe-Salpeter
equation,
i.e.\ at the {\em amputated} vertex.

%The outcome of the 
%projections on $A^{\mu \nu}$ and $B^{\mu \nu}$~\footnote{
%especially  eqs.~(\ref{BL}) and (\ref{BGF}) which are the average of
%the gap and anti-gap contribution of the projections on $A^{\mu \nu}$}
%shows that our use of the 
%Bethe-Salpeter equation {\em without} the projectors in 
%the generalized-pion-vertex (\ref{PSvertex}) 
%cannot be used  with the exact propagator
%(\ref{GluonPropExact}),
%but only with (\ref{GluonProp}) [and the latter {\em only}  
%in the case where the {\em particle}-gap is considered]. 

%The  Bethe-Salpeter equation {\em with} projectors in the
%generalized-pion-vertex (\ref{PSvertex})  can
%be used with the exact propagator (\ref{GluonPropExact}) or
%with the approximated (\ref{GluonProp}) with identical 
%outcome, provided that in the latter only the particle
%gaps are involved, which 

\end{appendix}

\end{document}